\documentclass[lettersize,journal]{IEEEtran}
\usepackage{amsmath,amsfonts}
\usepackage{algorithmic}
\usepackage{algorithm}
\usepackage{array}
\usepackage[caption=false,font=normalsize,labelfont=sf,textfont=sf]{subfig}
\usepackage{textcomp}
\usepackage{stfloats}
\usepackage{url}
\usepackage{soul}
\usepackage{siunitx}
\usepackage{booktabs} 
\usepackage{makecell}
\usepackage{verbatim}
\usepackage{graphicx}
\usepackage{cite}
\usepackage{longtable}  
\usepackage{enumitem}
\usepackage{afterpage}
\usepackage{supertabular}
\usepackage{multicol}
\usepackage{pdflscape}

\usepackage{hyperref}
\hypersetup{
    colorlinks=true,
    linkcolor=blue,
    filecolor=magenta,      
    urlcolor=blue,
}

\begin{document}


\title{URLLC for 6G Enabled Industry 5.0: A Taxonomy of Architectures, Cross Layer Techniques, and Time Critical Applications}

\author{
    Abdikarim Mohamed Ibrahim,
    Rosdiadee Nordin,~\IEEEmembership{Senior Member,~IEEE,}, Yahya S.M. Khamayseh, Angela Amphawan, ~\IEEEmembership{Senior Member,~IEEE,}, Muhammed Basheer Jasser, ~\IEEEmembership{Senior Member,~IEEE,}  
    \thanks{
        Abdikarim Mohamed Ibrahim, Rosdiadee Nordin (corresponding author), Angela Amphawan and Muhammed Basheer Jasser are with Sunway University, Malaysia.
    }
    \thanks{
        Manuscript received June 2025; revised July 2025.
    }
}

\markboth{Journal of \LaTeX\ Class Files,~Vol.~14, No.~8, August~2021}%
{Shell \MakeLowercase{\textit{et al.}}: A Sample Article Using IEEEtran.cls for IEEE Journals}

\IEEEpubid{0000--0000/00\$00.00~\copyright~2021 IEEE}

\maketitle

\begin{abstract}
The evolution from Industry 4.0 to Industry 5.0 introduces stringent requirements for ultra reliable low latency communication (URLLC) to support human centric, intelligent, and resilient industrial systems. Sixth-generation (6G) wireless networks aim to meet these requirements through sub-millisecond end-to-end delays, microsecond level jitter, and near perfect reliability, enabled by advances such as terahertz (THz) communication, reconfigurable intelligent surfaces (RIS), multi-access edge computing (MEC), and AI driven cross layer optimization. This paper presents a comprehensive review of URLLC solutions for 6G enabled industry 5.0, organized into a structured taxonomy including application domains, key technical enablers, design challenges, and performance enhancements. The survey examines emerging approaches, including digital twin integration, AI/ML based resource orchestration, Network Function Virtualization (NFV) enabled service function chaining, and cross domain networking, while mapping them to critical industrial scenarios such as smart manufacturing, connected healthcare, autonomous mobility, remote control, and next-generation mobile networks. Performance trade-offs between latency, reliability, scalability, and energy efficiency are analyzed in the context of representative state-of-the-art studies. Finally, the paper identifies open challenges and outlines future research directions to realize deterministic, secure, and sustainable URLLC architectures for Industry 5.0. 
 
\end{abstract}

\begin{IEEEkeywords}
6G, Industry 5.0, ultra reliable low latency communication (URLLC), cross-layer optimization, digital twin.
\end{IEEEkeywords}

\section{Introduction}\label{intro}

\IEEEPARstart{T}{he} transition from Industry 4.0 to Industry 5.0 marks a fundamental shift in manufacturing and industrial automation, emphasizing sustainability, human-centric design, and system level resilience \cite{tallat2023navigating}. Industry 4.0 introduced cyber physical systems, the Internet of Things (IoT), and cloud based automation but does not meet the strict latency, reliability, and synchronization demands of next-generation industrial applications. Specifically, it cannot satisfy requirements such as \SI{1}{\milli\second} end-to-end delay, \SI{99.9999}{\percent} reliability, and deterministic timing essential for collaborative robotics, closed loop motion control, and distributed digital twins \cite{asad2023human,langaas2025exploring}. Industry 5.0, by contrast, relies on ultra reliable low latency communication (URLLC) as a foundational service to coordinate real-time interaction between humans, machines, and artificial intelligence (AI) agents.

\begin{figure*}[t]
  \centering
  \includegraphics[width=\linewidth]{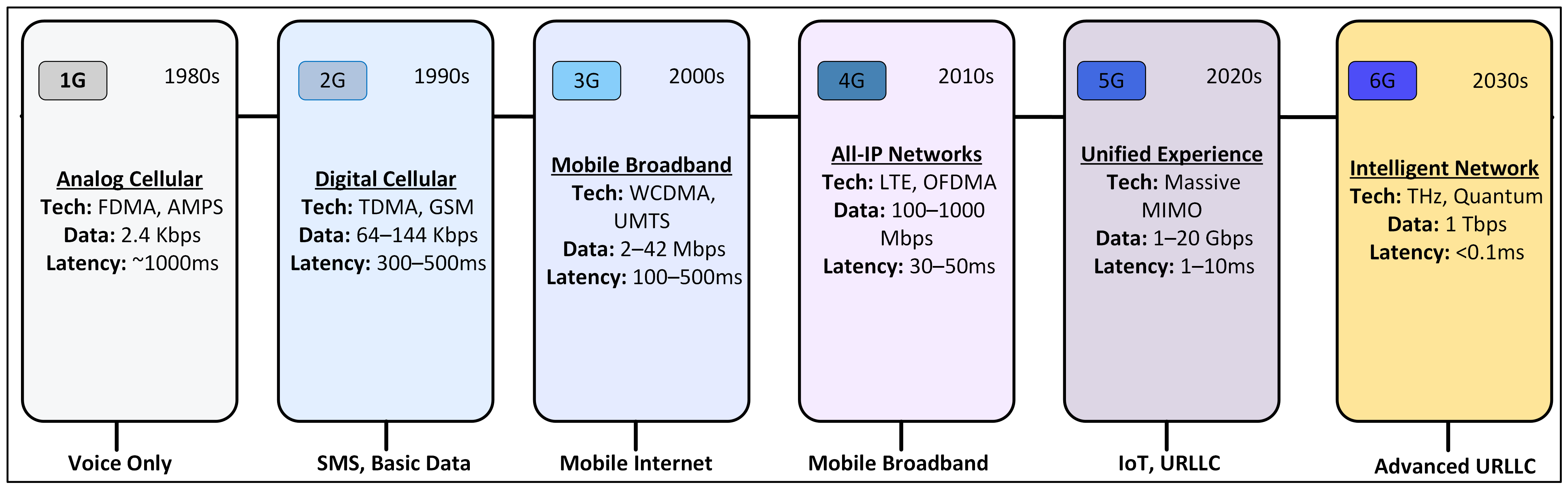}
  \caption{Evolution of mobile communication generations from 1G to 6G. Each generation reflects significant advances in latency, data rate, and service capabilities. 1G and 2G supported voice and basic data services. 3G and 4G enabled mobile internet and broadband experiences. 5G introduced unified support for IoT and URLLC. 6G aims to achieve sub-\SI{0.1}{\milli\second} latency and \SI{1}{\tera bps} throughput to support intelligent, real-time applications in Industry~5.0.}
  \label{fig:evolution_1g_6g}
\end{figure*}

Sixth-generation (6G) wireless networks aim to achieve ultra-low latency communication with over the air delays below \SI{0.1}{\milli\second}, synchronization accuracy better than \SI{100}{\nano\second}, and microsecond-level jitter \cite{shamsabadi2025exploring,kerboeuf2024design}. These goals are enabled by emerging technologies such as terahertz (THz) communication, reconfigurable intelligent surfaces (RIS), and AI-driven cross-layer optimization. As shown in Figure~\ref{fig:evolution_1g_6g}, 6G builds upon the progression from 1G to 5G, introducing new features to support applications with extreme performance requirements. These advancements are expected to support time critical industrial tasks, including sub-\SI{1}{\milli\second} emergency stops in human–robot collaboration, real-time remote control, and distributed motion coordination. However, meeting these strict requirements in complex industrial settings introduces trade-offs between latency, reliability, energy consumption, and security. For instance, edge device co-inference strategies can reduce perception latency by offloading computation to nearby servers, but may increase local power consumption due to the added overhead of data transmission and synchronization \cite{zhao2018deepthings}. Similarly, implementing physical layer security mechanisms introduces additional processing that can increase latency, which may interfere with the performance of time sensitive applications \cite{wang2019physical,yang2025towards}.

The convergence of Industry 5.0 and 6G communication systems is emerging as a foundational enabler for real-time, resilient, and human-centric industrial automation \cite{ghobakhloo2023behind,dang2020should}. Realizing the vision of Industry 5.0 centered on intelligent human machine collaboration and sustainable, adaptive manufacturing requires communication infrastructures that can deliver deterministic performance at scale. Technologies such as AI, IoT, digital twins, and time sensitive networking (TSN) rely on URLLC to ensure closed loop responsiveness, accurate synchronization, and predictive operations \cite{yang2025towards,chowdhury20206g}. These technologies form the shared backbone of 6G and Industry 5.0 systems, providing the capability to manage autonomous agents, synchronize distributed edge intelligence, and ensure uninterrupted safety and performance in real-world industrial environments.

However, integrating latency sensitive technologies such as AI and 6G into Industry 5.0 deployments introduces several practical challenges. These include increased complexity in network orchestration, uncertainty in data quality and availability, and stringent energy constraints at the device and edge levels \cite{verma2022blockchain,hazra20246g,jahid2023convergence}. In domains such as supply chain logistics, where near instantaneous traceability and responsiveness are critical, blockchain based ledgers have shown potential for secure tracking but suffer from latency and scalability limitations \cite{babaei2025assessing}. Meanwhile, predictive models deployed using federated learning or edge cloud AI pipelines must minimize inference delays and balance privacy and bandwidth costs \cite{chataut20246g}. Overcoming these issues is essential to build 6G enabled Industry 5.0 systems that meet the dual goals of ultra-low latency and system wide resilience.

With the increasing significance of URLLC in 6G-enabled industrial systems, several recent survey papers have explored its foundational concepts, technical enablers, and challenges. Hazra et al. \cite{hazra20246g} provide a comprehensive review of 6G enabled URLLC for Industry 5.0, emphasizing the importance of edge computing to minimize end-to-end latency and improve system reliability. Their study highlights the architectural challenges of deploying low latency services in data intensive industrial environments. Mahmood et al. \cite{mahmood2023ultra} provide a comprehensive overview of URLLC foundations, system design principles, and evolution toward 6G, including key enabling technologies such as flexible numerology, grant-free access, and network slicing. Their work distinguishes between physical and MAC layer techniques and emphasizes latency reliability trade-offs for mission critical scenarios. Haque et al. \cite{haque2024comprehensive} survey the transition from 5G URLLC to 6G, focusing on physical layer enhancements, MAC layer scheduling, and cross layer orchestration. Their analysis identifies future challenges in maintaining latency and reliability guarantees under dynamic industrial loads. Shamsabadi et al. \cite{shamsabadi2025exploring} introduce the concept of immersive, hyper reliable, low latency communication  as a candidate service class for 6G. This model integrates ultra reliability and deterministic latency with high data rate requirements. Salh et al. \cite{salh2021survey} present a dedicated review on the integration of deep learning (DL) techniques for addressing URLLC challenges in 6G wireless systems. They propose a multi level architecture where device, edge, and cloud intelligence collaboratively reduce delay and improve reliability through predictive learning, edge inference, and unsupervised optimization. Sefati and Halunga \cite{sefati2023ultra} focus on URLLC in the context of the IoT in industrial environments. Their work categorizes existing URLLC techniques into four main approaches: structure based, diversity based, metaheuristic based, and channel state information methods. The survey also highlights challenges such as energy efficiency, cost, and QoS trade-offs in latency sensitive IoT deployments.  

These review efforts collectively reflect the increasing momentum toward enabling low latency, high reliability communication in support of Industry 5.0. Each contributes insights into foundational concepts, enabling technologies, and architectural considerations for URLLC in future networks. Building on these contributions, this paper extends prior work by introducing a structured taxonomy of cross layer latency enabling attributes, examining trade-offs in security, reliability, and mobility management, and mapping these attributes to critical industrial domains including smart manufacturing, remote operations, connected healthcare, and next-generation mobile networks. Table~\ref{tab:comparison} summarizes the scope, focus, and contributions of recent URLLC surveys and situates the unique contribution of this paper within that landscape.

\begin{table*}[!t]
\caption{Comparison of Existing Surveys on 6G URLLC for Industrial Automation}
\label{tab:comparison}
\centering
\begin{tabular}{@{}p{2.5cm}p{3cm}p{4cm}p{6cm}@{}}
\toprule
\textbf{Study} & \textbf{Focus Area} & \textbf{Key Technologies Discussed} & \textbf{Unique Contribution} \\
\midrule
Shamsabadi et al. \cite{shamsabadi2025exploring} & Immersive URLLC in 6G & IHRLLC, THz, RIS, high data-rate links & Proposes new URLLC service class integrating high reliability and low latency \\
Hazra et al. \cite{hazra20246g} & 6G URLLC for Industry 5.0 & Edge computing, real-time analytics, network offloading & Overview of latency reduction strategies and reliability enhancement in smart factories \\
Mahmood et al. \cite{mahmood2023ultra} & URLLC foundations and evolution to 6G & Flexible numerology, slicing, latency reliability trade-offs & Detailed analysis of system-level enablers for ultra-low latency in 6G \\

Haque et al. \cite{haque2024comprehensive} & Evolution from 5G to 6G URLLC & Cross-layer design, MAC scheduling, link adaptation & Identifies physical and MAC layer bottlenecks in meeting URLLC requirements \\

Salh et al. \cite{salh2021survey} & Deep learning for 6G URLLC & Multi-level DL architectures, edge-cloud cooperation & Proposes DL-driven predictive learning architecture to reduce latency and improve reliability \\
Sefati and Halunga \cite{sefati2023ultra} & URLLC in IoT over 5G/6G & CSI, diversity-based, metaheuristic methods & Classifies URLLC techniques for IIoT, identifies energy/QoS trade-offs \\
Our Review & 
Cross-layer URLLC for Industry 5.0 & 
Latency performance trade-offs, deterministic scheduling, mobility management protocols, AI/edge orchestration & 
Structured taxonomy, cross domain attribute mapping, industrial application alignment (e.g., smart manufacturing, remote control, healthcare, mobile networks) \\

\bottomrule
\end{tabular}
\end{table*}

The remainder of this paper is organized as follows. Section~\ref{background} provides the background and key performance requirements of URLLC. Section~\ref{sec:taxonomy} introduces a structured taxonomy of URLLC solutions for 6G enabled Industry~5.0. Section~\ref{main} discusses emerging enablers including intelligent reflecting surfaces, Unmanned Aerial Vehicle (UAV), digital twins, AI/ML techniques, and cross-layer solutions. Section~\ref{sec:challenges} outlines open challenges and future research directions across semantic resource orchestration, adaptive digital twins, cross domain consistency, safe learning, and low latency control co-design. Section~\ref{conclusion} concludes the paper. To ensure clarity in the technical discussions that follow, Table~\ref{tab:terminology} provides an overview of key acronyms and  terms in alphabetical order used throughout the paper.

\begin{table}[htbp]
\centering
\caption{Table of Abbreviations}
\label{tab:terminology}
\begin{tabular}{ll}
\toprule
\textbf{Abbreviation} & \textbf{Full Term} \\
\midrule
6G & Sixth-Generation \\
AI & Artificial Intelligence \\
AR & Augmented Reality \\
CCAM & Connected, Cooperative, and Automated Mobility \\
CSI & Channel State Information \\
CUDU & Central Unit Distributed Unit \\
D2D & Device-to-Device \\
DDPG & Deep Deterministic Policy Gradient \\
DDQN & Double Deep Q Network \\
DL & Deep Learning \\
DNN & Deep Neural Network \\
DRL & Deep Reinforcement Learning \\
EC & Effective Capacity \\
eMBB & Enhanced Mobile Broadband \\
FBC & Finite-Blocklength Coding \\
F-RAN & Fog-Radio Access Network \\
HARQ & Hybrid Automatic Repeat reQuest \\
IoHT & Internet of Health Things \\
IoT & Internet of Things \\
JT-CoMP & Joint Transmission Coordinated Multipoint \\
MAC & Medium Access Control \\
MDP & Markov Decision Process \\
MEC & Multi-access Edge Computing \\
MIMO & Massive Multiple-Input Multiple-Output \\
MIoT & Maritime Internet of Things \\
ML & Machine Learning \\
NFV & Network Function Virtualization \\
OLT & Optical Line Terminal \\
QoE & Quality of Experience \\
QoS & Quality of Service \\
RB & Resource Block \\
RIS & Reconfigurable Intelligent Surfaces \\
RL & Reinforcement Learning \\
ROS & Robot Operating System \\
RTL & Round-Trip Latency \\
SFC & Service Function Chaining \\
SGZ & Secrecy Guard Zone \\
SDN & Software-Defined Networking \\
SNR & Signal-to-Noise Ratio \\
THz & Terahertz \\
TSN & Time Sensitive Networking \\
UAV & Unmanned Aerial Vehicle \\
UE & User Equipment \\
URLLC & Ultra-Reliable Low-Latency Communication \\
V2X & Vehicle-to-Everything \\
VNF & Virtual Network Function \\
\bottomrule
\end{tabular}
\end{table}


\section{Background on URLLC} \label{background}
URLLC is a foundational service class introduced in 5G and evolved further in 6G, designed to support mission critical applications with stringent end-to-end latency (sub-ms) and extremely high reliability (up to 99.9999\%) requirements \cite{husain20223gpp}. Table~\ref{tab:technical_overview} provides a summary of key enabling technologies that contribute to URLLC performance in terms of latency, reliability, and integration constraints. These techniques span multiple layers, from physical layer optimization to AI based control and cross layer designs. 

\begin{table*}[t]
\centering
\caption{Characteristics and Constraints of 6G Enabled URLLC Techniques}
\label{tab:technical_overview}
\begin{tabular}{@{}p{3.2cm}p{3.5cm}p{4.5cm}p{4.5cm}@{}}
\toprule
\textbf{Category} & \textbf{Technique} & \textbf{Key Benefits} & \textbf{Implementation Constraints} \\
\midrule
Physical layer design & Short packet transmission & Reduces transmission delay and buffering time \cite{mahmood2023ultra} & Requires new coding schemes and physical layer redesign  \\
Antenna systems & Massive/Cell-Free MIMO & Improves reliability and spatial diversity & Needs precise synchronization and inter-node coordination \\
High frequency access & THz communication & Supports ultra-high data rates \cite{sefati2023ultra} & Suffers from path loss and alignment sensitivity; benefits from RIS \\
Propagation control & RIS & Enhances NLoS coverage and improves THz reliability \cite{chen2019channel} & Real-time channel estimation remains a bottleneck \\
AI integration & Deep/RL based learning & Predicts traffic, adapts resource allocation, enhances security \cite{liu2024network,ibrahim2023multi} & Requires training data, raises computational load \\
Cross layer optimization & PHY–MAC–App coordination & Enables QoS aware control, reduces latency \cite{she2017cross} & Demands tight vertical integration and runtime adaptability \\
Security mechanisms & Physical layer authentication & Provides low layer trust via RF fingerprinting  & Computational cost can conflict with power budgets in mobile scenarios \\
\bottomrule
\end{tabular}
\end{table*}

\subsection{Foundations of URLLC: Latency, Reliability, and QoS Targets}

Unlike traditional mobile broadband services that prioritize data throughput, URLLC emphasizes two key performance requirements: a) extremely low end-to-end latency; and b) very high reliability. Specifically, URLLC targets latencies as low as 1~ms with reliability levels up to 99.999\%, making it essential for industrial automation, robotics, smart factories \cite{peng2022resource}, autonomous systems \cite{chang2021autonomous}, and remote healthcare \cite{zhou2020learning}. These performance requirements are based on critical application domains such as industrial automation, remote control, and autonomous systems, where any packet loss or delay could lead to safety hazards or operational failure. However, specific latency and reliability constraints can vary depending on the use case. For instance, most URLLC services aim for under \SI{1}{\milli\second} end-to-end latency; however, the physical layer itself, depending on the use case, may need to complete transmission within \SI{0.5}{\milli\second}, accounting for frame structure design, synchronization, and error checking \cite{le2020overview}. Similarly, end-to-end RTL which includes transmission, processing, and acknowledgment delays across all layers, is an important performance metric. For haptic feedback in the Tactile Internet, the RTL must stay below 20~ms to preserve user perception and control accuracy \cite{mourtzis2021smart}. Reducing such latency requires optimizing every component of the communication chain, including rapid packet acquisition, fast decoding, minimal retransmission cycles, and tight cross layer coordination.

\subsection{Design Tradeoffs and Short Packet Transmission}

Achieving ultra low latency and ultra high reliability simultaneously presents fundamental tradeoffs at both the physical and system layers. High reliability depends on robust mechanisms such as retransmissions and error correction, which inherently introduce delays. Conversely, reducing latency necessitates aggressive scheduling, minimized retransmission cycles, and streamlined protocol stacks, which comes at the cost of reduced redundancy and error tolerance. Therefore, URLLC design departs from traditional metrics based on average throughput and instead focuses on worst case or outage based reliability under strict time constraints. A core enabling mechanism is the use of short packet transmission. By reducing the size of each transmission, the system achieves faster delivery and reduced decoding delay, at the expense of coding efficiency and spectral utilization \cite{enenche2023network}. This tradeoff is significant in delay sensitive use cases, where every microsecond of savings is critical. Short packets also mitigate the time consumed by channel acquisition and control signaling, further supporting low latency operation. However, these benefits require revisiting the physical layer architecture in terms of the coding, modulation, and frame structures, to preserve reliability under tight latency budgets.

\subsection{Resource Allocation and QoS Aware Scheduling}

Resource allocation is central to maintaining QoS in URLLC systems under dynamic and resource constrained environments. Unlike traditional best effort traffic management, URLLC demands real-time responsiveness with strict latency and reliability guarantees. This necessitates proactive and context aware resource orchestration strategies \cite{tamim2023intelligent}. 

QoS aware scheduling must account for rapid variations in network traffic, channel conditions, and user mobility. MAC layer schedulers need to be tightly coupled with physical layer Channel State Information (CSI) to adapt transmission schedules in real-time. Conventional FIFO or round-robin scheduling mechanisms fall short under URLLC constraints. Instead, latency sensitive schedulers such as earliest-deadline-first or weighted fair queuing with dynamic priority adaptation are used to ensure bounded delay and minimal jitter \cite{queiroz2024new}. Furthermore, intelligent resource allocation strategies enable service continuity during network congestion or user contention. These strategies dynamically assign spectrum, computing power, and buffer space based on urgency, reliability class, and application requirements. The integration of machine learning (ML) based predictors can further improve efficiency by anticipating traffic bursts, link failures, or mobility induced handovers. In 6G oriented URLLC systems, emerging service classes such as extreme URLLC, scalable URLLC, and broadband URLLC require differentiated QoS provisioning. Supporting these diverse traffic profiles in parallel calls for advanced slicing mechanisms, multi objective optimization, and agile control loops to meet application specific latency, reliability, and throughput demands \cite{she2020deep, zeb2021edge}. These principles provide the basis for understanding how URLLC supports emerging real-time applications across industrial and societal domains.

\subsection{Machine Learning and Reinforcement Learning for URLLC}

URLLC’s stringent latency and reliability requirements challenge traditional static optimization methods, which rely on precomputed configurations and often fail to adapt to dynamic environments such as mobile nodes, interference, or fast fading channels. To address this, AI based techniques such as Deep Learning (DL) and Reinforcement Learning (RL), are being integrated into URLLC systems. DL models support channel prediction, anomaly detection, and traffic forecasting. By learning from historical data, DL can anticipate signal degradation, detect outliers, and optimize modulation and coding schemes in real time. This enables preemptive adjustments to communication parameters before service quality deteriorates, thereby minimizing retransmissions and delay. RL, on the other hand, is suited for adaptive resource allocation in highly dynamic settings. RL agents can be embedded in a centralized or distributed manners to learn optimal policies for spectrum scheduling, power control, or routing under uncertain and time varying network conditions \cite{ibrahim2022implications}. In URLLC systems, RL enables real-time trade-off management between latency and reliability, improving resource efficiency while maintaining service constraints.

However, these benefits come at a cost. Training large models requires extensive datasets and high computational resources. This overhead  conflicts with energy and time constraints inherent in edge based URLLC deployments. As such, lightweight model design, federated learning, and hardware acceleration (e.g., AI inference at the edge) are active research directions to ensure compatibility with URLLC’s stringent operational envelope. Table~\ref{tab:ai_urllc_mapping} summarizes key AI techniques and their specific advantages, limitations, and functional roles within URLLC deployments.

\begin{table*}[t]
\centering
\caption{AI Techniques and Their Roles in URLLC Systems}
\label{tab:ai_urllc_mapping}
\begin{tabular}{p{4.3cm} p{5.2cm} p{5.5cm}}
\toprule
\textbf{AI Technique} & \textbf{Primary Application in URLLC} & \textbf{Key Benefits and Constraints} \\
\midrule
Supervised learning (e.g., CNNs, RNNs) & Channel quality prediction, anomaly detection, traffic classification & High accuracy; requires labeled data and high training cost \\
Unsupervised learning (e.g., K-Means, Autoencoders) & Intrusion detection, clustering for resource optimization & No labels needed; lower interpretability and stability \\
DL & Dynamic channel estimation, beamforming prediction & Enables real-time adaptation; computationally intensive \\
RL (e.g., Q-learning, DQN, PPO) & Adaptive scheduling, power control, mobility management & Online learning under uncertainty; slow convergence, exploration risks \\
Federated learning & Privacy preserving collaborative model training across edge devices & Maintains data privacy; suffers from model drift and device heterogeneity \\
Transfer Learning & Accelerates model convergence in new URLLC scenarios & Reduces training cost; requires good source domain alignment \\
Online Learning & Lightweight real-time decision making in volatile environments & Low latency; limited by incremental learning stability \\
\bottomrule
\end{tabular}
\end{table*}

\subsection{Key Technology Enablers}

Several foundational technologies are critical to achieving the strict requirements of URLLC in 5G and future 6G networks. Advanced antenna systems, such as MIMO and cell-free massive MIMO, enable high spatial diversity and improved signal robustness through coordinated transmission, including Joint Transmission Coordinated Multipoint (JT-CoMP) \cite{ostman2021urllc, shi2022decentralized}. These systems reduce outage probabilities and enhance reliability, especially in dense industrial environments. Additionally, high frequency bands, such as millimeter-wave and THz, offer wide bandwidths necessary to support high data rates and bursty traffic associated with real-time control systems. However, THz signals suffer from high propagation loss and beam alignment sensitivity, requiring precise beamforming and robust link maintenance strategies \cite{farhad2023terahertz, alhulayil2025integrated}.

RIS have emerged as a key enabler to enhance wireless propagation, especially in non-line-of-sight or high frequency conditions. By dynamically adjusting the phase of reflected signals, RIS improve link quality and coverage \cite{amodu2024technical}. However, challenges such as real-time channel estimation and integration complexity remain \cite{chen2019channel}. For reliability enhancement, error control mechanisms such as Hybrid Automatic Repeat reQuest (HARQ) and short blocklength channel codes are employed. These approaches must be carefully tuned, as retransmissions can conflict with latency budgets when round trip delays exceed 1~ms in fast fading environments \cite{nadas2019performance}. Cooperative ARQ, utilizing relay nodes, can mitigate such effects in fading-prone or obstructed links \cite{nadas2019performance}.

Critically, URLLC operation falls within the finite blocklength scenarios, where the classical Shannon capacity model no longer holds. In this scenario (i.e., finite-blocklength coding (FBC)), the maximum achievable data rate is reduced due to short codeword lengths and stringent error constraints. This necessitates the use of information theoretic models, which account for decoding error probability and blocklength explicitly. These effects become significant in systems using short packets to minimize transmission delay.  To meet strict latency and reliability requirements under dynamic industrial conditions, cross layer design is increasingly adopted in URLLC systems. This approach integrates information across multiple protocol layers such as physical, MAC, and network layers, in order to make latency and reliability informed decisions in real time. For instance, physical layer CSI can be utilized at higher layers for rapid anomaly detection and adaptive QoS enforcement.

Despite significant advancements, fundamental trade-offs exist in URLLC system design. Enhancing reliability often leads to increased energy consumption and protocol overhead, complicating latency constraints. For example, security mechanisms such as physical layer authentication based on RF fingerprinting improve device level trust but introduce processing delays. This introduces the challenge of designing under the paradigm of zero slack security, where even microsecond level computation delays could violate URLLC’s latency envelope. In critical applications such as collaborative robotics or autonomous driving, where 99.9999\% availability is required, network slicing redundancy or hybrid frequency usage may be necessary to meet performance targets \cite{wu2024urllc}. Additionally, scaling these systems across thousands of nodes introduces significant challenges in resource coordination, decentralized learning, and cross layer optimization \cite{she2017cross}.

\section{Taxonomy of URLLC Solutions for 6G‑Enabled Industry 5.0}
\label{sec:taxonomy}

This taxonomy classifies the technological landscape of URLLC into four main attributes: a) application domains, b) key technical enablers, design challenges, and performance enhancements. Each item is coded for cross referencing in our analysis of studies. Table~\ref{tab:urlcc_summary} presents a summary of URLLC attributes applied to the state-of-the-art applications. In Table~\ref{tab:urlcc_summary}, each study is mapped to its corresponding attribute codes, enabling quick identification of research focus areas and facilitating comparative analysis across different works. Some of these mappings are based on the perceived benefits of the proposed approach to Industry~5.0, as certain performance enhancements are not explicitly reported by the original study but are inherently achieved through the nature of the proposed solution. This interpretative step ensures that the taxonomy reflects both documented outcomes and logical inferences drawn from the design and capabilities of each approach, thereby providing a more complete picture of their potential impact.

\begin{figure*}
    \centering
    \includegraphics[width=1.1\linewidth, trim=300 60 0 0, clip]{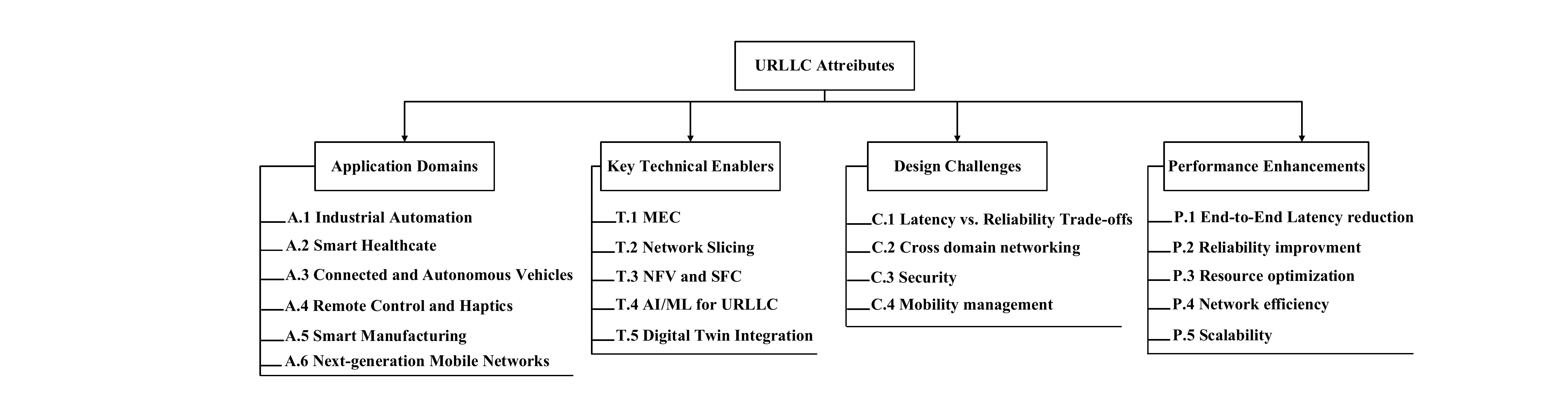}
    \caption{A taxonomy of URLLC attributes in enabling industry 5.0, categorized by applications, enabling technologies, design challenge, and performance enhancements.}
    \label{fig:taxonomy}
\end{figure*}

\subsection{Application Domains} \label{apps}
\begin{enumerate}[leftmargin=*,label=A.\arabic*]

\item \textbf{Industrial Automation} \label{IndustrialAutomation} refers to the use of URLLC in factory automation, robotics, UAVs, and process control where sub-1ms latency is needed for feedback loops. For instance, Mertes et al.~\cite{mertes2024evaluation} showed that 5G based motion control in CNC milling improved precision by minimizing delay variations. Additionally, Li et al.~\cite{li2024reliability}. addressed URLLC for industrial automation systems operating over multi connectivity 5G networks, focusing on resource allocation to satisfy latency and reliability constraints under varying channel and traffic conditions.

\item \textbf{Connected and Autonomous Vehicles} \label{ConnectedVehicles} involves Vehicle-to-Everything (V2X) communication for cooperative maneuvering, platooning, and perception sharing. For instance, Singh et al.~\cite{singh2024toward} designed a hybrid radio frequency-visible light communication protocol achieving sub-1ms delay in autonomous driving scenarios, enhancing safety and reliability. Rahimi et al.~\cite{rahimi2021design} also applied a hybrid edge computing architecture to support URLLC in connected vehicles by offloading computation close to user devices to minimize latency. Yuan et al.~\cite{yuan2023digital} proposed a digital twin vehicle platooning simulation system to enable real-time bidirectional communication between physical and digital entities for autonomous vehicle platoons.

  \item \textbf{Smart Healthcare} \label{SmartHealthcare} includes remote surgery, wearable health monitoring, and emergency response, all requiring ultra low latency and high reliability. For instance, Zhou et al.~\cite{zhou2020learning} proposed a federated Deep Reinforcement Learning (DRL) model to optimize computation offloading decisions under URLLC constraints for Internet of Health Things (IoHT) networks, improving average task latency and offloading success rate.

  \item \textbf{Remote Control and Haptics} \label{Haptics} enables tactile feedback in teleoperation, such as surgical robotics or industrial maintenance. For instance, Kokkinis et al.~\cite{kokkinis2025deep} developed a DRL  based slicing framework that increased haptic QoE under tight latency constraints by optimizing radio resource allocation between haptic and video streams. 

  \item \textbf{Smart Manufacturing} \label{SmartManufacturing} focuses on integrating digital twins, edge AI, and cyber physical systems into production environments. For instance, ~\cite{crespo2024flexible} implemented a real-time IoT edge cloud architecture that supports predictive maintenance using low latency digital twins. Noor-A-Rahim et al.~\cite{rahimi2021design} proposed using IRS to support URLLC in smart manufacturing by reconfiguring the radio environment for real-time wireless computation offloading. 

  \item \textbf{Next-generation Mobile Networks} \label{NextGen} refers to the evolving capabilities of 6G wireless networks that introduce new features to support applications with extreme performance requirements, including sub-0.1ms over the air delays and microsecond level jitter. These networks aim to support diverse traffic profiles through advanced slicing mechanisms, multi objective optimization, and agile control loops. For instance, She et al.~\cite{she2020deep} proposed a multi level DL architecture to meet URLLC performance requirements in non-stationary 6G networks, integrating device, edge, and cloud intelligence for mobility prediction and scheduling. 
\end{enumerate}

\subsection{Key Technical Enablers} \label{tech}
\begin{enumerate}[leftmargin=*,label=T.\arabic*]

   \item \textbf{Multi-access Edge Computing (MEC)} \label{MEC} refers to deploying compute and storage near the network edge to reduce transmission delay. For instance, Peng et al.~\cite{peng2022real} demonstrated that edge based task partitioning reduced control loop latency in industrial networks. Rahimi et al.~\cite{rahimi2021design} proposed a hybrid edge computing architecture integrating MEC, Fog computing, and Cloudlets to minimize latency by offloading computation close to user devices for applications like connected vehicles and smart manufacturing.

  \item \textbf{Network Slicing} \label{NetworkSlicing} creates isolated logical networks on shared infrastructure tailored to URLLC demands. For instance, Kalem et al.~\cite{kalem20245g} demonstrated that per-service network slices can be dynamically instantiated based on latency class, service priority, and device mobility profile, managed through A powered orchestration.

  \item \textbf{Network Function Virtualization (NFV) and Service Function Chaining (SFC)} \label{NFV} refers to the dynamic orchestration of virtualized network functions to meet URLLC key performance indicators. For instance, Erbati et al.~\cite{erbati2023service} proposed an SFC optimization algorithm that embeds SFCs while differentiating traffic based on latency criticality, which significantly reduces end-to-end delay and maximizes bandwidth utilization for Industry 5.0 services.

  \item \textbf{AI/ML for URLLC} \label{AI4URLLC} uses learning based policies for resource allocation, prediction, and control in dynamic environments. For instance, Li et al.~\cite{li2020deep} proposed an AI enhanced scheduling framework to minimize end-to-end latency and optimize spectrum utilization in URLLC scenarios for motion control and real-time haptic feedback.

  \item \textbf{Digital Twin Integration} \label{DigitalTwinIntegration} enables synchronized virtual models of physical systems for predictive control. For instance, Yuan et al.~\cite{yuan2023digital} proposed a digital twin platooning system for interactive simulation and real-time coordination of autonomous vehicle platoons using two-way feedback between physical and virtual entities. 
\end{enumerate}

\subsection{Design Challenges} \label{chall}
\begin{enumerate}[leftmargin=*,label=C.\arabic*]

  \item \textbf{Latency vs. Reliability Trade-offs} \label{LatencyVsReliability} refers to the fundamental tension between minimizing delay and ensuring ultra reliable delivery. For instance, Li et al.~\cite{li2024reliability} presented an analytical model to study this trade-off in low latency communication systems with FBC, demonstrating how improving one dimension (e.g., reliability) often compromises others (e.g., latency or throughput).
\
  \item \textbf{Cross-Domain Networking} \label{CrossDomain} involves interconnecting independently managed network domains with consistent low latency guarantees. For instance, Li et al.~\cite{li2023low} proposed a low latency terminal mutual access architecture based on 5G LAN cross domain networking, enabling real-time, ultra reliable communication across multiple 5G LAN domains managed by different operators. Shih et al.~\cite{shih2016enabling} also tackled cross domain resource allocation by relocating computational tasks from centralized cloud servers to localized Fog-Radio Access Network (F-RAN) nodes closer to end users.

  \item \textbf{Security} \label{Security} ensures confidentiality and integrity without violating strict latency budgets, often requiring zero slack security insertion. For instance, Ji et al.~\cite{ji2022massive} proposed a physical layer security enhancement framework for UAV enabled URLLC networks by integrating Massive MIMO transmission with a Secrecy Guard Zone (SGZ) mechanism, ensuring confidentiality without compromising ultra low latency requirements

  \item \textbf{Mobility Management} \label{Mobility} concerns handovers and service continuity in dynamic, non-stationary environments such as UAVs or autonomous vehicles. For instance, Kalem et al.~\cite{kalem20245g} addressed mobility management in dynamic UAV environments by dynamically instantiating per-service network slices and reassigning UAV gateways to low latency uplink paths when mobility events are detected. 
\end{enumerate}

\subsection{Performance Enhancements}  \label{performance}
\begin{enumerate}[leftmargin=*,label=P.\arabic*]

  \item \textbf{End-to-End Latency Reduction} \label{LatencyReduction} focuses on minimizing total transmission and processing delays across all layers of the communication chain, including rapid packet acquisition, fast decoding, and minimal retransmission cycles. For instance, Wang et al.~\cite{wang2019low} designed an integrated planning algorithm that achieved latency reduction in MEC enabled fiber-wireless networks by minimizing MEC server execution time, Optical Line Terminal (OLT) processing latency, and fiber propagation delay.

  \item \textbf{Reliability Improvement} \label{Reliability} aims to reduce packet loss and failure probability, ensuring deterministic performance even under worst-case scenarios. For instance, She et al.~\cite{she2017cross} integrated redundancy and adaptive modulation to achieve 99.999\% availability under wireless fading conditions for ultra reliable and low latency radio access networks. Li et al.~\cite{li2023cross} also proposed a cross layer optimization framework that uses physical layer packet replication and network-level path diversity to achieve lower delay violation probability, thereby enhancing reliability for industrial automation systems.

  \item \textbf{Resource Optimization} \label{ResourceOptimization} balances compute, bandwidth, and power to meet latency targets efficiently in dynamic and resource constrained environments. For instance, Ranjha et al.~\cite{ranjha2024consumer} proposed a low complexity optimization framework that jointly manages resource allocation and UAV trajectory planning to minimize energy consumption while satisfying URLLC constraints for multi UAV assisted MEC systems.

  \item \textbf{Network Efficiency} \label{NetworkEfficiency} refers to maximizing throughput and utilization under strict latency constraints while maintaining QoS or Quality of Experience (QoE). For instance, Kalem et al.~\cite{kalem20245g} showed that their AI enabled multi layer edge architecture ensures deterministic QoS for each URLLC application class while optimizing overall edge resource utilization in Industry 5.0 scenarios.

  \item \textbf{Scalability} \label{Scalability} ensures that URLLC systems can work effectively at a large scale, accommodating multiple users, diverse services, and increasing network nodes without degrading performance. For instance,  Narsani et al.~\cite{narsani2023leveraging} emphasized scalability by using distributed edge computing nodes on each autonomous vehicle to handle localized data processing and control, thus avoiding reliance on centralized cloud infrastructure. 
\end{enumerate}

\section{6G Low‑Latency Enablers for Industry 5.0}\label{main}

The evolution of wireless communication from 5G to 6G is fundamentally driven by the increasing demands of Industry 5.0, which brings a symbiotic collaboration between humans and intelligent machines, requiring new levels of connectivity and responsiveness \cite{kalem20245g}. At the core of this transformation is the need for URLLC, a critical enabler for advanced industrial applications. This section provides a state-of-the-art discussion of recent advancements in URLLC for Industry 5.0, focusing on how various technological attributes contribute to achieving the hard requirements of this new industrial era. We structure our discussion by first exploring general URLLC enablers and their fundamental characteristics, followed by an examination of recent approaches according to their latency objectives, the architecture, challenges to be solved, the application, and the performance enhancement.

\begin{figure*}
    \centering
    \includegraphics[width=0.9\linewidth, trim=0 950 0 0, clip]{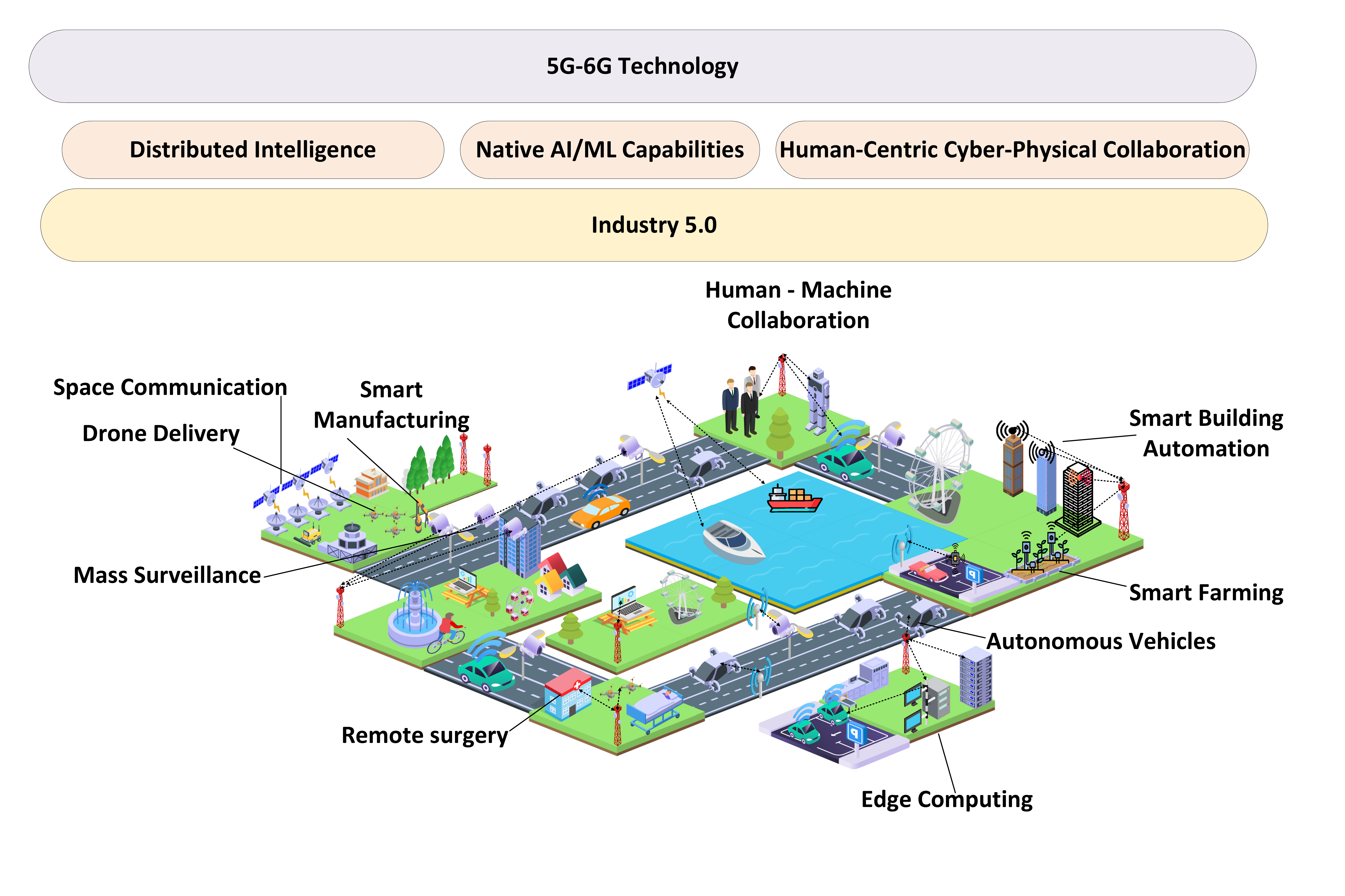}
    \caption{Industry~5.0 ecosystem empowered by 5G and 6G technologies. Enabling pillars such as distributed intelligence, native AI/ML, and human-centric cyber-physical systems drive the transition from traditional automation to collaborative autonomy. Applications span smart manufacturing, remote surgery, autonomous mobility, edge computing, and space-ground integration, all unified through real-time URLLC connectivity and semantic control.}
    \label{fig:5g_6g_industry_evolution}
\end{figure*}

Figure~\ref{fig:5g_6g_industry_evolution} shows the transformation of industrial systems enabled by 5G and 6G technologies on the emerging Industry~5.0 paradigm. Industry~4.0 established the groundwork with automation, real-time control, and cloud based data analytics enabled by URLLC, enhanced Mobile Broadband (eMBB), and massive Machine Type Communications (mMTC). The current evolution focuses on intelligence and collaboration. Industry~5.0 shifts from process digitization to distributed, human aware systems supported by 6G native capabilities. These include native AI/ML, distributed intelligence, and seamless human-machine collaboration. As shown in figure~\ref{fig:5g_6g_industry_evolution}, industrial domains such as smart manufacturing, edge computing, and drone logistics converge with human centric domains including remote surgery, autonomous driving, and XR-assisted training. This integration is managed through digital twins, semantic informed control, and multi-domain synchronization, all functioning under URLLC constraints. The shift from centralized automation to intelligent decentralization defines the extension of Industry~4.0 to Industry~5.0.

\subsection{Reliability-Latency-Rate Tradeoff and QoS in URLLC Systems}

In \cite{li2024reliability}, Li \textit{et al.} presented an analytical model to study the fundamental tradeoffs between reliability, latency, and data rate in low latency communication systems with FBC. Their work addresses the shortcomings of classical infinite block length assumptions in URLLC scenarios where strict latency and reliability requirements are imposed under limited bandwidth conditions (\ref{LatencyVsReliability}). The authors define and characterize the relationships between service rate gain, reliability gain, and real-time gain, providing insights into  efficient scheduling strategies for low latency systems. The authors developed a point-to-point communication model over both additive white Gaussian noise (AWGN) and Nakagami-\textit{m} fading channels with stochastic data arrivals. Their model accounts for random arrival processes and power limited transmission, thus modeling mission critical systems where deterministic assumptions break down. They propose a Laplace’s-method-based effective capacity approximation approach to derive an analytical expression for QoS constrained throughput.

\begin{figure}[!t]
  \centering
  \includegraphics[width=0.48\textwidth]{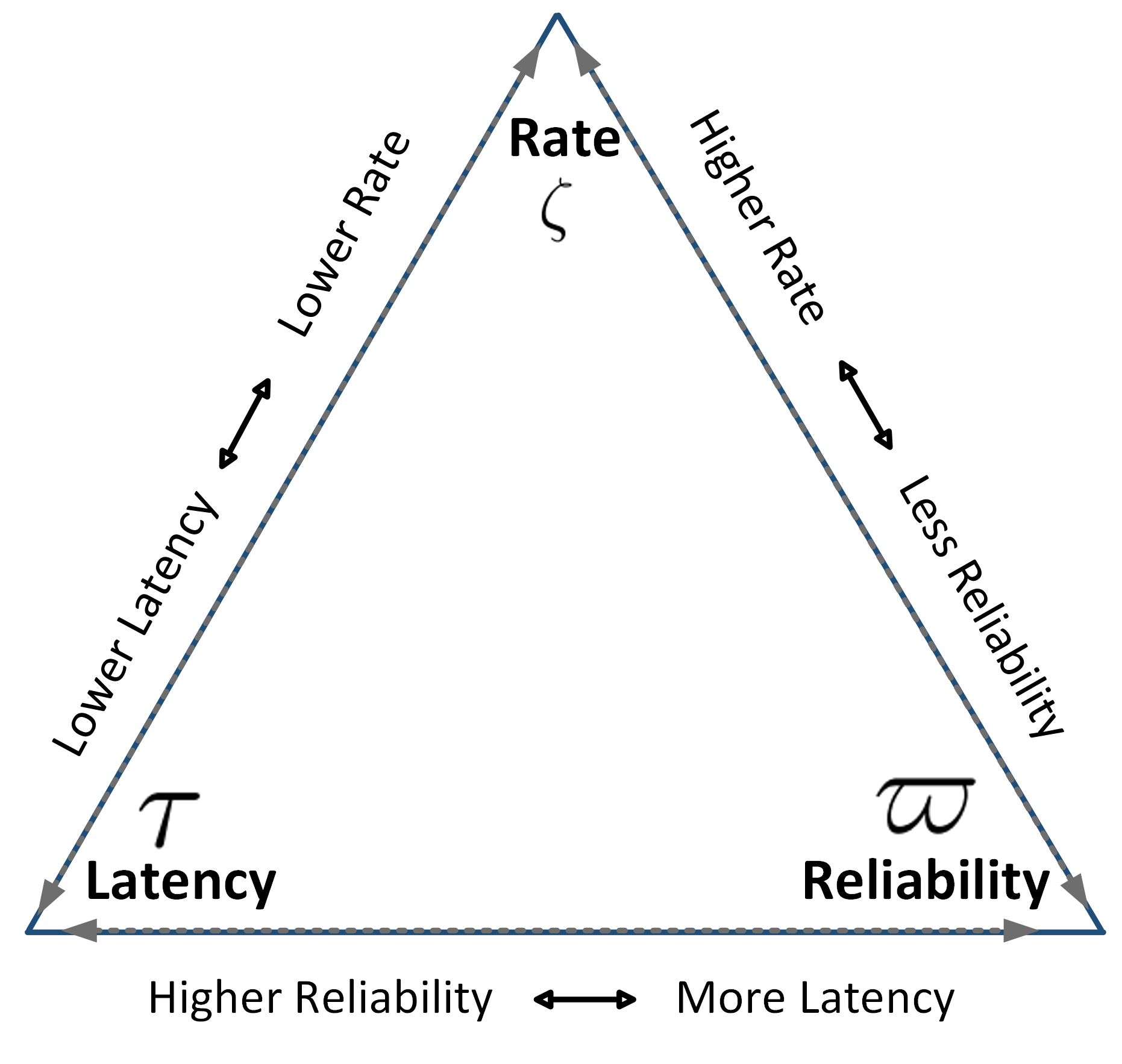}
  \caption{The tradeoff triangle among service rate gain ($\zeta$), reliability gain ($\varpi$), and real-time gain ($\tau$) under finite blocklength coding. As analyzed in~\cite{li2024reliability}, improving one metric often compromises others, especially under power and bandwidth constraints in URLLC scenarios. This graphical abstraction illustrates the conservation law guiding QoS-aware resource allocation in URLLC systems.}
  \label{fig:fbl_triangle}
\end{figure}

The central analytical tool is the EC, which quantifies the maximum sustainable arrival rate that ensures statistical QoS constraints such as queue-length-violation probability  and delay-violation probability. To better characterize performance under FBC, the authors introduced a set of gain conservation metrics, namely service rate gain ($\zeta$), reliability gain ($\varpi$), and real-time gain ($\tau$). These metrics together capture the tradeoffs between bandwidth efficiency, error probability, and queuing delay. As summarized in Table~\ref{tab:tradeoff_metrics}, each metric provides a different aspect for understanding how URLLC system constraints interplay under limited resources. Figure~\ref{fig:fbl_triangle} shows the tradeoff triangle where improving one dimension (e.g., reliability via tighter coding) can degrade others (e.g., latency or throughput). This triangle serves as a conceptual guide for URLLC resource orchestration, where tradeoff informed scheduling policies must navigate competing objectives and satisfy statistical QoS bounds. Furthermore,the authors showed that maximizing average transmission rate (e.g., via water filling) may not lead to optimal EC in finite blocklength scenarios. This highlights a critical gap between classical and modern low latency performance objectives This theoretical characterization contributes to fulfilling end-to-end latency reduction (\ref{LatencyReduction}) and reliability improvement (\ref{Reliability}) by enhancing network efficiency (\ref{NetworkEfficiency}).

\begin{table*}
\centering
\caption{Summary of Reliability vs. Latency Rate Tradeoff Metrics in URLLC under FBC}
\label{tab:tradeoff_metrics}
\renewcommand{\arraystretch}{1.2}
\begin{tabular}{p{3.1cm} p{4.8cm} p{6.4cm}}
\toprule
\textbf{Metric} & \textbf{Definition} & \textbf{Impact on URLLC Design} \\
\midrule
\textbf{Service-Rate Gain ($\zeta$)} &
Ratio of actual service rate to nominal channel capacity &
Higher $\zeta$ indicates efficient usage of available bandwidth, but may stress error control if blocklength is fixed. \\ \hline
\textbf{Reliability Gain ($\varpi$)} &
Ratio between achievable and required decoding success probability &
Quantifies how close the system operates to desired reliability; low $\varpi$ requires enhanced coding or diversity. \\ \hline
\textbf{Real-Time Gain ($\tau$)} &
Ratio between delay budget and actual system delay &
Reflects system’s capacity to meet real-time deadlines; smaller $\tau$ implies tighter scheduling and queue management. \\
\bottomrule
\end{tabular}
\end{table*}

In \cite{erbati2023service}, Erbati \textit{et al.} proposed a SFC optimization algorithm designed to enable URLLC across 5G, 6G, and IoT applications, to address the need for deterministic end-to-end delay guarantees for emerging Industry 5.0 services. This study falls under the architectural domain of NFV (\ref{NFV}) and targets delay sensitive vehicular (\ref{ConnectedVehicles}), smart healthcare (\ref{SmartHealthcare}), and remote control and haptics (\ref{Haptics}) use cases, including autonomous driving, teleoperated driving, virtual reality, augmented reality, and remote surgery.

Traditional hardware based network architectures are often rigid, expensive, and energy intensive, making them unsuitable for dynamic and latency sensitive services. The authors propose a shift toward NFV, where monolithic hardware functions are transformed into lightweight Virtual Network Functions (VNFs) running on general purpose computing platforms, enhancing flexibility and reducing operational costs for highly adaptive Industry 5.0 networks. The proposed approach focuses on optimal resource allocation by embedding SFCs while differentiating traffic based on latency criticality. A tunable priority coefficient ($\mu$) is introduced to prioritize ultra low latency traffic, enabling the reservation of dedicated bandwidth, CPU, and memory resources. The SFC embedding problem is modeled as an integer linear programming optimization that includes QoS constraints on latency and network resource consumption. The design ensures that the allocation of virtual and physical resources aligns with delay budgets and server capacity limits. The proposed scheme has been shown to reduce end-to-end delay (\ref{LatencyReduction}) while maintaining a high SFC acceptance rate (\ref{Reliability}) and maximizing bandwidth utilization (\ref{ResourceOptimization}).

\begin{figure}[!t]
    \centering
    \includegraphics[width=0.95\linewidth]{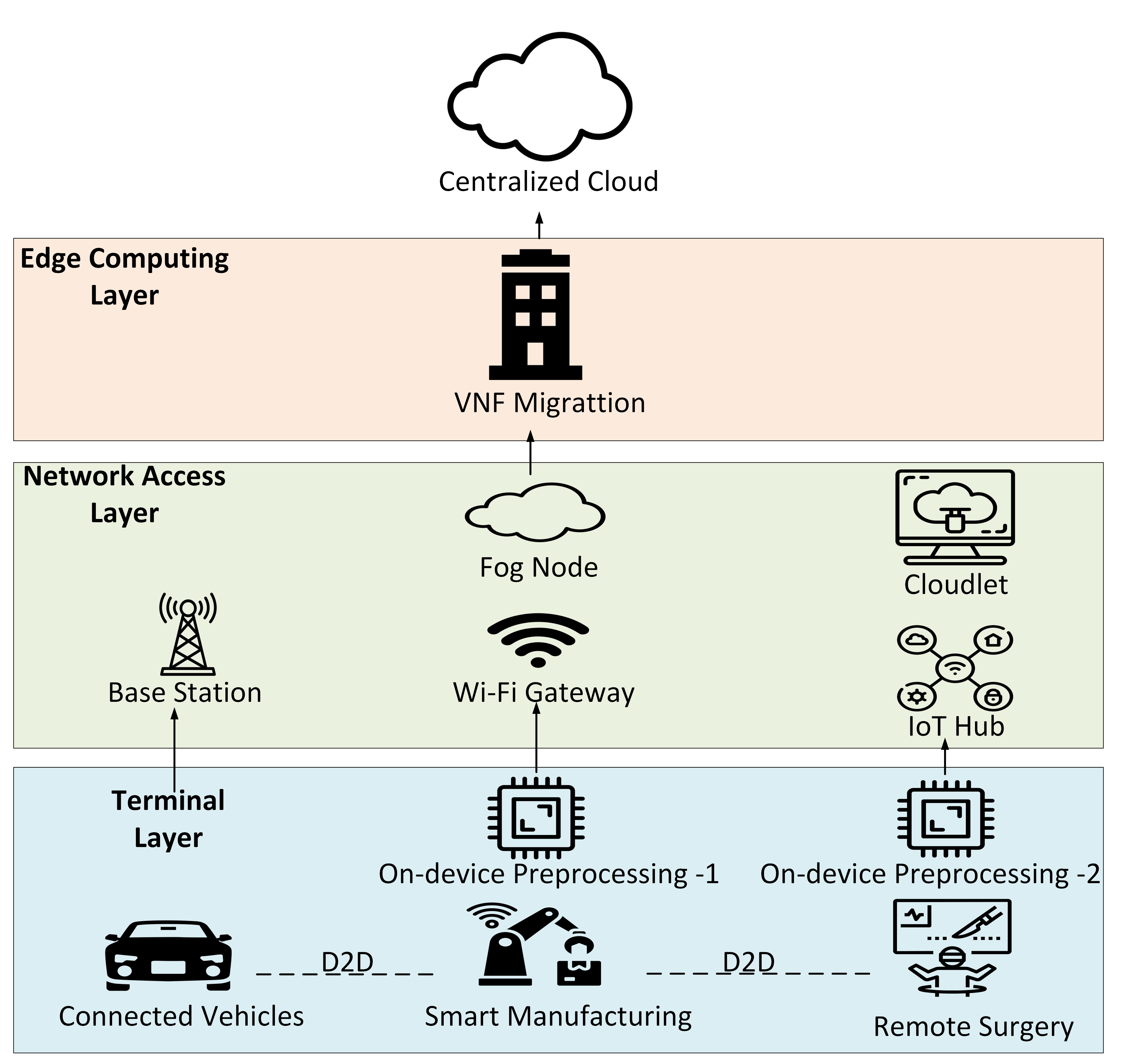}
    \caption{Rahimi’s hybrid MEC–Fog–Cloudlet architecture for URLLC in Industry 5.0. The architecture consists of a terminal layer (with on-device preprocessing and D2D communication), an edge computing layer (Fog and Cloudlets), and a network access layer supporting VNF migration across distributed nodes.}
    \label{fig:rahimi_architecture}
\end{figure}

 In \cite{rahimi2021design}, Rahimi \textit{et al.} applied a hybrid edge computing architecture that integrates MEC, Fog Computing, and Cloudlets to support URLLC in latency sensitive applications such as connected vehicles (\ref{ConnectedVehicles}), remote surgery (\ref{SmartHealthcare}), augmented reality (\ref{Haptics}), and smart manufacturing (\ref{SmartManufacturing}). The study addresses the latency vs. reliability trade-off (\ref{LatencyVsReliability}) and mobility management (\ref{Mobility}), which traditionally limit the performance of cloud centric networks. To mitigate these issues, the authors propose a multi layered architecture that consist of: a) a terminal layer that enables direct device-to-device (D2D) communication and client side data preprocessing; and b) a network access layer that integrates MEC, Fog nodes, and Cloudlets across heterogeneous access points (e.g., base stations, Wi-Fi gateways, IoT hubs). Each access point hosts virtualized compute resources capable of processing service requests locally. Workload allocation is performed dynamically based on traffic class, proximity to the user equipment (UE), and server load status. Delay sensitive requests are offloaded to the closest Fog or Cloudlet node with available compute capacity, while background tasks is routed to regional MEC servers (\ref{MEC}). This decision making process is supported by NFV (\ref{NFV}), which orchestrates VNFs across these distributed nodes, enabling real-time placement and migration of services as network conditions evolve.  This architecture enhances URLLC performance by reducing dependency on distant cloud servers, shortening round-trip times, and enabling context aware resource allocation. As shown in Figure~\ref{fig:rahimi_architecture}, the multi layered system integrates device level preprocessing, D2D communication, and edge cloud resource orchestration across heterogeneous access nodes. The study has been shown to better reduce latency compared to standard MEC setups (\ref{LatencyReduction}), provide higher reliability (\ref{Reliability}), and improve (\ref{Scalability}), to support critical Industry 5.0 services in real time.
 
Similarly, previous research on ultra-low latency architectures and approaches are timeless in enabling industry 5.0 and contribute to the current landscape of URLLC. In \cite{shih2016enabling}, Shih \textit{et al.} applied a F-RAN architecture to enable URLLC for ultra low latency IoT services, such as mobile augmented reality (AR) (\ref{Haptics}), smart surveillance, and real-time feedback control (\ref{SmartManufacturing}). The proposed system incorporates distributed edge cloud co-processing (\ref{MEC}) using fog nodes embedded in femtocell base stations and remote radio heads, allowing collaborative computing across multiple radio access points. This architecture tackles key design challenges in latency vs. reliability trade-offs (\ref{LatencyVsReliability}) and cross domain resource allocation (\ref{CrossDomain}) by relocating computational tasks from centralized cloud servers to localized F-RAN nodes closer to end-users. The system is composed of end-devices (e.g., smartphones, IoT sensors) connected to a master F-RAN node (FN), which coordinates task offloading across neighboring FNs. Their approach introduces the CART (Cooperative AR Tracking) algorithm that handles service decomposition, node selection, and joint optimization of task assignment and resource allocation.  Specifically, the CART algorithm selects candidate FNs based on their CPU load and wireless link quality, and then splits incoming AR data into fragments, assigning each fragment to a participating FN for parallel processing. The computed results are then gathered and merged by the master FN before being sent back to the user. This distributed design reduces computing latency at overloaded nodes and avoids overuse of communication links, minimizing round-trip delay in real time. The study showed that the CART based F-RAN system outperforms baseline methods in achieving the 20-ms latency target for mobile AR. The proposed work has been shown to improve end-to-end latency (\ref{LatencyReduction}), enhances compute communication tradeoff efficiency (\ref{ResourceOptimization}), and scales (\ref{Scalability}) across multiple fog nodes with minimal overhead.


In terms of the practical implementation of low latency networks for Industry 5.0, \cite{wang2019low}, Wang \textit{et al.} applied a latency oriented network planning approach for wavelength division multiplexing passive optical network based MEC enabled fiber-wireless access networks, aiming to support URLLC in latency sensitive Industry 5.0 services such as industrial automation (\ref{IndustrialAutomation}), autonomous vehicles (\ref{ConnectedVehicles}), and intelligent transport systems. The system uses MEC (\ref{MEC}) as a key enabler and applies centralized network planning to minimize propagation and processing delays across optical and wireless segments.

The study addresses two main challenges: the latency vs. reliability trade-off (\ref{LatencyVsReliability}) and physical constraints such as maximum fiber length and power budget. To meet these constraints, the authors develop a mathematical model that explicitly minimizes total transmission latency which is defined as the sum of MEC server execution time, central unit distributed unit (CUDU) OLT processing latency, and fiber propagation delay. The model incorporates practical constraints on fronthaul latency (e.g., 100 $\mu s$), fiber distance, and available power budget. To solve this model, the authors proposed a heuristic algorithm called latency minimized integrated multi associated positioning and routing algorithm. This algorithm iteratively identifies optimal placements of MEC servers, CUDU-OLTs and AWGs, and configures the fiber paths such that delay sensitive requests are routed through the lowest latency configurations. The system uses the Warshall-Floyd algorithm for optimal path finding and adjusts routing decisions dynamically under bandwidth and power constraints. The proposed method has been shown to better reduce latency in sparse and in dense networks  compared to traditional enumeration based approaches (\ref{LatencyReduction}). It significantly improves resource efficiency (\ref{ResourceOptimization}) and ensures end-to-end latency within 5–10 ms bounds required by mission critical Industry 5.0 services (\ref{Reliability}).

\subsection{Intelligent Reflecting Surfaces (IRS) for Industry 5.0 Smart Manufacturing}\label{main2}

In \cite{noor2022toward}, Noor-A-Rahim \textit{et al.} conceptualized the integration of IRSs into (\ref{SmartManufacturing}) to enable URLLC. The authors presented multiple IRS-assisted use cases, including blockage mitigation, mmWave/THz communications, wireless energy transfer, sensing and localization, and MEC (\ref{MEC}). 

The authors addressed two central challenges: the high mobility and dynamicity of smart factory environments (\ref{Mobility}) and the need for reliable URLLC in highly reflective and obstructed environments, which creates tension between reliability and latency (\ref{LatencyVsReliability}). To overcome these, they conceptualized an IRS assisted MEC architecture wherein IRS elements are deployed between field devices and edge servers or access points to create reconfigurable links that bypass physical obstacles and improve channel gain. These links are used to offload real-time tasks to edge nodes, enabling distributed computation even under fluctuating connectivity. Specifically, the IRS elements use tunable passive components (e.g., PIN diodes, varactor diodes) to dynamically control the phase and direction of incoming signals. These components are managed by a dedicated controller that configures reflection coefficients in real time based on channel conditions. This architecture enables a smart manufacturing setting (\ref{SmartManufacturing}) where mobile devices benefit from low latency and high throughput wireless links for offloading data, receiving commands, and powering devices wirelessly through IRS enhanced energy harvesting. 

The proposed system has been shown to reduce path loss (\ref{Reliability}) and latency (\ref{LatencyReduction}). The study also outlines unresolved challenges including IRS user association in dynamic settings, radio resource management under multi reflection paths, and scalable channel estimation for large IRS arrays. The proposed architecture supports low latency closed-loop motion control, collaborative robotics, and sensing/localization, which are fundamental aspects of Industry 5.0 realization.

\subsection{UAVs and Digital Twins in Industry 5.0 URLLC}\label{main3}

In \cite{kalem20245g}, Kalem \textit{et al.} presented a 5G/6G-enabled edge network architecture designed to support URLLC in three critical Industry 5.0 application scenarios: a) XR-aided industrial co-design (\ref{Haptics}); b) UAV based real-time inspection in oil refineries (\ref{ConnectedVehicles}); and c) AI driven smart manufacturing fault detection (\ref{SmartManufacturing}). Their architecture utilizes secure private 5G infrastructure and AI powered orchestration to achieve sub-10~ms latency for time sensitive services, using MEC and distributed AI layers (\ref{MEC}). 

\begin{figure}[ht]
    \centering
    \includegraphics[width=0.95\linewidth]{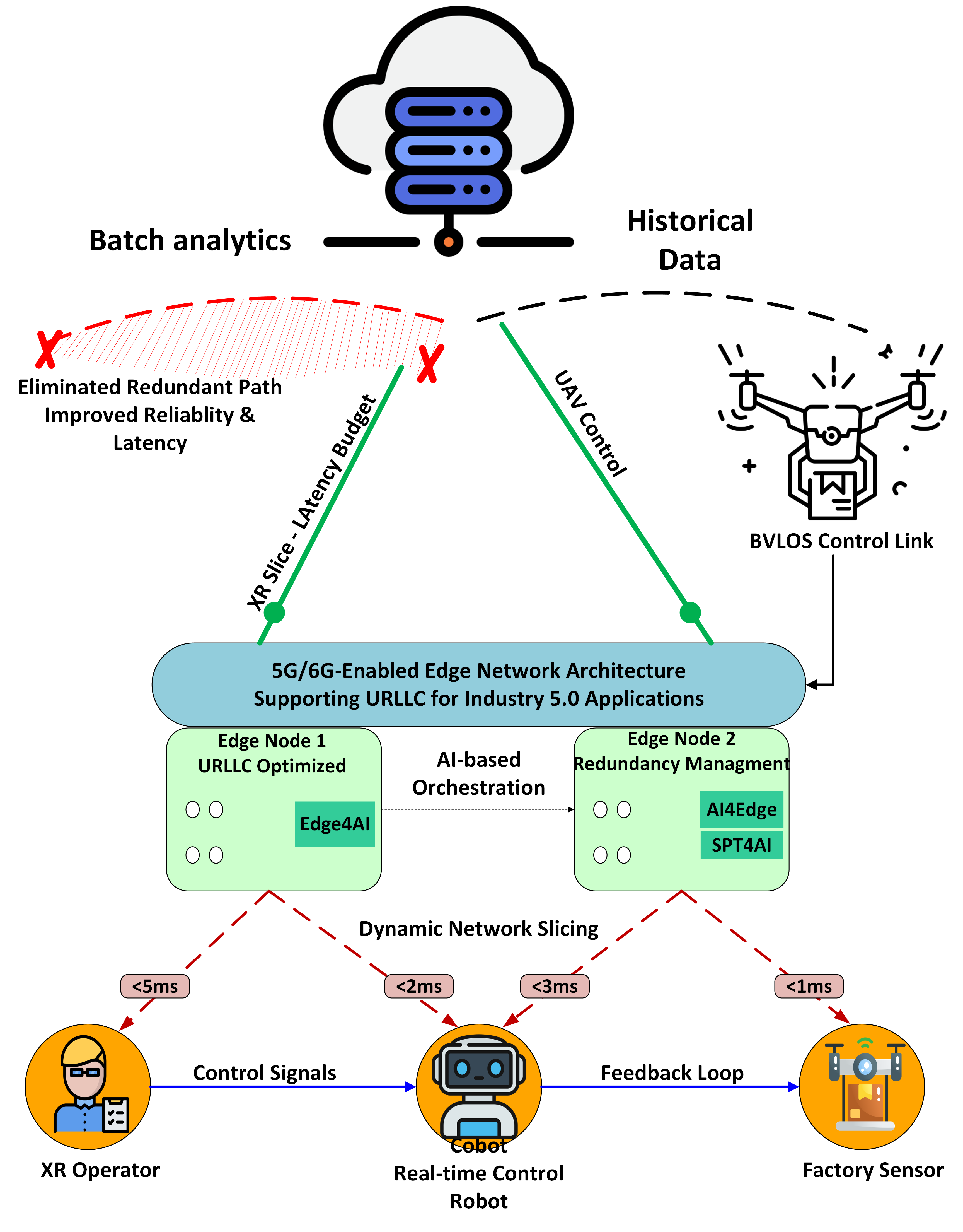}
    \caption{AI-enabled multi-layer edge architecture for URLLC in Industry~5.0 showing per-application slice isolation and real-time control over XR, UAVs, and smart factory sensors.}
    \label{fig:KalemEdgeArchitecture}
\end{figure}

The study addresses key design challenges such as mobility management in dynamic UAV environments (\ref{Mobility}), security and privacy protection in distributed intelligent IoT systems (\ref{Security}), and the fundamental latency vs. reliability trade-off in separating control and payload communications (\ref{LatencyVsReliability}). To overcome these, the authors introduce a layered architecture composed of three components: a) Edge4AI for XR rendering and streaming; b) AI4Edge for predictive AI-based resource management; and c) SPT4AI for privacy preserving analytics and secure federated learning. Each layer contributes to adaptive traffic steering, dynamic slicing, and optimized compute offloading across heterogeneous edge nodes (\ref{NetworkSlicing}). Specifically, the system separates control plane traffic (e.g., BVLOS UAV command-and-control links) from data plane traffic (e.g., XR streams or UAV video feeds) to avoid congestion and ensure latency isolation. Figure \ref{fig:KalemEdgeArchitecture} shows the proposed architecture in which, per-service network slices are dynamically instantiated based on latency class, service priority, and device mobility profile. This slicing decision is executed through the AI4Edge layer, which employs predictive analytics to monitor traffic patterns, infer application requirements, and trigger automated orchestration workflows. These workflows interact with the SDN controller to allocate bandwidth, route paths, and bind edge compute resources in real time. The orchestration logic uses a feedback loop based on performance metrics (e.g., delay jitter, throughput stability, mobility handover status) to continuously adapt slice allocation. For instance, UAV gateways are dynamically reassigned to low latency uplink paths when mobility events are detected, while XR services are moved closer to micro edge nodes for reduced frame rendering latency. Meanwhile, smart manufacturing tasks are assigned to edge inference engines based on factory sensor load and anomaly probability estimates generated by local AI models. This closed-loop AI based orchestration ensures deterministic QoS (\ref{NetworkEfficiency}) for each URLLC application class while optimizing overall edge resource utilization.

The proposed architecture improves end-to-end URLLC performance by offloading latency sensitive processing to the edge, enabling real-time coordination between machines and operators in hybrid industrial environments.  The proposed architecture has been shown to reduce latency (\ref{LatencyReduction}), improve reliability (\ref{Reliability}) by reducing redundant processing paths, and

\clearpage
\onecolumn

\begin{landscape}

\begingroup
  \noindent
  \begin{longtable}{p{4.2cm} p{4.1cm} p{3.2cm} p{2.5cm} p{2.5cm} p{2.5cm}}
\caption{Summary of URLLC studies and attributes.}
\label{tab:urlcc_summary} \\

  \toprule
  \textbf{Study} & \textbf{Reference} & \textbf{Application Domains}
  & \textbf{Key Technical Enablers} & \textbf{Design Challenges}
  & \textbf{Performance Enhancements} \\
  \midrule
  \endfirsthead

  \multicolumn{6}{c}{\tablename\ \thetable{} -- continued from previous page} \\
  \toprule
  \textbf{Study} & \textbf{Reference} & \textbf{Application Domains}
  & \textbf{Key Technical Enablers} & \textbf{Design Challenges}
  & \textbf{Performance Enhancements} \\
  \midrule
  \endhead

  \midrule \multicolumn{6}{r}{Continued on next page} \\
  \endfoot

  \bottomrule
  \endlastfoot
Li’s FBC tradeoff analysis & Li et al. (2024)~\cite{li2024reliability} & \textbf{--} & \textbf{--} & \ref{LatencyVsReliability} & \ref{LatencyReduction}, \ref{Reliability}, \ref{NetworkEfficiency}\\

Erbati’s NFV-based SFC & Erbati et al. (2023)~\cite{erbati2023service} & \ref{ConnectedVehicles}, \ref{SmartHealthcare}, \ref{Haptics} & \ref{NFV} & \ref{LatencyVsReliability} & \ref{LatencyReduction}, \ref{Reliability}, \ref{ResourceOptimization} \\

Rahimi’s hybrid MEC–Fog–Cloudlet architecture & Rahimi et al. (2021)~\cite{rahimi2021design} & \ref{ConnectedVehicles}, \ref{SmartHealthcare}, \ref{Haptics}, \ref{SmartManufacturing} & \ref{MEC}, \ref{NFV} & \ref{LatencyVsReliability}, \ref{Mobility} & \ref{LatencyReduction}, \ref{Reliability}, \ref{ResourceOptimization}, \ref{Scalability}\\

Shih’s F-RAN architecture with CART for AR offloading & Shih et al. (2016)~\cite{shih2016enabling} & \ref{Haptics}, \ref{SmartManufacturing} & \ref{MEC} & \ref{LatencyVsReliability}, \ref{CrossDomain} & \ref{LatencyReduction}, \ref{ResourceOptimization}, \ref{Scalability} \\

Wang’s latency optimized FiWi planning using LMI-MAPRA & Wang et al. (2019)~\cite{wang2019low} & \ref{IndustrialAutomation}, \ref{ConnectedVehicles}, \ref{SmartManufacturing} & \ref{MEC} & \ref{LatencyVsReliability} & \ref{LatencyReduction}, \ref{Reliability}, \ref{ResourceOptimization} \\

IRS-aided MEC for Smart Manufacturing & Noor-A-Rahim et al. (2022)~\cite{noor2022toward} 
& \ref{SmartManufacturing} 
& \ref{MEC} 
& \ref{LatencyVsReliability}, \ref{Mobility} 
& \ref{LatencyReduction}, \ref{Reliability} \\

AI-enhanced 5G/6G secure slicing for XR, UAV, and factories & Kalem et al. (2024)~\cite{kalem20245g} 
& \ref{ConnectedVehicles}, \ref{Haptics}, \ref{SmartManufacturing} 
& \ref{MEC}, \ref{NetworkSlicing}, \ref{AI4URLLC} 
& \ref{LatencyVsReliability}, \ref{Security}, \ref{Mobility} 
& \ref{LatencyReduction}, \ref{Reliability},  \ref{NetworkEfficiency} \\

Consumer-centric sustainability: URLLC in multi-UAV-assisted MEC & Ranjha et al. (2024)~\cite{ranjha2024consumer} &  \ref{SmartManufacturing} & \ref{MEC} &  \ref{LatencyVsReliability}, \ref{Mobility} &   \ref{LatencyReduction}, \ref{Reliability}, \ref{ResourceOptimization} \\

Digital twin enabled 6G MEC in IIoT & Imtiaz et al. (2025)~\cite{imtiaz2025digital} &  \ref{SmartManufacturing} &  \ref{MEC},  \ref{DigitalTwinIntegration} & \ref{LatencyVsReliability} &  \ref{LatencyReduction},  \ref{Reliability}, \ref{ResourceOptimization} \\

Li’s 5G LAN-based terminal mutual access architecture & Li et al. (2023)~\cite{li2023low} 
& \ref{SmartHealthcare}, \ref{NextGen} 
&  \ref{MEC}, \ref{NetworkSlicing}, \ref{NFV} 
& \ref{CrossDomain}, \ref{Security} , \ref{Mobility}
& \ref{LatencyReduction}, \ref{Reliability}, \ref{NetworkEfficiency} \\

Digital Twin Platooning Simulation System with VR  & Yuan et al. (2023)~\cite{yuan2023digital} 
& \ref{ConnectedVehicles} 
&  \ref{MEC}, \ref{DigitalTwinIntegration} 
&  \ref{LatencyVsReliability},  \ref{Mobility} 
&  \ref{LatencyReduction} \\

Digital Twin Platform for CCAM Teleoperation & Forrai et al. (2024)~\cite{forrai2024application} 
&  \ref{ConnectedVehicles}, \ref{SmartManufacturing} 
& \ref{MEC}, \ref{DigitalTwinIntegration}  
&  \ref{LatencyVsReliability}, \ref{CrossDomain}, \ref{Mobility} 
&  \ref{LatencyReduction},  \ref{Reliability} \\

Digital Twin Reputation System for UAV Networks  & Qu et al. (2023)~\cite{qu2023digital} 
&  \ref{IndustrialAutomation} 
& \ref{DigitalTwinIntegration}, \ref{MEC} 
& \ref{LatencyVsReliability}, \ref{Security} 
& \ref{LatencyReduction}, \ref{Reliability}  \\

Distributed DT-based UAV Cluster Control & Xiao et al. (2023)~\cite{xiao2023research} 
& \ref{IndustrialAutomation} 
&  \ref{MEC}, \ref{DigitalTwinIntegration}
&  \ref{LatencyVsReliability},  \ref{Mobility} 
&  \ref{LatencyReduction}, \ref{Scalability}  \\

Massive MIMO and SGZ for Secure UAV URLLC 
 & Ji et al. (2022)~\cite{ji2022massive} 
& \ref{IndustrialAutomation}, \ref{ConnectedVehicles} 
&  \textbf{-} 
& \ref{LatencyVsReliability}, \ref{Security}  
&  \ref{Reliability}, \ref{Scalability} \\

UAV assisted Physical Layer Security & Narsani et al. (2023)~\cite{narsani2023leveraging} 
& \ref{ConnectedVehicles} 
&  \textbf{-}
& \ref{LatencyVsReliability}, \ref{Security}, \ref{Mobility} 
& \ref{LatencyReduction}, \ref{Reliability} \\
Cross layer physical-MAC optimization for factory URLLC & Ramly et al. (2021)~\cite{ramly2021cross} & \ref{SmartManufacturing} & -- &  \ref{LatencyVsReliability}, \ref{Mobility} & \ref{LatencyReduction}, \ref{Reliability},  \ref{NetworkEfficiency}, \ref{Scalability} \\

Cross layer resource allocation over multi-connectivity & Li et al. (2023)~\cite{li2023cross} & \ref{IndustrialAutomation}, \ref{NextGen} & -- & \ref{LatencyVsReliability}, \ref{Mobility} &  \ref{Reliability}, \ref{NetworkEfficiency}, \ref{Scalability} \\

Gao’s multi-objective optimization for URLLC Metaverse & Gao et al. (2023)~\cite{gao2023multi} 
& \ref{Haptics} 
& \ref{MEC} 
& \ref{LatencyVsReliability}, \ref{Mobility} 
& \ref{LatencyReduction}, \ref{Reliability},  \ref{ResourceOptimization}, \ref{NetworkEfficiency} \\

Deterministic cross layer scheduling for maritime URLLC & Wang et al. (2024)~\cite{wang2024deterministic} & \ref{IndustrialAutomation}, \ref{NextGen}  & \ref{AI4URLLC} & \ref{LatencyVsReliability}, \ref{Mobility} & \ref{LatencyReduction}, \ref{Reliability}, \ref{NetworkEfficiency}, \ref{Scalability} \\

 Peng’s uplink resource allocation for Cell-Free mMIMO in smart factories  & Peng et al. (2022)~\cite{peng2022resource} 
& \ref{SmartManufacturing}, \ref{NextGen} 
& \textbf{--}
& \ref{LatencyVsReliability}, \ref{Mobility} 
& \ref{LatencyReduction}, \ref{Reliability}, \ref{ResourceOptimization}, \ref{NetworkEfficiency}, \ref{Scalability} \\

Li’s DRL Scheduler for eMBB and URLLC & Li et al. (2020)~\cite{li2020deep} 
& \ref{NextGen} 
& \ref{AI4URLLC} 
& \ref{LatencyVsReliability}, \ref{Mobility} 
& \ref{LatencyReduction}, \ref{Reliability}, \ref{ResourceOptimization} \\

Hierarchical DL based Slicing for URLLC & Setayesh et al. (2022)~\cite{setayesh2022resource} 
& \ref{NextGen} 
& \ref{AI4URLLC} 
& \ref{LatencyVsReliability} , \ref{Mobility}
& \ref{LatencyReduction}, \ref{Reliability}, \ref{ResourceOptimization} \\

Multi-Level DL Framework with Federated Learning & She et al. (2020)~\cite{she2020deep} 
& \ref{NextGen} 
& \ref{MEC}, \ref{AI4URLLC}, \ref{DigitalTwinIntegration} 
& \ref{LatencyVsReliability}, \ref{Security}, \ref{Mobility} 
& \ref{LatencyReduction}, \ref{Reliability}, \ref{ResourceOptimization}, \ref{NetworkEfficiency}, \ref{Scalability} \\

NOMA-DQN Resource Allocation for Smart Factories & Gengtian et al. (2025)~\cite{gengtian2025deep} 
& \ref{SmartManufacturing}, \ref{NextGen} 
& \ref{AI4URLLC} 
& \ref{LatencyVsReliability}, \ref{Mobility} 
& \ref{LatencyReduction}, \ref{Reliability}, \ref{ResourceOptimization}, \ref{NetworkEfficiency}, \ref{Scalability} \\

Federated DRL-based Task Offloading for IoHT & Zhou et al. (2020)~\cite{zhou2020learning} 
& \ref{SmartHealthcare}, \ref{NextGen} 
& \ref{AI4URLLC} 
& \ref{LatencyVsReliability},  \ref{Security}, \ref{Mobility} 
& \ref{LatencyReduction}, \ref{Reliability} \\

DRL-based RAN slicing for haptic–video synchronization & Kokkinis et al. (2025)~\cite{kokkinis2025deep} 
& \ref{Haptics} 
& \ref{AI4URLLC} 
& \ref{LatencyVsReliability}, \ref{Mobility} 
& \ref{LatencyReduction}, \ref{Reliability}, \ref{ResourceOptimization}, \ref{NetworkEfficiency}, \ref{Scalability} \\
\end{longtable}
\endgroup
\end{landscape}
\clearpage
\twocolumn
\noindent ensure efficient resource allocation over multi-scale workloads. This layered edge architecture provides a strong foundation for Industry 5.0 scenarios requiring human-machine collaboration, situational awareness, and time critical decision-making.

In \cite{ranjha2024consumer}, Ranjha \textit{et al.} proposed a sustainable multi-UAV assisted MEC architecture designed to support URLLC services in AIoT enabled Industry 5.0 intelligent manufacturing environments (\ref{SmartManufacturing}). The system aims to minimize energy consumption while meeting the hard latency and reliability requirements imposed by URLLC use cases. Their proposed word integrates key technical enablers such as MEC deployment for localized processing (\ref{MEC}).

The study addresses two major challenges: a) the trade-off between latency and energy efficiency in power constrained UAV platforms (\ref{LatencyVsReliability}, \ref{ResourceOptimization}); and b) maintaining ultra reliability during UAV trajectory changes  (\ref{Mobility}) and remote AIoT device operations (\ref{Reliability}). To overcome these issues, the authors developed a low complexity optimization framework that jointly manages resource allocation and UAV trajectory planning.  The optimization is based on a successive convex approximation algorithm combined with a random restart method to escape local minima and improve convergence. The model aims to minimize the weighted sum of energy consumption across UAVs and ground AIoT devices while satisfying URLLC level constraints, such as sub-millisecond latency and ultra low packet error rates. Specifically, smaller blocklengths and tighter error tolerances are incorporated into the QoS formulation to ensure deterministic transmission bounds. The proposed work has been shown to reduces total energy consumption (\ref{ResourceOptimization}) compared to benchmark fixed trajectory and static resource allocation strategies. This work enhances URLLC support in dynamic edge environments by ensuring end-to-end reliability (\ref{Reliability}), achieving latency constrained energy savings (\ref{ResourceOptimization}). These capabilities make the proposed framework suitable for real-time Industry 5.0 deployments involving mobile AIoT networks and mission critical edge applications.

\begin{figure}[!t]
  \centering
  \includegraphics[width=0.47\textwidth]{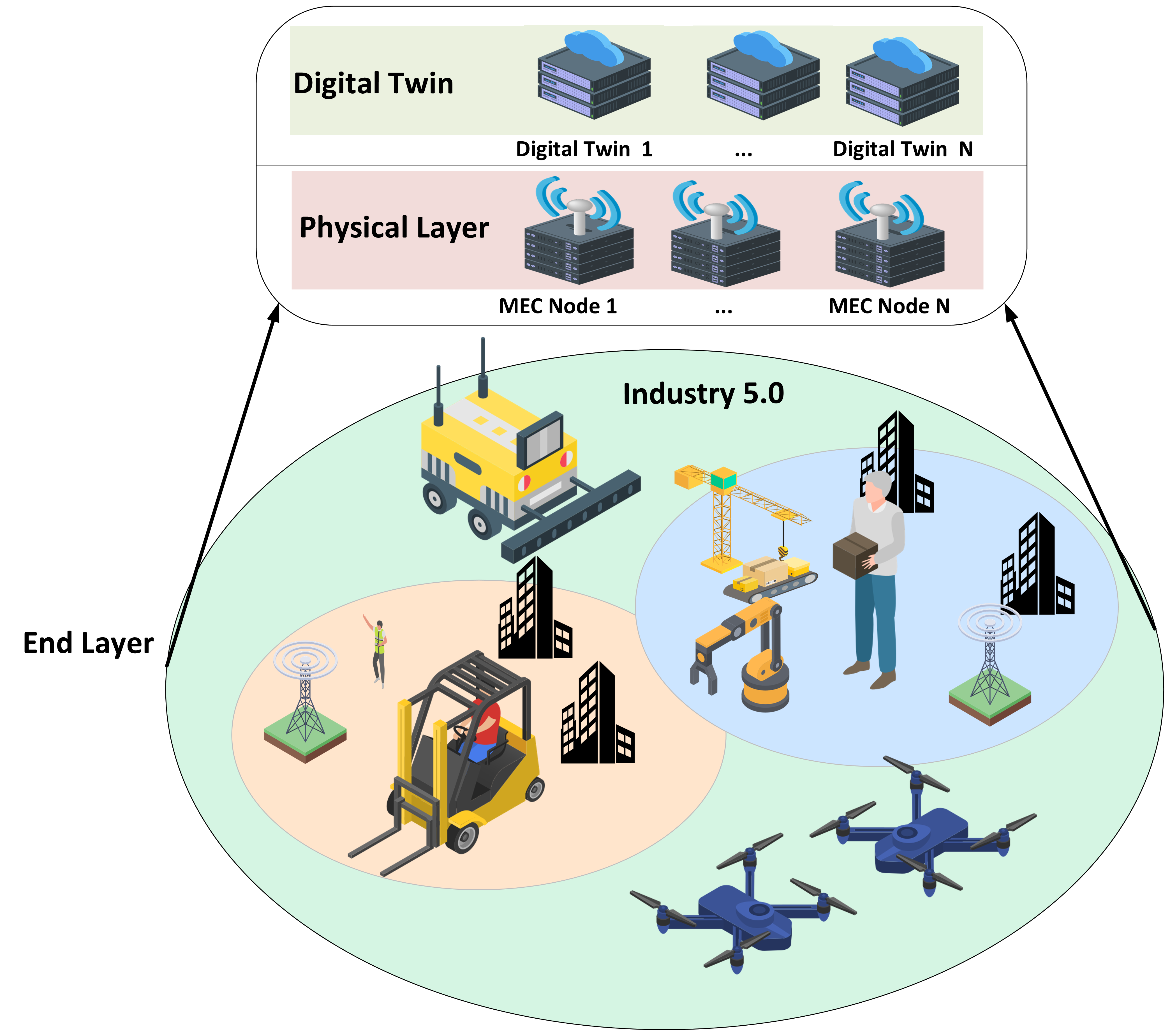}
  \caption{Three-layer predictive edge architecture for computationally efficient task offloading in Industry 5.0 smart manufacturing environments. Digital twins at the top layer continuously synchronize with edge computing nodes in the physical layer, enabling proactive resource forecasting and reliability-aware offloading decisions. The end layer comprises industrial robots and human operators executing latency-sensitive IIoT tasks. This architecture leverages predictive intelligence to reduce service outage probability, improve energy efficiency, and satisfy URLLC requirements across distributed 6G-enabled factories.}
  \label{fig:dt_predictive_architecture}
\end{figure}

In \cite{imtiaz2025digital}, Imtiaz \textit{et al.} proposed a DT enabled MEC framework aimed at improving computational energy efficiency in latency sensitive consumer Industrial IoT (IIoT) applications for Industry 5.0 smart manufacturing environments (\ref{SmartManufacturing}) to  balance latency vs. reliability trade-offs (\ref{LatencyVsReliability}). The system as shown in Figure~\ref{fig:dt_predictive_architecture} introduces predictive digital twin agents that continuously synchronize with physical edge computing nodes, enabling proactive decision making for task offloading. Unlike traditional reactive MEC systems, the proposed framework predicts resource bottlenecks in advance and dynamically schedules tasks to minimize latency and energy consumption simultaneously (\ref{DigitalTwinIntegration}, \ref{MEC}).

The key optimization challenge is modeled as a mixed integer non-linear programming  problem that jointly minimizes energy consumption and task latency while satisfying strict URLLC constraints. To achieve computational tractability, the authors apply piecewise linear approximation to relax non-linear constraints. Task assignment decisions are made based on predicted CPU availability, wireless channel quality, and residual energy of nodes, allowing latency aware offloading while ensuring reliability. This predictive scheduling framework enhances both the responsiveness and sustainability of real-time IIoT applications by allocating edge resources more efficiently (\ref{ResourceOptimization}). The proposed work has been shown to improve computational energy efficiency compared to traditional edge only models and reduces the service outage probability  under strict latency and reliability requirements. These improvements contribute to end-to-end latency reduction (\ref{LatencyReduction}), reliability improvement (\ref{Reliability}), and resource optimization (\ref{ResourceOptimization}) which are, key performance enhancements required in URLLC systems. The proposed DT assisted optimization approach provides energy efficient foundation for deploying URLLC enabled smart manufacturing services in Industry 5.0.

\subsection{Digital Twins and Autonomous Systems for Industry 5.0}\label{main4}

Li \textit{et al.}~\cite{li2023low} proposed a low latency terminal mutual access architecture based on 5G LAN cross-domain networking designed to support real-time communication demands in dynamic industrial scenarios. The architecture addresses the shortcomings of traditional communication frameworks that route data through central servers, introducing latency and single points of failure. Instead, their solution enables D2D communication using 5G LAN capabilities, which is important for latency sensitive applications such as remote surgeries~(\ref{SmartHealthcare}),  and supporting the evolution toward next-generation mobile networks (\ref{NextGen})

The proposed architecture addresses two main challenges: a) enabling real-time, ultra reliable communication across multiple 5G LAN domains managed by different operators~(\ref{CrossDomain}); and b) maintaining privacy and security while bypassing centralized cores~(\ref{Security}). The architecture integrates MEC (\ref{MEC}) to host computation intensive tasks near end users, network slicing (\ref{NetworkSlicing}) to logically separate services, and NFV and SFC (\ref{NFV}) to flexibly deploy and interconnect virtual network functions across domains. To achieve this, the authors use a distributed mesh topology with user plane function based routing (i.e., decentralized traffic anchoring at the edge to enable direct data forwarding between terminals without passing through the core network) supported by secure access service edge and edge offloading~(\ref{MEC}). These mechanisms enable traffic to bypass base stations and core networks  (\ref{Mobility}), thereby achieving latency isolation and privacy preserving communication under a multi-domain setup~(\ref{Security}).

The architecture introduces intelligent grouping of terminal devices using the virtual network group (VN Group) concept, which is a logical abstraction that binds related terminal devices under a common group ID. Group level policies stored in the unified data management function are shared with the session and access management functions, enabling edge level UPFs to directly assign intra group sessions without redundant control plane signaling. This allows low latency inter terminal scheduling even across geographically remote nodes. Moreover, routing paths are dynamically constructed over the mesh using peer aware heuristics, enabling real-time rerouting in the event of node mobility or link failure. This is effective in emergency or remote settings, offering resilient mesh connectivity through D2D protocols. The proposed architecture has been shown to improve end-to-end delay performance~(\ref{LatencyReduction}), enhance overall communication robustness and reliability~(\ref{Reliability}), and reduce dependency on centralized cloud infrastructure (\ref{NetworkEfficiency}). It provides a strong foundation for Industry~5.0 ecosystems involving autonomous systems, digital twins, and real-time control tasks in both public and private 5G networks, as well as in scenarios requiring seamless cross-domain mobility and deterministic performance guarantees.


Forrai \textit{et al.}~\cite{forrai2024application} present an application of digital twins in the context of Connected, Cooperative, and Automated Mobility (CCAM) within Industry~5.0. Their work emphasizes how digital twins can elevate autonomous driving systems by enabling safer, more efficient, and more user centric transportation aligned with the broader goals of human–machine collaboration and situational awareness in Industry~5.0. The authors explore three CCAM use cases of Industry~5.0: a) periodic vehicle data collection for predictive maintenance, b) mixed reality testing of automated systems~(\ref{SmartManufacturing}), and c) safe teleoperation after emergency stops~(\ref{ConnectedVehicles}). These services critically depend on real-time sensing, 5G/6G connectivity, and IoT infrastructure~(\ref{MEC}, \ref{DigitalTwinIntegration}).

The study introduces a distributed driver-in-the-loop test environment, which mixes virtual and physical elements through LAN and cloud based communication~(\ref{CrossDomain}). To maintain deterministic latency bounds for teleoperation, the system incorporates latency mitigation mechanisms including image compression, Smith Predictor based delay compensation (i.e., a model based control strategy that predicts future system outputs to counteract known communication delays)~(\ref{LatencyVsReliability}), and a priority aware queuing strategy. This enables stable performance across different CCAM scenarios with varying delay sensitivity.

For safety critical control scenarios, the digital twin maintains synchronized state across virtual replicas and physical sensors, supporting real-time feedback and low latency actuation. MQTT-based messaging (i.e., a lightweight publish subscribe protocol optimized for low bandwidth and unreliable networks) ensures interoperability and system scalability. In addition, this architecture accommodates both long horizon workflows (e.g., predictive maintenance) and ultra reliable real-time control (e.g., emergency braking), which leads to a balanced performance trade-offs in terms of reliability~(\ref{Reliability}) and latency~(\ref{LatencyReduction}). This research contributes a scalable and latency aware digital twin platform that advances CCAM testing and deployment for Industry~5.0 in domains demanding remote control, resilience, and intelligent coordination among autonomous systems.

Yuan \textit{et al.}~\cite{yuan2023digital} propose a digital twin vehicle platooning simulation system for the interactive simulation of autonomous vehicle platoons~(\ref{ConnectedVehicles}), aimed at advancing Industry~5.0 applications that require coordination and real-time responsiveness. Unlike traditional simulation approaches that rely purely on hardware in the loop setups, their architecture uses digital twin models integrated with a VR driven environment to enable real-time bidirectional communication between physical and digital entities~(\ref{DigitalTwinIntegration}). This addresses the need for agile, low latency \ref{LatencyVsReliability} platforms for testing and verifying platooning algorithms under realistic mobility conditions.

The system is composed of three core layers: a) the physical layer consisting of microcontroller unit equipped with various sensors, b) an edge layer responsible for two way communication via WiFi 6 and cloud platforms (\ref{MEC}), and c) a unity powered virtual environment that replicates physical fleets through detailed 3D models and real-time rendering. Unity is a widely used game engine that supports immersive simulation through physics modeling and environment dynamics, while the microcontroller unit provides real-time onboard processing to support sensing and actuation. The system includes advanced digital modeling, including environmental variation support (e.g., lighting, humidity), Unity based terrain simulation, and physical interaction modeling using built in physics engines. The proposed system has been shown to improve latency in real to virtual and virtual to real settings, showing its suitability for latency sensitive validation tasks~(\ref{LatencyReduction}). This work presents a cost effective simulation platform that advances the development and testing of autonomous driving and vehicle platooning algorithms in Industry~5.0 scenarios. By integrating digital twin modeling, edge computing, and virtual simulation, it enables a robust foundation for scalable URLLC applications in cooperative mobility and intelligent transportation systems.

To synthesize how digital twin technologies are operationalized across various URLLC systems, Table~\ref{tab:dt_urlcc_comparison} provides a comparative summary of representative digital twin driven architectures. Each study highlights a unique scope of digital twin synchronization, from device level state prediction to distributed UAV coordination, and targets diverse optimization goals such as latency reduction, energy efficiency, and trust evaluation. The corresponding system layers (e.g., edge, cloud, or simulation environments) reflect where digital twin are embedded and synchronized. This table extends the taxonomy by clarifying how digital twin materially contribute to specific URLLC performance gains, including end-to-end latency reduction (\ref{LatencyReduction}), reliability enhancement (\ref{Reliability}), and improved scalability across dynamic, real-time industrial environments (\ref{Scalability}).

\begin{table*}[!t]
\centering
\caption{Comparison of Digital Twin Architectures for URLLC in Industry 5.0}
\label{tab:dt_urlcc_comparison}
\renewcommand{\arraystretch}{1.2}
\begin{tabular}{p{2.1cm} p{2.1cm} p{2.8cm} p{2.5cm} p{2.5cm} p{3.0cm}}
\toprule
\textbf{Study} & \textbf{Application Domain} & \textbf{Digital Twin Synchronization Scope} & \textbf{Optimization Target} & \textbf{System Layer} & \textbf{URLLC Performance Gains} \\
\midrule
Imtiaz \textit{et al.}~\cite{imtiaz2025digital} & Smart Manufacturing & Edge-to-Device CPU State, Channel Conditions & Latency and Energy Minimization & MEC Edge & Lower service outage probability, reduced computation latency \\
Qu \textit{et al.}~\cite{qu2023digital} & UAV-based Inspection & UAV Identity, Reputation, Blockchain Ledger & Trust and Attack Mitigation & Cloud + DT + Blockchain & Improved reliability, reduced verification latency \\
Yuan \textit{et al.}~\cite{yuan2023digital} & Vehicle Platooning & Vehicle Dynamics and Control Commands & Control Update Synchronization Delay Minimization & VR Simulation + Controller & Lower latency in platoon command exchange, improved reliability \\
Forrai \textit{et al.}~\cite{forrai2024application} & CCAM Remote Driving & Real-time Sensor-Actuator State & Teleoperation Stability & DT + Edge Server & Reduced jitter, improved tactile response time \\
Xiao \textit{et al.}~\cite{xiao2023research} & UAV Cluster Control & UAV Status + Control Command Aggregation & Cooperative Control Latency Minimization & Distributed Twin Mesh & Higher scalability, reduced control command latency \\
\bottomrule
\end{tabular}
\end{table*}

\subsection{Security, Privacy, and Trust Mechanisms in URLLC for Industry 5.0} \label{main5}

 Mission critical services such as autonomous vehicle coordination, remote surgery, and aerial swarm control demand real-time integrity, confidentiality, and trust without sacrificing performance. This requires zero slack security insertion, where protection mechanisms must be embedded directly into the data plane without violating timing constraints~(\ref{LatencyVsReliability}). This section presents recent architectures that address these challenges using lightweight physical layer security~(\ref{Security}), privacy preserving machine learning, blockchain enhanced trust systems, and digital twin enabled safety models. The focus is on designs that ensure secure and scalable operation of cyber physical systems under decentralized, mobile, and dynamic conditions.


Qu \textit{et al.}~\cite{qu2023digital} introduce a layered architecture called the digital twins enabled reputation system to ensure reliable, secure ~(\ref{Security}), and privacy preserving operations in microchain based UAV networks (\ref{IndustrialAutomation}). The system is designed to tackle main challenges such as data integrity, device authentication, and behavioral monitoring under the constraints of dynamic UAV environments where traditional centralized architectures fail to provide timely (\ref{LatencyVsReliability}) and robust security guarantees~(\ref{Security}). The architecture integrates three core components: a) a physical UAV network equipped with a lightweight blockchain layer (referred to as a microchain), a digital twin platform (\ref{DigitalTwinIntegration}) that mirrors UAVs in virtual space, and a decentralized reputation system that continuously evaluates participant reliability based on real-time and historical activity.

\begin{figure}[ht]
    \centering
    \includegraphics[width=0.95\linewidth]{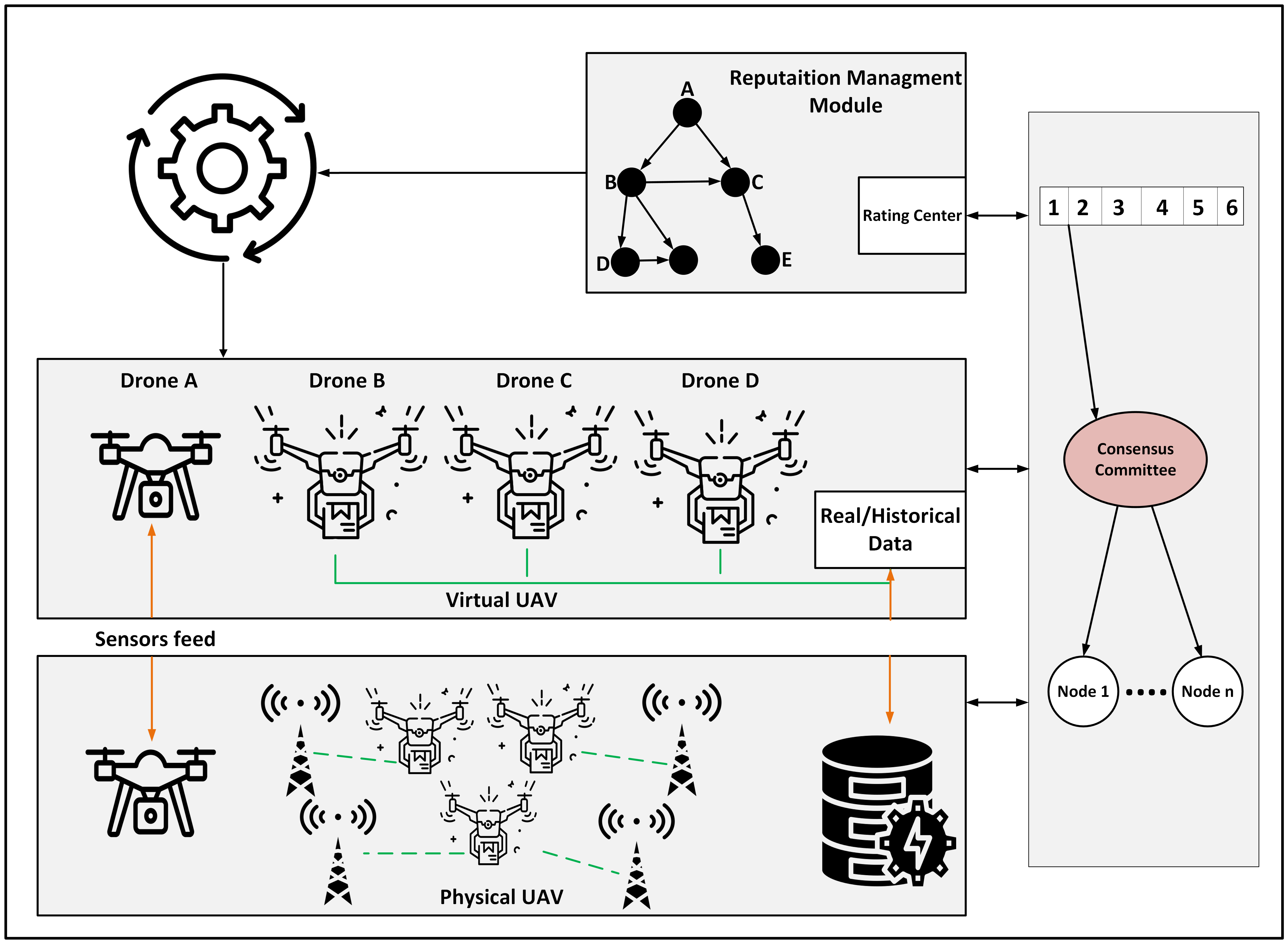}
    \caption{Qu \textit{et al.} proposed system architecture showing the integration of physical UAVs, their virtual digital twin counterparts, a microchain based consensus mechanism, and a reputation evaluation module for decentralized, secure, and real-time UAV operations.}
    \label{fig:QuDITER}
\end{figure}

Each UAV is equipped with sensors for real-time monitoring and transmits its data to a UAV server, which in turn updates the corresponding digital twin. This mirrored digital twin enables predictive analysis and behavior modeling across the fleet. In addition, a permissioned blockchain infrastructure authenticates UAV identities and maintains tamper-proof records of transactions using smart contracts. The reputation layer computes reliability scores for each UAV based on various operational metrics such as task completion, transmission quality, and behavioral consistency. These scores are accessible to all parties and updated automatically through smart contract logic, ensuring autonomous access control and promoting network trust without centralized authorities~(\ref{Security}). As shown in Figure~\ref{fig:QuDITER}, the proposed architecture connects physical UAV data streams to their digital twin representations, while a microchain and consensus committee maintain secure state updates and reputation tracking.  This framework contributes to the ongoing development of secure, intelligent, and reliable (\ref{Reliability}) UAV systems for Industry~5.0 in critical services such as urban air mobility, delivery systems, and aerial surveillance. The proposed system establishes a solid foundation for scalable and secure digital twin driven UAV infrastructure with real-time feedback loops and zero slack security insertion.

In \cite{xiao2023research}, Xiao \textit{et al.} propose a distributed cooperative control architecture for UAV clusters using digital twin modeling to improve safety and environmental adaptability in dynamic scenarios (\ref{DigitalTwinIntegration}). The work addresses limitations of centralized UAV control approaches, including communication bottlenecks (\ref{LatencyVsReliability}), single-point failure, and rigid formation dynamics (\ref{Mobility}), which are especially problematic for UAV deployment in Industry~5.0 smart city and industrial environments~(\ref{IndustrialAutomation}).

The proposed system utilizes the Robot Operating System (ROS), which is a modular middleware framework commonly used for robotic system development and communication to enable distributed networking between UAV subsystems. Each UAV is equipped with a ROS node that publishes its state and subscribes to others via LAN communication, using high range Wi-Fi for seamless peer to peer data exchange~(\ref{DigitalTwinIntegration}). Real-time simulation models are constructed using historical flight data to build adaptive digital twins, which are updated continuously during operation for proactive flight planning and coordination. To model environmental influences such as wind resistance and gravity, the authors define a quadrotor power model that is integrated into the digital twin layer, enabling realistic trajectory prediction and control. For path planning, the architecture employs the Ego-Planner algorithm, which is an optimization based local planner that constructs safe, time parameterized trajectories by minimizing jerk and avoiding dynamic obstacles using real-time environmental perception. 

Trajectory tracking is managed by a cascaded control framework composed of an outer loop position controller and an inner loop attitude controller, ensuring that high-level trajectory commands are accurately translated into motor level actuation. This control framework is implemented using a compatible interface that bridges between ROS topics and MAVLink protocol used by flight controllers, enabling accurate and real-time command execution. The proposed work has been shown to maintain accurate flight behavior under dynamic conditions, with digital twin predictions exhibiting less than 0.2~m Euclidean deviation from actual UAV positions. This validates the twin's synchronization accuracy and its capacity for real-time adaptation. The distributed digital twin design supports latency sensitive coordination~(\ref{LatencyReduction}), improves resilience through decentralized autonomy~(\ref{Scalability}), and enhances adaptability for dynamic factory or urban environments in Industry~5.0.

In~\cite{ji2022massive}, Ji \textit{et al.} propose a physical layer security enhancement framework for UAV enabled URLLC networks for critical Industry~5.0 applications such as autonomous factory inspection~(\ref{IndustrialAutomation}) and wireless sensor offloading in dynamic industrial environments~(\ref{ConnectedVehicles}). The proposed solution integrates Massive MIMO transmission with a SGZ mechanism, which is a spatial region around the transmitter within which any potential eavesdropper is proactively suppressed to prevent confidential information leakage.  To achieve this, the study addresses two key URLLC challenges: a) ensuring physical layer confidentiality without compromising ultra low latency requirements~(\ref{LatencyVsReliability}, \ref{Security}); and b) preserving robustness under short packet transmissions, which are typical for UAV telemetry and mission critical updates. The SGZ based protocol actively nullifies signal energy in the direction of unauthorized receivers (i.e., beamforming nulls), using destructive interference to ensure these users receive no decodable signal. At the same time, legitimate UAVs are served through CSI guided precoding, in order to maximize their reception while minimizing leakage elsewhere. Massive MIMO plays a critical role by using its large number of antenna elements to achieve spatial resolution to simultaneously direct signals toward legitimate users while suppressing leakage toward unauthorized ones. This spatial separation  improves security performance without introducing significant delay or complexity. 

The proposed work has been shown to improve secrecy throughput under latency bounds and reduce decoding error probabilities in blocklength limited scenarios. This framework contributes to URLLC in Industry~5.0 by enabling secure UAV communication with real-time performance and minimal cryptographic overhead. The combined use of Massive MIMO and SGZ offers a scalable (\ref{Scalability}), low-complexity solution to protect airborne industrial networks, leading to an enhanced reliability~(\ref{Reliability}).


In \cite{narsani2023leveraging}, Narsani \textit{et al.} examined the use of an UAV (\ref{ConnectedVehicles}) to enhance physical layer security in URLLC scenarios where a legitimate ground transmitter receiver pair is threatened by an eavesdropper (\ref{Security}). The study considered two operational modes for the UAV: a) acting as a decode-and-forward relay; or b) functioning as a friendly jammer. The secrecy rate, defined as the positive difference between the achievable rates of the legitimate and interception links, was derived under finite blocklength constraints, taking into account the blocklength $M$ and target error probability $\epsilon$. The analysis incorporated realistic FBL achievable rate expressions for both the legitimate and eavesdropper channels, enabling accurate evaluation of security latency trade-offs. To carry out the evaluation, the authors modeled the geometry of the transmitter, receiver, eavesdropper, and UAV, then computed link distances and channel gains as functions of UAV altitude and horizontal position. For each configuration, the FBL rates were calculated, and the secrecy rate was determined for both UAV modes. The system performance was then compared over a range of altitudes, Tx–Rx separations, and channel conditions to identify the optimal UAV role and placement for maximizing secrecy rate. Simulation results demonstrated that placing the UAV in jammer mode directly above the eavesdropper generally achieved the highest secrecy rate, especially at moderate altitudes, while relay mode was more beneficial for longer transmitter receiver distances. The work addresses URLLC design challenges concerning latency reliability trade-offs~(\ref{LatencyVsReliability}) and security~(\ref{Security}), showing how UAV mobility~(\ref{Mobility}) can be leveraged to dynamically optimize positioning and operational role without violating strict delay budgets (\ref{LatencyReduction}, \ref{Reliability}). By explicitly modeling performance under finite blocklength conditions, the study provides practical insight into securing aerial assisted URLLC systems against eavesdropping threats while preserving latency and reliability guarantees.



\subsection{Cross-Layer Networking Solutions}\label{main6}

\begin{figure}[ht]
    \centering
    \includegraphics[width=0.95\linewidth]{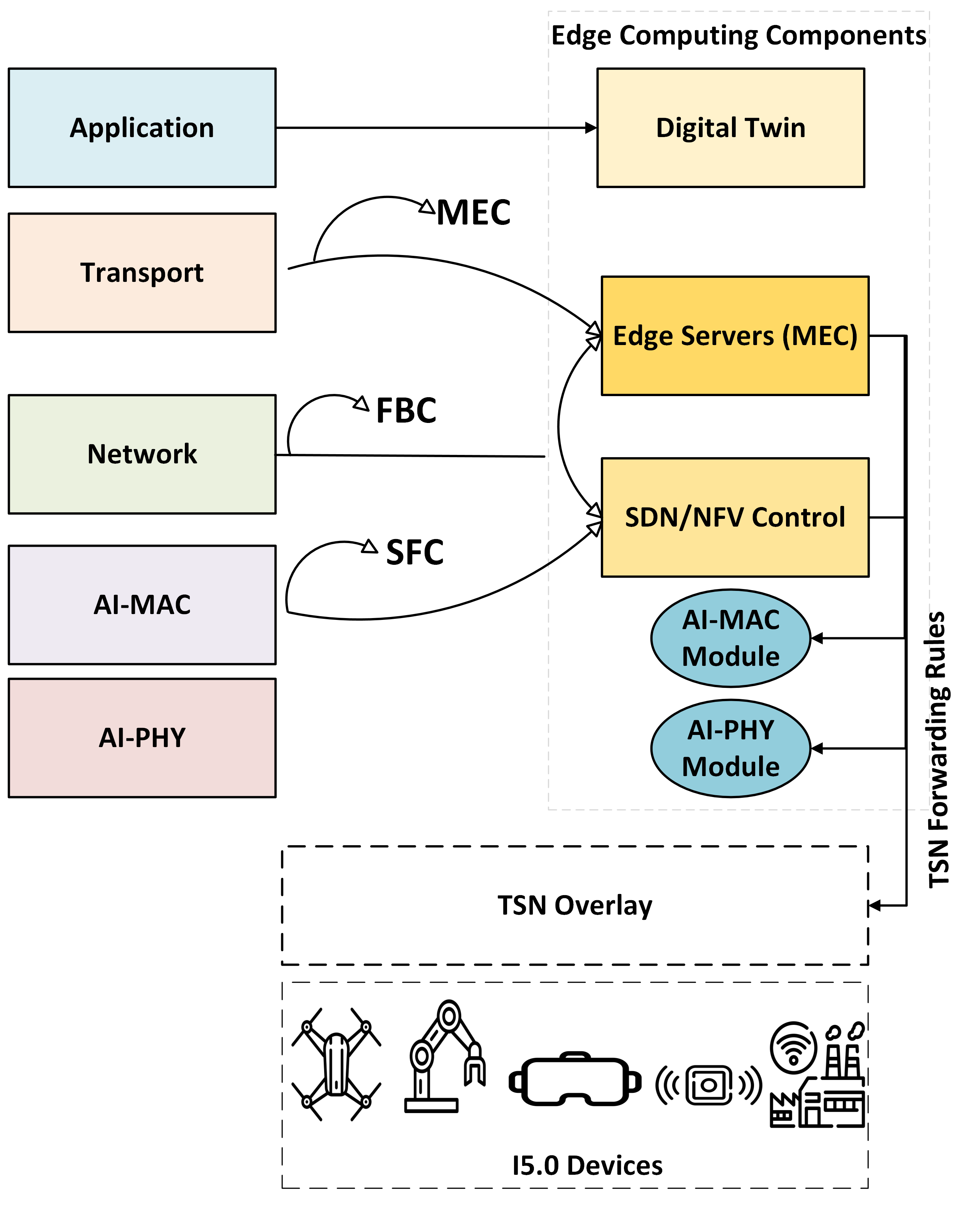}
    \caption{
        An example of cross-layer networking architecture for URLLC in Industry~5.0. The design integrates traditional protocol layers with AI embedded in MAC and PHY modules and uses edge computing components such as digital twin systems, SDN/NFV control, and MEC servers. Resource allocation strategies (MEC, FBC, and SFC) coordinate flows between layers and edge intelligence.  TSN overlays ensure deterministic connectivity between control logic and Industry 5.0 devices, including drones, industrial robots, and AR/VR applications.
    }
    \label{fig:cross_layer_architecture}
\end{figure}

To realize the hard reliability and latency demands of URLLC in Industry~5.0, network architectures must extend the traditional isolated optimization layers. Cross layer design strategies such as integrating physical, MAC, and network layers. This includes AI in physical/MAC designs, TSN overlays, and resource aware protocol stacks that collectively reduce end-to-end delays and ensure predictable reliability. Figure~\ref{fig:cross_layer_architecture} shows an example of cross layer networking architecture for URLLC in Industry~5.0, where protocol layers are coupled with edge intelligence and deterministic control paths. Figure~\ref{fig:cross_layer_architecture} shows integrated AI modules at both the MAC and PHY layers to optimize link adaptation, scheduling, and interference mitigation. FBC bridges the network layer and SDN/NFV (\ref{NFV}) control plane to ensure short packet transmission reliability and SFC connects the MAC layer to edge servers, to ensure dynamic orchestration of URLLC services. MEC (\ref{MEC}) facilitates low latency offloading from the transport layer to digital twin platforms for predictive analytics and cyber physical synchronization. At the lower plane, TSN  overlays propagate deterministic forwarding rules from the edge stack toward I5.0 devices such as industrial robots, drones, and AR/VR systems, maintaining bounded delay and end-to-end reliability. This section presents similar architectural advances and approaches that optimize multiple protocol layers to meet the performance guarantees required by URLLC use cases such as real-time control, autonomous systems, and tactile internet.

In~\cite{ramly2021cross}, Ramly \textit{et al.} presented a cross layer design for achieving URLLC in 5G enabled smart factory environments~ (\ref{SmartManufacturing}). A primary challenge in such settings is the guarantee of deterministic delay and error performance in heterogeneous factory devices, dynamic interference, and varying user densities (\ref{LatencyVsReliability}, \ref{Mobility}). They address these strict requirements, including reliability as low as $10^{-9}$ for applications such as robotic arms and conveyor systems.

To tackle these challenges, the authors proposed a joint optimization framework across the physical and MAC layers. At the physical layer, they utilized Non-Orthogonal Multiple Access and adaptive modulation and coding based on link reliability and SINR feedback. At the MAC layer, a semi-persistent scheduling scheme was integrated, designed to minimize retransmission delay and manage both periodic and sporadic traffic, providing latency advantages by exploiting prior traffic characteristics. The joint design is formulated as a constrained optimization problem, with the objective of minimizing end-to-end latency while satisfying reliability constraints under FBC. The work investigated the performance with various 5G spectrums (3.5 GHz and 28 GHz), speeds (0 km/h, 3 km/h, 7 km/h, and 10 km/h), and frequency diversity. The proposed framework demonstrated significant latency reduction under bursty traffic and dynamic topologies (\ref{LatencyReduction}). The layered adaptation enabled scalable URLLC provisioning, contributing to scalability (\ref{Scalability}), while also maintaining energy efficiency and throughput (\ref{NetworkEfficiency}). These enhancements collectively contribute to reliability improvement (\ref{Reliability}). This work shows the potential of coordinated cross layer design for delivering URLLC services in complex smart manufacturing settings. By jointly optimizing physical-MAC layers, the framework aligns with Industry 5.0 goals of reliable and responsive automation.


Similar to~\cite{ramly2021cross}, Li \textit{et al.}~\cite{li2023cross} addressed the critical challenge of achieving URLLC for industrial automation systems operating over multi-connectivity 5G networks (\ref{IndustrialAutomation}, \ref{NextGen}). A key design challenge lies in jointly allocating cross layer resources to satisfy stringent latency and reliability constraints under varying channel and traffic conditions (\ref{LatencyVsReliability}, \ref{Mobility}) given scarce spectrum resources and dynamic wireless links. The target QoS metrics for URLLC demand 99.999\% reliability for transmitting a 32-byte packet within 1 ms user plane latency.

To solve this, the authors proposed a cross-layer optimization framework that integrates efforts across multiple protocol layers. The framwork utilized techniques such as physical layer packet replication, MAC layer queue state tracking, and network level path diversity. The model leverages multi-connectivity by considering multiple wireless links between the controller and actuators to enhance link robustness. Additionally, they incorporate other cross layer methods such as grant-free contention based access, data replication, and broadcasting, alongside the use of a processor-sharing server in the queue scheduling to reduce pressure on bandwidth requirements. The system jointly optimizes the number of replicated packets per link, the link scheduling order, and transmission power, all aimed at minimizing delay violation probability under URLLC constraints. They derive an expression for the packet loss probability, considering both collision due to the contention based access scheme and decoding error due to dynamic wireless channels. This derivation is used to provide and compare the overall reliability across different transmission types and prove reliability enhancement through multi-connectivity.

The proposed approach has been shown to achieve higher reliability compared to single layer baselines and fixed link assignments (\ref{Reliability}). It adapts transmission redundancy and path allocation dynamically based on channel and queue states. This work enhances system scalability (\ref{Scalability}) and network utilization~(\ref{NetworkEfficiency}) under complex industrial traffic demands by minimizing total bandwidth.


Gao \textit{et al.}~\cite{gao2023multi} addressed the challenge of delivering URLLC compliant services for latency sensitive metaverse applications such as real-time holographic communication~(\ref{Haptics}). The primary difficulty lies in simultaneously satisfying ultra-low latency, high reliability, and immersive QoE requirements under limited wireless and computing resources~(\ref{LatencyVsReliability}, \ref{Mobility}). To address this, the authors proposed a multi-objective optimization framework for joint allocation of communication, computation, and storage resources in edge assisted networks (\ref{MEC}). The system model considers a heterogeneous user base with varying latency budgets, traffic demands, and task data sizes. Reliability is enforced using probabilistic bounds on decoding errors ~(\ref{LatencyReduction}, \ref{Reliability}). The resulting non-convex, integer constrained problem is solved using a decomposition based evolutionary algorithm to approximate the Pareto front, enabling designers to select operating points that best balance URLLC constraints and QoE. Simulation results demonstrate improved delay guarantees (\ref{LatencyReduction}) and higher adaptability~(\ref{ResourceOptimization}).


Wang \textit{et al.}~\cite{wang2024deterministic} proposed a deterministic cross layer framework to support reliable and low latency data delivery in Maritime Internet of Things (MIoT) environments for smart ocean services (\ref{IndustrialAutomation}). The study addresses fundamental URLLC challenges in Industry 5.0, including wide area coverage, multi-hop reliability, and end-to-end delay guarantees over highly variable topologies~(\ref{LatencyVsReliability}, \ref{Mobility}). Key challenges in MIoT include managing link contention and transmission delays due to dynamic and harsh marine meteorological conditions, sea clutter, and complex, rapidly changing network topologies.

The proposed system integrates scheduling at the MAC layer and routing at the network layer through a cross layer design. For deterministic scheduling of mixed traffic flows, the mechanism minimizes end-to-end delay and network resource utilization while ensuring deterministic transmission. It calculates mixed flow delay, channel contention delay, and transmission conflict delay bounds using network delay analysis methods (e.g., WirelessHART). High priority flows always receive transmission precedence, and for flows with identical priority, a network resource proportion minimization algorithm is employed to select available network resources (i.e., slots, channels, wireless interfaces). For high reliability and low delay routing, the authors designed a redundant multi-path protocol based on reliability aware link weights, where each link is evaluated for delay, packet loss, and expected lifetime. The routing mechanism employs a Double Deep Q Network (DDQN) for path selection. To optimize the DDQN's action space and accelerate training, a forwarding node classification algorithm is used to classify and filter candidate nodes based on transmission performance indicators such as link rate, number of neighboring nodes, and node forwarding success rate~(\ref{AI4URLLC}). The routing problem is conceptualized as a Markov Decision Process (MDP), and the reward function is formulated using link transmission rate, node utilization, and data flow delay. The scheduling and routing layers are optimized jointly using an iterative algorithm that minimizes the maximum latency across flows while maintaining redundancy. The algorithms are designed to operate within Software-Defined Networking (SDN) controllers.

 The proposed deterministic scheduling algorithm has been shown to better improve flow schedulability compared to traditional algorithms and consistently maintains a low average end-to-end delay. Similarly, the proposed routing algorithm has been shown to increases the network packet delivery rate and reduces the average end-to-end delay. These enhancements contribute to reliability improvement~(\ref{Reliability}) and end-to-end latency reduction (\ref{LatencyReduction}). The layered design ensures better adaptation to topology dynamics, contributing to scalability~(\ref{Scalability}) and energy efficient routing decisions~(\ref{NetworkEfficiency}). This work highlights the feasibility of deterministic cross layer protocols for URLLC over dynamic wireless infrastructures in next-generation mobile networks (\ref{NextGen}).

Similar to~\cite{gengtian2025deep}, Peng \textit{et al.}~\cite{peng2022resource} addressed the challenge of meeting uplink URLLC demands in cell-free Massive MIMO smart factory environments~(\ref{SmartManufacturing}, \ref{NextGen}). One major design issue is ensuring strict latency and reliability under spatially distributed network topologies and interference limited uplink conditions~(\ref{LatencyVsReliability}, \ref{Mobility}). To tackle this, the authors formulated a resource allocation problem that jointly optimizes user association, access point selection, and power control for uplink transmissions. The goal is to minimize the maximum latency across all users while satisfying reliability constraints~(\ref{LatencyReduction}, \ref{Reliability}).

\begin{figure}[ht]
    \centering
    \includegraphics[width=0.9\linewidth]{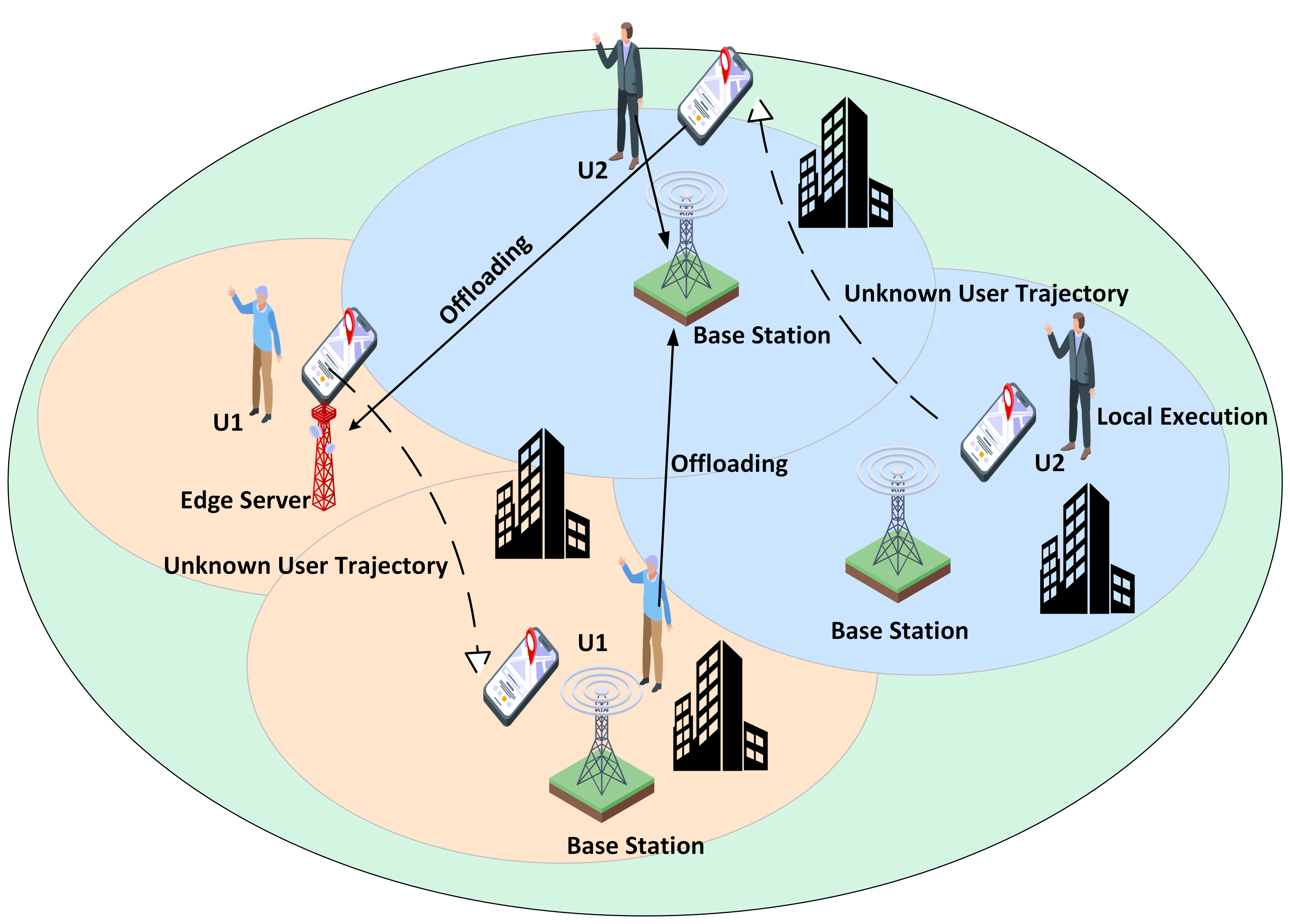}
    \caption{System model of URLLC-aware task offloading in IoHT with user mobility and edge-assisted execution.}
    \label{fig:IoHT_Offloading}
\end{figure}

 The system model as shown in figure \ref{fig:IoHT_Offloading} adopts a cell free architecture with multiple distributed APs coordinated via a CPU. Each UE transmits short packets using FBC, with reliability quantified as the decoding error probability under a target latency threshold. The optimization incorporates reliability via blocklength dependent outage probability bounds and uses a successive convex approximation framework to solve the resulting non-convex problem. The authors propose an iterative two stage algorithm: a) the first stage performs access point selection using user centric heuristics. Due to the practical complexities of implementation, each device is not served by all available access points. Instead, the system is designed so that multiple nearby APs are chosen to provide service to each device. This selection process is based on a threshold driven mechanism. Specifically, the large scale fading parameters ($\beta_{m,k}$), which indicate the signal strength between individual access points and a given device, are sorted in descending order. access points are then sequentially chosen until a predefined condition is satisfied. This condition is expressed as, the sum of the large scale fading parameters for the selected access points ($M_k$) divided by the sum of parameters for all access points ($M$) must be greater than or equal to a certain threshold ($Th$). This approach ensures that each user is served by a specific subset of APs that offer the most robust channel gains, dynamically adapting to the user's location and wireless environment. The second stage optimizes transmit power and rate allocation using convex relaxation~(\ref{ResourceOptimization}). 
 
 The proposed approach has been shown to improve the average weighted sum rate compared to centralized Massive MIMO systems and outperforms existing benchmark algorithms in various scenarios for devices with lower energy budgets. The proposed approach adapts to heterogeneous delay constraints and fluctuating channel states, showing strong scalability~(\ref{Scalability}) and network utilization efficiency~(\ref{NetworkEfficiency}). By using distributed Massive MIMO architectures similar to~\cite{ji2022massive} and latency constrained optimization, this work contributes to realizing uplink URLLC in dynamic industrial automation settings~(\ref{SmartManufacturing}).

\subsection{AI/ML for Industry 5.0}\label{main7}

In dynamic next-generation wireless networks (\ref{NextGen}), accurate channel prediction is fundamental for optimizing communication and ensuring low latency, reliability and security of data transmission. AI, Including DL and RL, offer great advantages for URLLC use cases. For instance, Li \textit{et al.}~\cite{li2020deep} proposed a DRL based joint scheduler that dynamically balances QoS trade-offs between eMBB and URLLC (\ref{NextGen}). Unlike static or risk sensitive models, the authors formulated a multi-objective optimization problem using reliability~(\ref{Reliability}) and service satisfaction level as key metrics. Reliability is quantified through the probability of successful URLLC packet decoding under finite blocklength constraints, where decoding error probabilities exceeding a target threshold (e.g., $10^{-5}$) adds penalties in the reward. The environment is modeled such that the state includes channel quality indicators and queue lengths, the action consists of selecting resource block assignments and puncturing decisions for URLLC packets, and the reward function jointly optimizes service satisfaction level fairness for eMBB and packet error rate constraints for URLLC traffic. They used Deep Deterministic Policy Gradient (DDPG) with prioritized experience replay to train the scheduler on real-time traffic variations. The proposed approach has been shown to increase service satisfaction level~(\ref{ResourceOptimization}) and maintain reliability requirements for URLLC transmissions~(\ref{Reliability}). By using DRL~(\ref{AI4URLLC}), the scheduler learns to reallocate resources under stochastic traffic demands and fluctuating wireless channel conditions~(\ref{Mobility}), thereby enabling latency aware, priority preserving access in mixed traffic scenarios. This supports the goals of URLLC by ensuring intelligent resource orchestration, reduced queuing delays, and improved service continuity under hard latency and reliability constraints~(\ref{LatencyReduction}).


Similar to~\cite{li2020deep}, Setayesh \textit{et al.}~\cite{setayesh2022resource} proposed a hierarchical DL framework for dynamic radio access resource slicing between eMBB and URLLC traffic flows in 5G networks~(\ref{NextGen}). The architecture consists of two levels: a) a Deep Q Network (DQN) for slicing at the base station level; and b) DDPG model for intra-slice resource allocation~(\ref{AI4URLLC}). The system aims to address three major challenges in multi-service slicing: a) maintaining low latency for URLLC~(\ref{LatencyReduction}); b) achieving fairness and throughput for eMBB~(\ref{ResourceOptimization}); and c) adapting to variable traffic conditions~(\ref{Mobility}).
At the first stage, the DQN observes global traffic states such as average packet arrival rate, channel quality, and queue length to allocate bandwidth between eMBB and URLLC slices. At the second stage, the DDPG refines resource block scheduling within each slice based on real-time feedback, queue dynamics, and link quality. The URLLC delay violation probability is included in the reward to guide learning based prioritization. Compared to static slicing and conventional Q-learning baselines, this hierarchical model achieves better delay performance for URLLC and maintaining eMBB throughput under traffic fluctuations. The proposed approach has been shown to reduce average URLLC latency (\ref{LatencyReduction}) and improves eMBB spectral efficiency (\ref{ResourceOptimization}). This work shows how combining layered AI models enables scalable and adaptive slicing strategies to support URLLC's dual demands of reliability and latency~(\ref{Reliability}, \ref{LatencyVsReliability}).

\begin{figure}[htbp]
    \centering
    \includegraphics[width=0.9\linewidth, trim=620 500 200 150, clip]{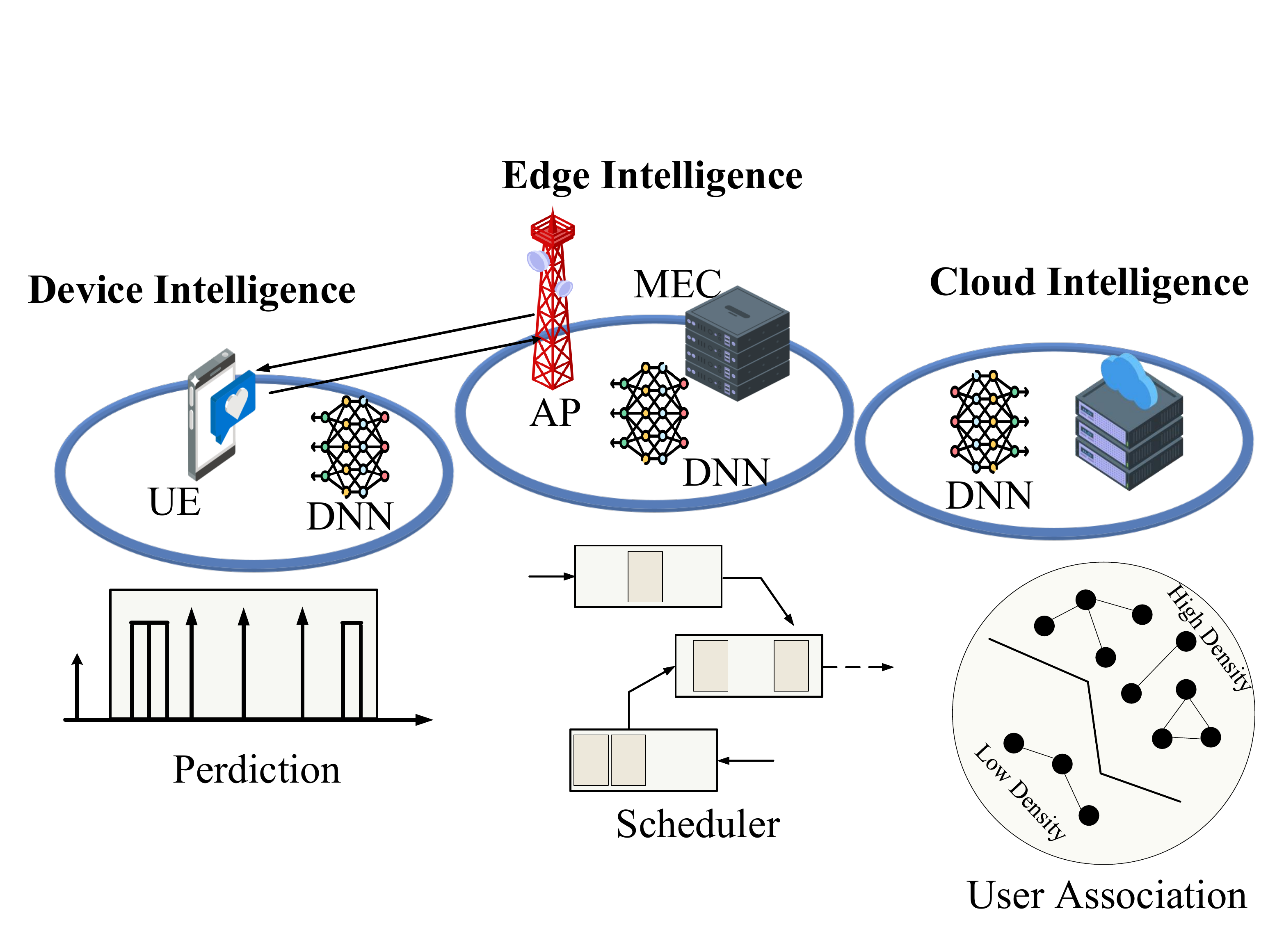}
    \caption{Isometric representation of the multi-level DL architecture integrating device, edge, and cloud intelligence for URLLC in 6G networks.}
    \label{fig:multi_level_dl}
\end{figure}

Similar to~\cite{li2020deep} and~\cite{setayesh2022resource}, She \textit{et al.}~\cite{she2020deep} proposed a multi-level DL architecture that addresses the challenge of meeting URLLC performance requirements (\ref{Mobility}) in non-stationary 6G networks~(\ref{NextGen}). Figure~\ref{fig:multi_level_dl} shows the proposed framework, which integrates device level intelligence for mobility prediction, edge level (\ref{MEC}) intelligence for scheduler design, and cloud level intelligence for user association. At the device level, they use fully connected DNNs trained on tactile mobility traces to predict user movements, achieving a prediction error probability below $10^{-5}$, thereby enhancing reliability guarantees~(\ref{Reliability}). At the cell level, a DRL based actor critic scheduler is developed and evaluated in a live LTE testbed. The scheduler is trained using two DNNs (i.e., actor and critic), each with multiple hidden layers, to minimize packet loss and ensure delay and jitter bounds. They demonstrated real-time performance under strict constraints using pre-training in a digital twin, then fine tuning in real-world deployments~(\ref{DigitalTwinIntegration}, \ref{LatencyReduction}). At the network level, the system uses DNN based user association, trained with large scale channel gains and packet arrival rates, and compared against optimal and baseline models. To address learning in dynamic environments, the authors employed deep transfer learning to adapt pre-trained models with limited new data~(\ref{AI4URLLC}), and applied hierarchical federated learning to enable distributed training while preserving privacy (\ref{Security}) and reducing communication overhead~(\ref{Mobility}). The proposed approach has been shown to improve resource utilization (\ref{ResourceOptimization}) and energy efficient user association across varying service types and user distributions~(\ref{NetworkEfficiency}, \ref{Scalability}). By extending the learning from device to edge, and cloud layers, this study supports the main principles of URLLC by enabling ultra low latency task scheduling, accurate user mobility tracking, and proactive resource orchestration in highly dynamic environments which ensures reliable performance across heterogeneous traffic scenarios~(\ref{LatencyVsReliability}).

Gengtian \textit{et al.}~\cite{gengtian2025deep} proposed a DQN based resource allocation framework for non-orthogonal multiple access smart factory environments~(\ref{SmartManufacturing}, \ref{NextGen}). The study addresses two main URLLC challenges: a) real-time adaptability under traffic fluctuations~(\ref{Mobility}); and b) maintaining strict latency and reliability requirements for heterogeneous industrial nodes~(\ref{LatencyVsReliability}). The goal is to maximize overall system throughput while satisfying URLLC constraints by assigning sub-channels and transmission power levels to factory robots, sensors, and controllers. Specifically, the system allocates lower latency sub-channels and higher transmission priority to critical URLLC nodes such as sensors and controllers, while allocating remaining resources to best effort eMBB devices such as robots~(\ref{ResourceOptimization}).

The environment is modeled as a MDP, where the state includes channel gains and current power levels, and the action defines a selected pair of sub-channel and power allocation for a given user. The reward function is designed to jointly maximize throughput and penalize delay violations through a tunable weight parameter $\lambda$, allowing the agent to flexibly prioritize delay sensitive URLLC flows over throughput driven ones~(\ref{LatencyReduction}). The agent interacts with the environment by observing current network state, selecting resource power assignments, receiving feedback based on reward, and updating its policy accordingly using DQN~(\ref{AI4URLLC}). The network is trained using experience replay and a target Q-network to stabilize learning, which allows it to generalize across varying traffic and interference levels~(\ref{Mobility}).

The proposed framework achieves URLLC level performance by dynamically selecting optimal resource configurations in a shared NOMA spectrum. It has been shown to maintain sub-10~ms latency for URLLC nodes and adaptively reconfigure power channel allocations based on fluctuating demands and interference~(\ref{Reliability}). This study contributes to enabling URLLC by showing how DL can be used to coordinate NOMA scheduling and power control in complex industrial IoT settings. It addresses key challenges of real-time resource allocation, energy-efficient spectrum use~(\ref{NetworkEfficiency}), and scalable policy learning under diverse QoS demands~(\ref{Scalability}).

Zhou \textit{et al.}~\cite{zhou2020learning} addressed the challenge of supporting URLLC for smart healthcare services in Internet of Health Things (IoHT) networks~(\ref{SmartHealthcare}, \ref{NextGen}). Ensuring ultra low latency and high reliability for tasks such as remote health monitoring and emergency diagnosis is challenging due to device heterogeneity, dynamic channel conditions, constrained computation~(\ref{LatencyVsReliability}, \ref{Mobility}), and the need for privacy preserving solutions~(\ref{Security}). The authors proposed a learning based task offloading approach that uses federated DRL model to optimize computation offloading decisions under URLLC constraints. The model is based on  the exponential weight algorithm for exploration and exploitation to optimize computation offloading decisions under URLLC constraints. The proposed approach adopts a hybrid architecture where IoHT devices decide whether to locally execute or offload tasks to edge servers. A federated DQN is used to train local agents collaboratively without sharing raw data, preserving patient privacy~(\ref{Security}). Each agent observes a state composed of local queue length, channel quality, and device energy. The action includes binary offloading decisions and edge server selections. The reward function includes task success rate and delay violations in order to enforce URLLC compliance. Agents update local policies using prioritized experience replay and synchronize with a global model to adapt to environmental changes~(\ref{AI4URLLC}). The proposed approach has been shown to reduce average task latency (\ref{LatencyReduction}) and improve offloading success rate (\ref{Reliability}) compared to traditional approaches. This enables URLLC support for delay sensitive healthcare applications such as real-time ECG monitoring and AR assisted surgery. By integrating privacy aware learning, energy-efficient offloading, and delay sensitive scheduling, the work contributes to reliable and low latency healthcare delivery in dense IoHT environments.

Kokkinis \textit{et al.}~\cite{kokkinis2025deep} addressed the challenge of synchronizing video and haptic streams under URLLC  constraints in tactile Internet environments, which falls under the application domain of Remote Control and Haptics~(\ref{Haptics}). A key issue they focused on was meeting distinct service level requirements for haptic and video traffic, such as sub-10 ms latency for haptic control loops and desynchronization thresholds below 50 ms for user perception, while maintaining spectral efficiency and robustness amidst fluctuating wireless conditions. These challenges inherently involve latency vs. reliability trade-offs~(\ref{LatencyVsReliability}) and mobility management~(\ref{Mobility}). The authors proposed a soft actor-critic based Deep Reinforcement Learning (DRL) framework, leveraging AI/ML for URLLC~(\ref{AI4URLLC}), to optimize radio resource allocation between haptic (URLLC slice) and video (eMBB slice) modalities. The DRL environment models state by capturing average latency, packet loss, data rate, spectrum efficiency, and buffer occupancy per slice, with actions corresponding to discrete allocations of resource blocks. Their reward function penalizes violations of service-specific constraints, including slice-wise data rate~(\ref{ResourceOptimization}), packet loss~(\ref{Reliability}), and inter-modal synchronization~(\ref{LatencyReduction}). The system reduces computational complexity in edge deployments by performing inter-slice scheduling every 50 ms instead of 1 ms. The proposed framework has been shown to achieve a higher satisfaction rate~(\ref{NetworkEfficiency}) under high user density~(\ref{Scalability}) and spectrum efficiency fluctuations compared to traditional approaches. By modeling heterogeneous service demands and optimizing resource allocation via DRL, their approach enables stable, low latency, and synchronized teleoperation services, contributing to the evolving capabilities of next-generation mobile networks (\ref{NextGen}).

The above studies show that AI and ML techniques are increasingly central to enabling URLLC in 6G enabled Industry~5.0 applications. From optimizing dynamic slicing and uplink scheduling to managing delay constrained offloading and haptic video synchronization, AI agents provide scalable and adaptive control policies under stochastic traffic and wireless dynamics~(\ref{AI4URLLC}). These learning based approaches: a) improve latency reliability trade-offs~(\ref{LatencyVsReliability}); and b) supports intelligent orchestration across mobility intensive~(\ref{Mobility}) and heterogeneous service environments. This is done by bridging real-time perception, decision-making, and control, in order to  contribute to fulfilling the end-to-end URLLC objectives of latency reduction~(\ref{LatencyReduction}), reliability improvement~(\ref{Reliability}), and resource efficiency~(\ref{ResourceOptimization}, \ref{NetworkEfficiency}), which in turn makes it one of the main enablers of intelligent and responsive Industry~5.0 systems.

\section{Open Challenges and Future Research Directions}\label{sec:challenges}

The realization of URLLC in Industry 5.0 applications (\ref{apps}), faces several complex design challenges (\ref{chall}), that necessitate continued research across various domains on the main enablers of industry 5.0 (\ref{tech}), to achieve several significant performance enhancements (\ref{performance}). While significant progress has been made in addressing the stringent demands of low latency (\ref{LatencyReduction}), high reliability (\ref{Reliability}), and other performance metrics (\ref{ResourceOptimization},\ref{NetworkEfficiency},\ref{Scalability}), the practical deployment and widespread adoption of URLLC enabled systems still present significant challenges.

\subsection{Semantic informed Resource Orchestration for URLLC}\label{openissue1}

Semantic informed resource orchestration, as investigated in~\cite{shokrnezhad2024semantic}, represents a promising future direction in achieving URLLC for mission critical applications in (\ref{SmartManufacturing}), (\ref{ConnectedVehicles}), and (\ref{SmartHealthcare}). Traditional URLLC systems rely on physical layer metrics such as SNR, packet error rate, and blocklength aware reliability guarantees~\cite{li2024reliability}. However, these do not account for the meaning or contextual importance of the transmitted data. In many Industry~5.0 use cases, the semantic content such as anomaly reports in a factory, collision warnings in autonomous driving, or abnormal vitals in IoHT, is more important than other routine data. Transmitting all data with equal latency and reliability requirements is both inefficient and unsustainable for real-time systems operating under resource constraints.

AI enabled schedulers~\cite{li2020deep,setayesh2022resource} and cross layer solutions~\cite{ramly2021cross,li2023cross} have demonstrated the ability to reduce delay and improve fairness through channel and traffic prediction. However, they are fundamentally unaware of the semantic value of each packet. Future research should develop semantic informed scheduling frameworks that learn to prioritize transmission based on the inferred utility of the content alongside the traditional predictions of channel state and queue dynamics. This would involve integrating semantic encoders that map data into task informed embeddings~\cite{gao2023multi}, and building semantic importance estimators within the RL agent~(\ref{AI4URLLC}). Resource allocation policies then, would maximize raw throughput as well task level performance metrics, such as prediction accuracy, control loop stability, or diagnostic latency.

To enable this, new training paradigms must optimize communication and inference objectives jointly. For example, the reward in a semantic informed DRL scheduler can include downstream task loss or missed detection penalties to reflect semantic degradation. Moreover, standard reliability definitions (e.g., packet delivery ratio) must be extended to semantic reliability, quantifying the effect of dropped packets on task utility~(\ref{Reliability}, \ref{LatencyVsReliability}). By aligning URLLC objectives with application level semantics, future systems can achieve intelligent resource orchestration that improves both performance and efficiency under tight latency constraints~(\ref{LatencyReduction}, \ref{ResourceOptimization}, \ref{NetworkEfficiency}).

\begin{figure}[!t]
  \centering
  \includegraphics[width=0.47\textwidth]{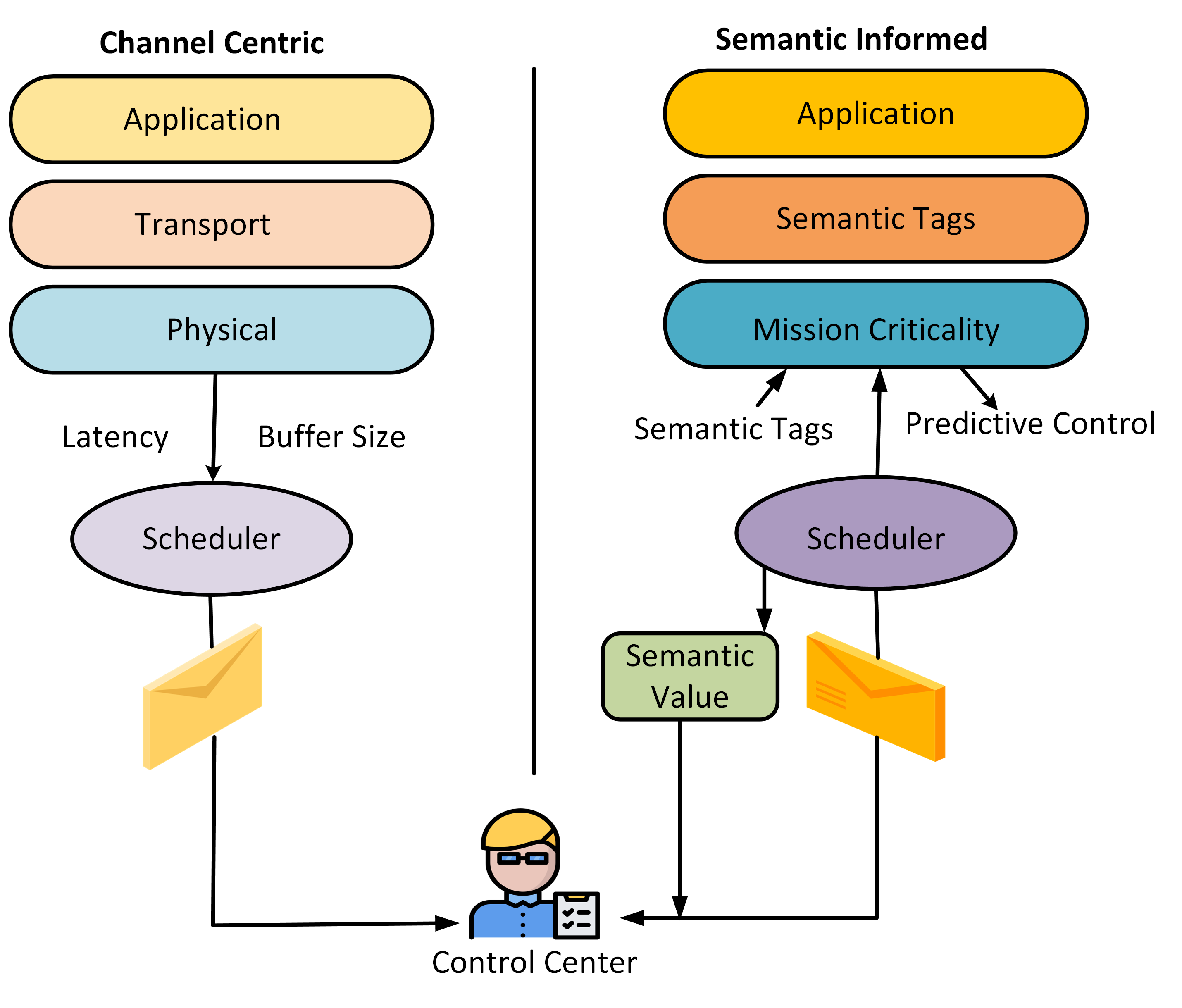}
  \caption{Comparison between traditional channel centric resource scheduling and semantic informed URLLC orchestration. The left model allocates resources based on physical metrics such as latency and buffer size, while the right integrates semantic tags, task criticality, and predictive control to enhance mission aware decision making. This enables selective transmission of high value data, reducing bandwidth demands and maintain reliability and responsiveness.}
  \label{fig:semantic_scheduler}
\end{figure}


\begin{figure*}[!t]
    \centering
    \includegraphics[width=0.9\linewidth, trim=450 0 0 150, clip]{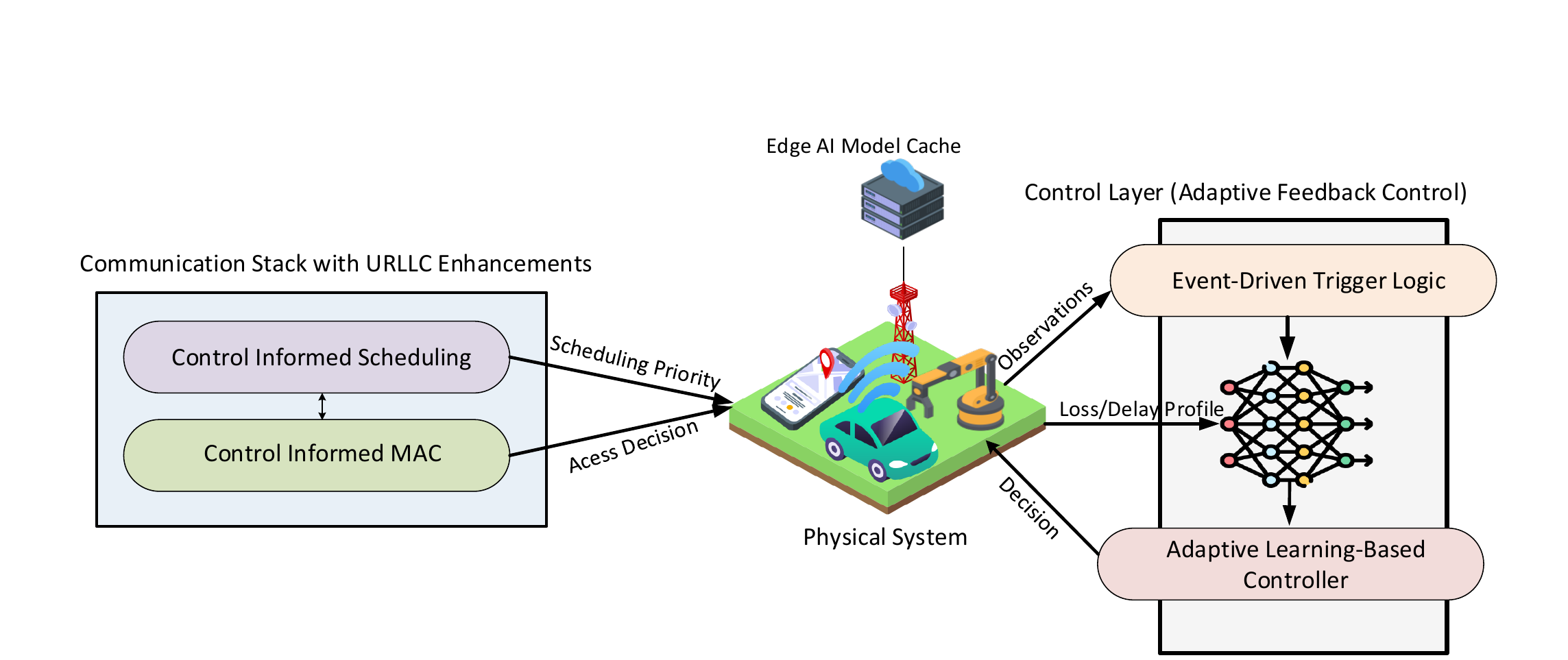}
    \caption{A communication control co-design framework for distributed low latency systems in Industry~5.0. The framework captures the closed loop interaction between intelligent controllers and latency informed communication modules. Control informed scheduling, adaptive MAC, and event triggered communication minimize transmission overhead and preserves feedback stability. Network informed controllers utilize probabilistic models to tolerate bounded delay and stochastic loss, jointly optimizing communication cost, control accuracy, and actuation latency.}
    \label{fig:comm_control_codesign}
\end{figure*}

\subsection{Dynamic Digital Twin Adaptation under Uncertainty}\label{openissue2}

Dynamic digital twin adaptation under uncertainty is an open challenge for enabling closed loop URLLC in complex cyber physical environments such as (\ref{SmartManufacturing}), (\ref{ConnectedVehicles}), and (\ref{Haptics}). Recent studies have demonstrated the potential of digital twin based frameworks to synchronize physical and virtual environments~\cite{xiao2023research,qu2023digital}. However, their effectiveness deteriorates in non-stationary settings where system dynamics, network conditions, or environmental variables evolve rapidly. Existing digital twin implementations assume static model parameters, ideal sensing conditions, and deterministic update intervals~\cite{imtiaz2025digital}, which limits their use in real-time mission critical operations requiring sub-10~ms response and 99.999\% reliability.

The key technical challenge is in maintaining twin fidelity and control relevance in the presence of unpredictable factors such as sensor noise, packet loss, link handovers, and partial observability~(\ref{Mobility}, \ref{LatencyVsReliability}). Most digital twins rely on deterministic model driven simulations that are insufficient to capture such uncertainty. Moreover, the update policies for digital twin synchronization are static or periodic, ignoring the urgency or relevance of the physical changes being mirrored. This mismatch between reality and its virtual representation leads to suboptimal decision making, violating both latency and safety guarantees~(\ref{Reliability}, \ref{LatencyReduction}).

To address this, future research must explore uncertainty aware twin adaptation using probabilistic models and event driven synchronization schemes. For instance, Bayesian neural networks or ensemble learning could be used to model the confidence of twin predictions. This will ensure adaptive trust calibration and fallback to physical sensing when needed~(\ref{AI4URLLC}). In addition, event triggered update policies should replace periodic digital twin synchronization to reduce latency and network overhead. These policies would prioritize updates based on deviation thresholds, control instability risk, or semantic urgency of state changes. Furthermore, the control plane must be co-designed with the digital twin, in which, the decision logic dynamically accounts for both twin fidelity and actuation deadlines. DRL frameworks that jointly optimize twin accuracy, update cost, and control latency can provide robust performance under uncertainty. As digital twins scale in complexity and interact with heterogeneous devices, a layered architecture that integrates fast edge level twins with slower cloud level models could offer better scalability~(\ref{Scalability}, \ref{NetworkEfficiency}). The goal is to develop adaptive digital twins that self-assess, self-correct, and self-prioritize their update logic. This will be essential to enable reliable, low latency operation in next-generation Industry~5.0 networks~(\ref{NextGen}).


\subsection{Cross-Domain Consistency and Time Synchronization in Federated URLLC}\label{openissue3}

Cross domain consistency and time synchronization in federated URLLC remains a challenge to reliable, low latency service delivery in decentralized industrial environments such as \ref{SmartManufacturing}, \ref{Haptics}, and \ref{IndustrialAutomation}. Emerging URLLC use cases consist of multiple administrative and physical domains, requiring cooperative behavior across factory floors, mobile edge nodes, remote teleoperators, and cloud based analytics platforms~\cite{li2023low,forrai2024application}. However, inconsistencies in data semantics, control loops, and timing granularity between these domains lead to performance degradation, including latency jitter, state mismatch, and actuation conflicts~(\ref{LatencyVsReliability}, \ref{CrossDomain}).

A key challenge is maintaining accurate time alignment and state consistency across loosely synchronized subsystems operating under heterogeneous constraints. Most current systems rely on simplified time protocols such as network time protocol and global positioning system, which are insufficient for sub-millisecond URLLC requirements. In addition, semantic uncertainty, where local interpretations of shared state evolve independently, undermine deterministic control. These limitations are critical in federated architectures, where entities exchange processed features or learned policies instead of raw data, such as in federated DRL based task offloading for healthcare~\cite{zhou2020learning,she2020deep}.

To address this, future work must integrate time synchronization mechanisms such as IEEE~1588 precision time protocol into URLLC control planes and extend them across domains using boundary clocks and transparent gateways. Semantic consistency must also be enforced. Promising directions include the use of semantic metadata channels and vector clocks to tag control messages with causal context and freshness~(\ref{Security}, \ref{Mobility}). Furthermore, distributed consensus mechanisms that are latency bounded, such as real-time variants of Raft or Paxos which are fault tolerant protocols used in cloud systems to achieve consensus among distributed nodes without relying on a single leader, could help ensure that multi-domain agents agree on shared states within the constraints of URLLC~(\ref{Reliability}). Although these are traditional protocols; real-time adaptations of these protocols could be optimized for bounded latency and minimal communication overhead, making them suitable for time critical URLLC applications across federated domains.
 Similarly, hybrid synchronization models combining time driven and event driven updates may also provide greater flexibility, where non-critical updates use delay tolerant strategies and mission critical tasks rely on tightly synchronized transactions. As federated URLLC networks become more common in Industry~5.0, ensuring temporal and semantic alignment across trust boundaries will be essential for stable and trustworthy system orchestration~(\ref{NextGen}).


\subsection{Learning under Sparse Feedback and Safety Constraints}\label{openissue4}

In URLLC systems designed for Industry~5.0, RL and DL based agents are employed to optimize scheduling, routing, and task offloading~(\ref{AI4URLLC}). However, these agents require dense, trial and error interaction with the environment to learn effective policies, which conflicts with the stringent reliability and safety demands of critical applications~(\ref{Reliability}). In applications such as remote surgery~(\ref{SmartHealthcare}) or autonomous factory robotics~(\ref{SmartManufacturing}), allowing an RL agent to explore suboptimal actions can lead to severe performance degradation or operational failure. Moreover, safety critical systems often provide sparse or delayed feedback, where the impact of decisions (e.g., a failure in end-to-end haptic coordination~\cite{kokkinis2025deep}) is only observable after a cascade of dependent tasks, complicating reward shaping and convergence. 

To address these limitations, future research should focus on constrained RL frameworks, where policies are learned under strict safety envelopes defined by probabilistic bounds on latency, reliability, or energy usage~(\ref{LatencyVsReliability}). Safe policy optimization methods such as constrained policy optimization approaches offer promising directions. Additionally, model based RL that simulates worst case transitions using a digital twin~(\ref{DigitalTwinIntegration}) can reduce unsafe exploration during early learning stages. Another viable research direction is offline RL, where the agent is trained from high quality datasets collected from expert demonstrations or prior logs, minimizing risky exploration. These learning paradigms must also adapt to variable feedback sparsity by incorporating memory based credit assignment and temporal abstraction to better propagate delayed outcomes. Achieving sample efficient, safety aware learning under sparse feedback will be essential to scaling AI driven URLLC systems in domains where failure is costly and information is limited.

\subsection{Communication Control Co-Design for Distributed Low Latency Feedback Systems}\label{openissue5}

As URLLC systems become embedded in complex industrial cyber physical environments, the interaction between control theory and communication theory must be revisited through a unified co-design perspective. Classical control design assumes ideal communication links, while communication protocols treat control messages as generic data. This isolated approach fails under the tight timing and reliability demands of distributed real-time control systems in Industry~5.0~(\ref{IndustrialAutomation}, \ref{SmartManufacturing}). For instance, in wireless feedback control loops used in high speed robotics or drone swarm coordination, even a single delayed or dropped packet can lead to control instability~(\ref{LatencyVsReliability}, \ref{Reliability}). In addition, traditional control theory does not account for time varying latency or probabilistic packet error rates inherent to wireless environments.

To overcome this, future research must develop communication control co-design frameworks that explicitly model the mutual dependencies between control loop stability and network resource behavior. One promising direction is the integration of control informed MAC and scheduling algorithms, where communication decisions prioritize system states with higher control sensitivity. Similarly, learning based controllers can be trained to tolerate bounded stochastic delay or loss profiles by incorporating network statistics into their internal models~\cite{li2020deep}. Another future direction is to extend event triggered control strategies to network aware triggering, where transmissions occur only when system state deviations exceed both control thresholds and network cost functions. Such approaches can reduce bandwidth and preserve control quality under tight latency budgets~(\ref{ResourceOptimization}). Figure~\ref{fig:comm_control_codesign} shows this interaction: on the left, the communication stack includes an adaptive MAC and a control informed scheduler, which prioritize time critical transmissions and suppress non-impactful updates using event triggered logic. On the right, controllers use network informed feedback, such as channel state estimates and delay statistics, to maintain loop stability under imperfect feedback. These controllers are further supported by an Edge AI model cache, which enables acceptable degradation by predicting actuator inputs during transient disconnections. Together, these pathways minimize feedback uncertainty and optimize end-to-end latency by enabling joint adaptation across the control and communication layers. The integration of adaptive event triggered control, low latency communication, and real-time learning across the stack forms a foundational pillar for robust cyber-physical operation in Industry 5.0 (\ref{NextGen}).

\section{Conclusion}\label{conclusion}
This paper has reviewed URLLC solutions for 6G enabled Industry 5.0 through a structured taxonomy linking applications, enabling technologies, design challenges, and performance enhancements. Across diverse domains including industrial automation, autonomous vehicles, remote control, healthcare, smart manufacturing, and next-generation mobile networks state-of-the-art approaches have been analyzed in terms of latency reduction, reliability improvement, resource optimization, network efficiency, and scalability. Key enabling technologies such as MEC, network slicing, NFV/SFC, AI/ML frameworks, and digital twin integration have been shown to significantly enhance URLLC performance, while addressing challenges in cross domain networking, security, mobility management, and latency reliability trade-offs. Representative studies demonstrate the role of cross layer optimization, predictive intelligence, and reconfigurable environments in meeting stringent end-to-end QoS targets under dynamic industrial conditions. However, open issues remain in areas such as zero slack security, scalable channel estimation for RIS, adaptive orchestration across heterogeneous edge resources, and energy informed mobility management. Addressing these challenges will be essential for delivering deterministic, sustainable, and human centric URLLC systems, thereby enabling the full realization of Industry 5.0’s vision of intelligent, collaborative, and resilient industrial ecosystems.

\section*{Acknowledgments}
This work was supported by Malaysia's Ministry of Higher Education, under Fundamental Research Grant Scheme (Ref: FRGS/1/2022/ICT09/SYUC/03/1). The authors gratefully acknowledge the financial support that made this research possible.

\bibliographystyle{IEEEtran}

\bibliography{references} 

\begin{thebibliography}{10}
\providecommand{\url}[1]{#1}
\csname url@samestyle\endcsname
\providecommand{\newblock}{\relax}
\providecommand{\bibinfo}[2]{#2}
\providecommand{\BIBentrySTDinterwordspacing}{\spaceskip=0pt\relax}
\providecommand{\BIBentryALTinterwordstretchfactor}{4}
\providecommand{\BIBentryALTinterwordspacing}{\spaceskip=\fontdimen2\font plus
\BIBentryALTinterwordstretchfactor\fontdimen3\font minus \fontdimen4\font\relax}
\providecommand{\BIBforeignlanguage}[2]{{%
\expandafter\ifx\csname l@#1\endcsname\relax
\typeout{** WARNING: IEEEtran.bst: No hyphenation pattern has been}%
\typeout{** loaded for the language `#1'. Using the pattern for}%
\typeout{** the default language instead.}%
\else
\language=\csname l@#1\endcsname
\fi
#2}}
\providecommand{\BIBdecl}{\relax}
\BIBdecl

\bibitem{tallat2023navigating}
R.~Tallat, A.~Hawbani, X.~Wang, A.~Al-Dubai, L.~Zhao, Z.~Liu, G.~Min, A.~Y. Zomaya, and S.~H. Alsamhi, ``Navigating industry 5.0: A survey of key enabling technologies, trends, challenges, and opportunities,'' \emph{IEEE Communications Surveys \& Tutorials}, vol.~26, no.~2, pp. 1080--1126, 2023.

\bibitem{asad2023human}
U.~Asad, M.~Khan, A.~Khalid, and W.~A. Lughmani, ``Human-centric digital twins in industry: A comprehensive review of enabling technologies and implementation strategies,'' \emph{Sensors}, vol.~23, no.~8, p. 3938, 2023.

\bibitem{langaas2025exploring}
E.~F. Lang{\aa}s, M.~H. Zafar, and F.~Sanfilippo, ``Exploring the synergy of human-robot teaming, digital twins, and machine learning in industry 5.0: A step towards sustainable manufacturing,'' \emph{Journal of Intelligent Manufacturing}, pp. 1--24, 2025.

\bibitem{shamsabadi2025exploring}
A.~A. Shamsabadi, A.~Yadav, Y.~Gadallah, and H.~Yanikomeroglu, ``Exploring the 6g potentials: Immersive, hyperreliable, and low-latency communication,'' \emph{IEEE Vehicular Technology Magazine}, 2025.

\bibitem{kerboeuf2024design}
S.~Kerboeuf, P.~Porambage, A.~Jain, P.~Rugeland, G.~Wikstr{\"o}m, M.~Ericson, D.~T. Bui, A.~Outtagarts, H.~Karvonen, P.~Alemany \emph{et~al.}, ``Design methodology for 6g end-to-end system: Hexa-x-ii perspective,'' \emph{IEEE Open Journal of the Communications Society}, 2024.

\bibitem{zhao2018deepthings}
Z.~Zhao, K.~M. Barijough, and A.~Gerstlauer, ``Deepthings: Distributed adaptive deep learning inference on resource-constrained iot edge clusters,'' \emph{IEEE Transactions on Computer-Aided Design of Integrated Circuits and Systems}, vol.~37, no.~11, pp. 2348--2359, 2018.

\bibitem{wang2019physical}
N.~Wang, P.~Wang, A.~Alipour-Fanid, L.~Jiao, and K.~Zeng, ``Physical-layer security of 5g wireless networks for iot: Challenges and opportunities,'' \emph{IEEE internet of things journal}, vol.~6, no.~5, pp. 8169--8181, 2019.

\bibitem{yang2025towards}
L.~Yang, S.~Naser, A.~Shami, S.~Muhaidat, L.~Ong, and M.~Debbah, ``Towards zero touch networks: Cross-layer automated security solutions for 6g wireless networks,'' \emph{IEEE Transactions on Communications}, 2025.

\bibitem{ghobakhloo2023behind}
M.~Ghobakhloo, M.~Iranmanesh, M.-L. Tseng, A.~Grybauskas, A.~Stefanini, and A.~Amran, ``Behind the definition of industry 5.0: a systematic review of technologies, principles, components, and values,'' \emph{Journal of Industrial and Production Engineering}, vol.~40, no.~6, pp. 432--447, 2023.

\bibitem{dang2020should}
S.~Dang, O.~Amin, B.~Shihada, and M.-S. Alouini, ``What should 6g be?'' \emph{Nature Electronics}, vol.~3, no.~1, pp. 20--29, 2020.

\bibitem{chowdhury20206g}
M.~Z. Chowdhury, M.~Shahjalal, S.~Ahmed, and Y.~M. Jang, ``6g wireless communication systems: Applications, requirements, technologies, challenges, and research directions,'' \emph{IEEE Open Journal of the Communications Society}, vol.~1, pp. 957--975, 2020.

\bibitem{verma2022blockchain}
A.~Verma, P.~Bhattacharya, N.~Madhani, C.~Trivedi, B.~Bhushan, S.~Tanwar, G.~Sharma, P.~N. Bokoro, and R.~Sharma, ``Blockchain for industry 5.0: Vision, opportunities, key enablers, and future directions,'' \emph{Ieee Access}, vol.~10, pp. 69\,160--69\,199, 2022.

\bibitem{hazra20246g}
A.~Hazra, A.~Munusamy, M.~Adhikari, L.~K. Awasthi \emph{et~al.}, ``6g-enabled ultra-reliable low latency communication for industry 5.0: challenges and future directions,'' \emph{IEEE Communications Standards Magazine}, vol.~8, no.~2, pp. 36--42, 2024.

\bibitem{jahid2023convergence}
A.~Jahid, M.~H. Alsharif, and T.~J. Hall, ``The convergence of blockchain, iot and 6g: Potential, opportunities, challenges and research roadmap,'' \emph{Journal of Network and Computer Applications}, vol. 217, p. 103677, 2023.

\bibitem{babaei2025assessing}
A.~Babaei, E.~B. Tirkolaee, and S.~S. Ali, ``Assessing the viability of blockchain technology in renewable energy supply chains: A consolidation framework,'' \emph{Renewable and Sustainable Energy Reviews}, vol. 212, p. 115444, 2025.

\bibitem{chataut20246g}
R.~Chataut, M.~Nankya, and R.~Akl, ``6g networks and the ai revolution—exploring technologies, applications, and emerging challenges,'' \emph{Sensors}, vol.~24, no.~6, p. 1888, 2024.

\bibitem{mahmood2023ultra}
N.~H. Mahmood, I.~Atzeni, E.~A. Jorswieck, O.~L. Alcaraz~Lopez \emph{et~al.}, ``Ultra-reliable low-latency communications: Foundations, enablers, system design, and evolution towards 6g,'' \emph{Foundations and Trends® in Communications and Information Theory}, vol.~20, no. 5-6, pp. 512--747, 2023.

\bibitem{haque2024comprehensive}
M.~Haque, F.~Tariq, M.~Hossain, M.~Khandaker, M.~Imran, and K.~Wong, ``A comprehensive survey of 5g urllc and challenges in the 6g era,'' \emph{IEEE Communications Surveys \& Tutorials}, 2024.

\bibitem{salh2021survey}
A.~Salh, L.~Audah, N.~S.~M. Shah, A.~Alhammadi \emph{et~al.}, ``A survey on deep learning for ultra-reliable and low-latency communications challenges on 6g wireless systems,'' \emph{IEEE Access}, vol.~9, pp. 55\,098--55\,131, 2021.

\bibitem{sefati2023ultra}
S.~S. Sefati and S.~Halunga, ``Ultra-reliability and low-latency communications on the internet of things based on 5g network: literature review, classification, and future research view,'' \emph{Transactions on Emerging Telecommunications Technologies}, vol.~34, no.~6, p. e4770, 2023.

\bibitem{husain20223gpp}
S.~Husain, A.~Kunz, and J.~Song, ``3gpp 5g core network: An overview and future directions,'' 2022.

\bibitem{chen2019channel}
J.~Chen, Y.-C. Liang, H.~V. Cheng, and W.~Yu, ``Channel estimation for reconfigurable intelligent surface aided multi-user mimo systems,'' \emph{arXiv preprint arXiv:1912.03619}, vol. 122, 2019.

\bibitem{liu2024network}
R.~Liu, J.~Shi, X.~Chen, and C.~Lu, ``Network anomaly detection and security defense technology based on machine learning: A review,'' \emph{Computers and Electrical Engineering}, vol. 119, p. 109581, 2024.

\bibitem{ibrahim2023multi}
A.~M. Ibrahim, M.~H. Ling, and K.-L.~A. Yau, ``Multi-agent deep reinforcement learning for resource allocation in 5g and 6g networks,'' in \emph{2023 IEEE International Conference on Computing (ICOCO)}.\hskip 1em plus 0.5em minus 0.4em\relax IEEE, 2023, pp. 225--231.

\bibitem{she2017cross}
C.~She, C.~Yang, and T.~Q. Quek, ``Cross-layer optimization for ultra-reliable and low-latency radio access networks,'' \emph{IEEE Transactions on Wireless Communications}, vol.~17, no.~1, pp. 127--141, 2017.

\bibitem{peng2022resource}
Q.~Peng, H.~Ren, C.~Pan, N.~Liu, and M.~Elkashlan, ``Resource allocation for uplink cell-free massive mimo enabled urllc in a smart factory,'' \emph{IEEE Transactions on Communications}, vol.~71, no.~1, pp. 553--568, 2022.

\bibitem{chang2021autonomous}
B.~Chang, L.~Li, G.~Zhao, Z.~Chen, and M.~A. Imran, ``Autonomous d2d transmission scheme in urllc for real-time wireless control systems,'' \emph{IEEE Transactions on Communications}, vol.~69, no.~8, pp. 5546--5558, 2021.

\bibitem{zhou2020learning}
Z.~Zhou, Z.~Wang, H.~Yu, H.~Liao, S.~Mumtaz, L.~Oliveira, and V.~Frascolla, ``Learning-based urllc-aware task offloading for internet of health things,'' \emph{IEEE Journal on Selected Areas in Communications}, vol.~39, no.~2, pp. 396--410, 2020.

\bibitem{le2020overview}
T.-K. Le, U.~Salim, and F.~Kaltenberger, ``An overview of physical layer design for ultra-reliable low-latency communications in 3gpp releases 15, 16, and 17,'' \emph{IEEE access}, vol.~9, pp. 433--444, 2020.

\bibitem{mourtzis2021smart}
D.~Mourtzis, J.~Angelopoulos, and N.~Panopoulos, ``Smart manufacturing and tactile internet based on 5g in industry 4.0: Challenges, applications and new trends,'' \emph{Electronics}, vol.~10, no.~24, p. 3175, 2021.

\bibitem{enenche2023network}
P.~Enenche, D.~H. Kim, and D.~You, ``Network coding as enabler for achieving urllc under tcp and udp environments: A survey,'' \emph{IEEE Access}, vol.~11, pp. 76\,647--76\,674, 2023.

\bibitem{tamim2023intelligent}
I.~Tamim, S.~Aleyadeh, and A.~Shami, ``Intelligent o-ran traffic steering for urllc through deep reinforcement learning,'' \emph{arXiv preprint arXiv:2303.01960}, 2023.

\bibitem{queiroz2024new}
G.~A. Queiroz and E.~R. da~Silva, ``A new channel and qos aware scheduler algorithm for real time and non real time traffic in 5g heterogeneous networks,'' \emph{IEEE Latin America Transactions}, vol.~22, no.~8, pp. 659--669, 2024.

\bibitem{she2020deep}
C.~She, R.~Dong, Z.~Gu, Z.~Hou, Y.~Li, W.~Hardjawana, C.~Yang, L.~Song, and B.~Vucetic, ``Deep learning for ultra-reliable and low-latency communications in 6g networks,'' \emph{IEEE network}, vol.~34, no.~5, pp. 219--225, 2020.

\bibitem{zeb2021edge}
S.~Zeb, M.~A. Rathore, A.~Mahmood, S.~A. Hassan, J.~Kim, and M.~Gidlund, ``Edge intelligence in softwarized 6g: Deep learning-enabled network traffic predictions,'' \emph{arXiv preprint arXiv:2108.00332}, 2021.

\bibitem{ibrahim2022implications}
A.~M. Ibrahim, K.-L.~A. Yau, and L.~M. Hong, ``Implications of centralized and distributed multi-agent deep reinforcement learning in dynamic spectrum access,'' in \emph{2022 IEEE 6th International Symposium on Telecommunication Technologies (ISTT)}.\hskip 1em plus 0.5em minus 0.4em\relax IEEE, 2022, pp. 62--67.

\bibitem{ostman2021urllc}
J.~{\"O}stman, A.~Lancho, G.~Durisi, and L.~Sanguinetti, ``Urllc with massive mimo: Analysis and design at finite blocklength,'' \emph{IEEE Transactions on Wireless Communications}, vol.~20, no.~10, pp. 6387--6401, 2021.

\bibitem{shi2022decentralized}
E.~Shi, J.~Zhang, J.~Zhang, D.~W.~K. Ng, and B.~Ai, ``Decentralized coordinated precoding design in cell-free massive mimo systems for urllc,'' \emph{IEEE Transactions on Vehicular Technology}, vol.~72, no.~2, pp. 2638--2642, 2022.

\bibitem{farhad2023terahertz}
A.~Farhad and J.-Y. Pyun, ``Terahertz meets ai: The state of the art,'' \emph{Sensors}, vol.~23, no.~11, p. 5034, 2023.

\bibitem{alhulayil2025integrated}
M.~Alhulayil, M.~A. Aqoulah, M.~L{\'o}pez-Ben{\'\i}tez, M.~F. Al-Mistarihi, M.~Alammar, and A.~Al~Ayidh, ``Integrated thz/mmwave transmission method for enhanced urllc communications,'' \emph{IEEE Access}, 2025.

\bibitem{amodu2024technical}
O.~A. Amodu, R.~Nordin, N.~F. Abdullah, S.~A. Busari, I.~Otung, M.~Ali, M.~Behjati \emph{et~al.}, ``Technical advancements towards ris-assisted ntn-based thz communication for 6g and beyond,'' \emph{IEEE Access}, 2024.

\bibitem{nadas2019performance}
J.~P.~B. Nadas, O.~Onireti, R.~D. Souza, H.~Alves, G.~Brante, and M.~A. Imran, ``Performance analysis of hybrid arq for ultra-reliable low latency communications,'' \emph{IEEE Sensors Journal}, vol.~19, no.~9, pp. 3521--3531, 2019.

\bibitem{wu2024urllc}
Q.~Wu, W.~Wang, P.~Fan, Q.~Fan, J.~Wang, and K.~B. Letaief, ``Urllc-awared resource allocation for heterogeneous vehicular edge computing,'' \emph{IEEE Transactions on Vehicular Technology}, 2024.

\bibitem{mertes2024evaluation}
J.~Mertes, M.~Schmitz, D.~Lindenschmitt, C.~Schellenberger, M.~Klar, B.~Ravani, H.~D. Schotten, and J.~C. Aurich, ``Evaluation of 5g-based closed-loop control on part quality for milling processes,'' \emph{Digital Engineering}, vol.~3, p. 100024, 2024.

\bibitem{li2024reliability}
L.~Li, W.~Chen, P.~Popovski, and K.~B. Letaief, ``Reliability-latency-rate tradeoff in low-latency communications with finite-blocklength coding,'' \emph{IEEE Transactions on Information Theory}, 2024.

\bibitem{singh2024toward}
G.~Singh, A.~Srivastava, V.~A. Bohara, M.~Noor-A-Rahim, Z.~Liu, and D.~Pesch, ``Toward 6g-v2x: Aggregated rf-vlc for ultra-reliable and low-latency autonomous driving,'' \emph{IEEE Communications Standards Magazine}, vol.~8, no.~4, pp. 80--87, 2024.

\bibitem{rahimi2021design}
H.~Rahimi, Y.~Picaud, K.~D. Singh, G.~Madhusudan, S.~Costanzo, and O.~Boissier, ``Design and simulation of a hybrid architecture for edge computing in 5g and beyond,'' \emph{IEEE Transactions on Computers}, vol.~70, no.~8, pp. 1213--1224, 2021.

\bibitem{yuan2023digital}
X.~Yuan, J.~Du, Y.~Yao, J.~Xu, G.~Wu, N.~Culligan, R.~Bierig, and T.~Bi, ``Digital twins: A vehicle platooning simulation system for vr,'' in \emph{2023 IEEE Smart World Congress (SWC)}.\hskip 1em plus 0.5em minus 0.4em\relax IEEE, 2023, pp. 1--6.

\bibitem{kokkinis2025deep}
G.~Kokkinis, A.~Iosifidis, and Q.~Zhang, ``Deep reinforcement learning-based video-haptic radio resource slicing in tactile internet,'' \emph{arXiv preprint arXiv:2503.14066}, 2025.

\bibitem{crespo2024flexible}
M.~Crespo-Aguado, R.~Lozano, F.~Hernandez-Gobertti, N.~Molner, and D.~Gomez-Barquero, ``Flexible hyper-distributed iot--edge--cloud platform for real-time digital twin applications on 6g-intended testbeds for logistics and industry,'' \emph{Future Internet}, vol.~16, no.~11, p. 431, 2024.

\bibitem{peng2022real}
Y.~Peng, A.~Jolfaei, Q.~Hua, W.-L. Shang, and K.~Yu, ``Real-time transmission optimization for edge computing in industrial cyber-physical systems,'' \emph{IEEE Transactions on Industrial Informatics}, vol.~18, no.~12, pp. 9292--9301, 2022.

\bibitem{kalem20245g}
G.~Kalem, S.~Kosu, and M.~Basaran, ``5g/6g technology capabilities designed for secure edge network: Smart city use cases of turkcell,'' in \emph{2024 6th International Conference on Blockchain Computing and Applications (BCCA)}.\hskip 1em plus 0.5em minus 0.4em\relax IEEE, 2024, pp. 807--811.

\bibitem{erbati2023service}
M.~M. Erbati, M.~M. Tajiki, and G.~Schiele, ``Service function chaining to support ultra-low latency communication in nfv,'' \emph{Electronics}, vol.~12, no.~18, p. 3843, 2023.

\bibitem{li2020deep}
J.~Li and X.~Zhang, ``Deep reinforcement learning-based joint scheduling of embb and urllc in 5g networks,'' \emph{IEEE Wireless Communications Letters}, vol.~9, no.~9, pp. 1543--1546, 2020.

\bibitem{li2023low}
W.~Li, M.~Yan, and C.~A. Chan, ``A low-latency terminal mutual access architecture based on 5g lan cross-domain networking,'' in \emph{2023 IEEE 11th International Conference on Computer Science and Network Technology (ICCSNT)}.\hskip 1em plus 0.5em minus 0.4em\relax IEEE, 2023, pp. 69--74.

\bibitem{shih2016enabling}
Y.-Y. Shih, W.-H. Chung, A.-C. Pang, T.-C. Chiu, and H.-Y. Wei, ``Enabling low-latency applications in fog-radio access networks,'' \emph{IEEE network}, vol.~31, no.~1, pp. 52--58, 2016.

\bibitem{ji2022massive}
Y.~Ji, K.~Yu, J.~Qiu, and J.~Yu, ``Massive mimo and secrecy guard zone based improving physical layer security in uav-enabled urllc networks,'' \emph{IEEE Transactions on Vehicular Technology}, vol.~72, no.~4, pp. 4553--4567, 2022.

\bibitem{wang2019low}
X.~Wang, Y.~Ji, J.~Zhang, L.~Bai, and M.~Zhang, ``Low-latency oriented network planning for mec-enabled wdm-pon based fiber-wireless access networks,'' \emph{IEEE Access}, vol.~7, pp. 183\,383--183\,395, 2019.

\bibitem{li2023cross}
K.~Li, P.~Zhu, Y.~Wang, J.~Wang, and X.~You, ``Cross-layer resource allocation for urllc industrial automation over multi-connectivity,'' \emph{IEEE Transactions on Wireless Communications}, vol.~23, no.~7, pp. 7334--7348, 2023.

\bibitem{ranjha2024consumer}
A.~Ranjha, D.~Naboulsi, M.~El~Emary, and F.~Gagnon, ``Consumer-centric sustainability: Empowering urllc in multi-uav-assisted mec systems for industry 5.0,'' \emph{IEEE Transactions on Consumer Electronics}, 2024.

\bibitem{narsani2023leveraging}
H.~K. Narsani, A.~Ranjha, K.~Dev, F.~H. Memon, and N.~M.~F. Qureshi, ``Leveraging uav-assisted communications to improve secrecy for urllc in 6g systems,'' \emph{Digital Communications and Networks}, vol.~9, no.~6, pp. 1458--1464, 2023.

\bibitem{noor2022toward}
M.~Noor-A-Rahim, F.~Firyaguna, J.~John, M.~O. Khyam, D.~Pesch, E.~Armstrong, H.~Claussen, and H.~V. Poor, ``Toward industry 5.0: Intelligent reflecting surface in smart manufacturing,'' \emph{IEEE Communications Magazine}, vol.~60, no.~10, pp. 72--78, 2022.

\bibitem{imtiaz2025digital}
H.~H. Imtiaz, W.~U. Khan, A.~K. Bashir, M.~Obayya, N.~Negm, A.~E. Yahya, and A.~K. Dutta, ``Digital twin enabled 6g mec networks: Computational energy efficiency for consumer iiot in industry 5.0,'' \emph{IEEE Transactions on Consumer Electronics}, 2025.

\bibitem{forrai2024application}
A.~Forrai, A.~Gali, and I.~Barosan, ``Application of digital twins for connected, cooperative and automated mobility,'' in \emph{2024 IEEE 3rd Conference on Information Technology and Data Science (CITDS)}.\hskip 1em plus 0.5em minus 0.4em\relax IEEE, 2024, pp. 1--7.

\bibitem{qu2023digital}
Q.~Qu, S.~Ogunbunmi, M.~Hatami, R.~Xu, Y.~Chen, G.~Chen, and E.~Blasch, ``A digital twins enabled reputation system for microchain-based uav networks,'' in \emph{2023 IEEE 12th International Conference on Cloud Networking (CloudNet)}.\hskip 1em plus 0.5em minus 0.4em\relax IEEE, 2023, pp. 428--432.

\bibitem{xiao2023research}
H.~Xiao, Y.~Zhang, C.~Tang, L.~Zhang, W.~Lv, and S.~Xu, ``Research on cooperative control of uav cluster based on digital twin technology,'' in \emph{2023 IEEE International Conference on Signal Processing, Communications and Computing (ICSPCC)}.\hskip 1em plus 0.5em minus 0.4em\relax IEEE, 2023, pp. 1--5.

\bibitem{ramly2021cross}
A.~M. Ramly, N.~F. Abdullah, and R.~Nordin, ``Cross-layer design and performance analysis for ultra-reliable factory of the future based on 5g mobile networks,'' \emph{IEEE Access}, vol.~9, pp. 68\,161--68\,175, 2021.

\bibitem{gao2023multi}
X.~Gao, W.~Yi, Y.~Liu, and L.~Hanzo, ``Multi-objective optimization of urllc-based metaverse services,'' \emph{IEEE Transactions on Communications}, vol.~71, no.~11, pp. 6745--6761, 2023.

\bibitem{wang2024deterministic}
C.~Wang, Y.~Peng, J.~Wu, L.~Liu, S.~Mumtaz, M.~Dong, and M.~Guizani, ``Deterministic scheduling and reliable routing for smart ocean services in maritime internet of things: A cross-layer approach,'' \emph{IEEE Transactions on Services Computing}, 2024.

\bibitem{setayesh2022resource}
M.~Setayesh, S.~Bahrami, and V.~W. Wong, ``Resource slicing for embb and urllc services in radio access network using hierarchical deep learning,'' \emph{IEEE Transactions on Wireless Communications}, vol.~21, no.~11, pp. 8950--8966, 2022.

\bibitem{gengtian2025deep}
S.~Gengtian, J.~Liu, and S.~Shimamoto, ``Deep q-network for optimizing noma-aided resource allocation in smart factories with urllc constraints,'' in \emph{2025 IEEE Wireless Communications and Networking Conference (WCNC)}.\hskip 1em plus 0.5em minus 0.4em\relax IEEE, 2025, pp. 1--6.

\bibitem{shokrnezhad2024semantic}
M.~Shokrnezhad, H.~Mazandarani, T.~Taleb, J.~Song, and R.~Li, ``Semantic revolution from communications to orchestration for 6g: Challenges, enablers, and research directions,'' \emph{IEEE Network}, vol.~38, no.~6, pp. 63--71, 2024.

\end{thebibliography}

\vspace{-33pt}

\begin{IEEEbiography}[{\includegraphics[width=1in,height=1.25in,clip,keepaspectratio]{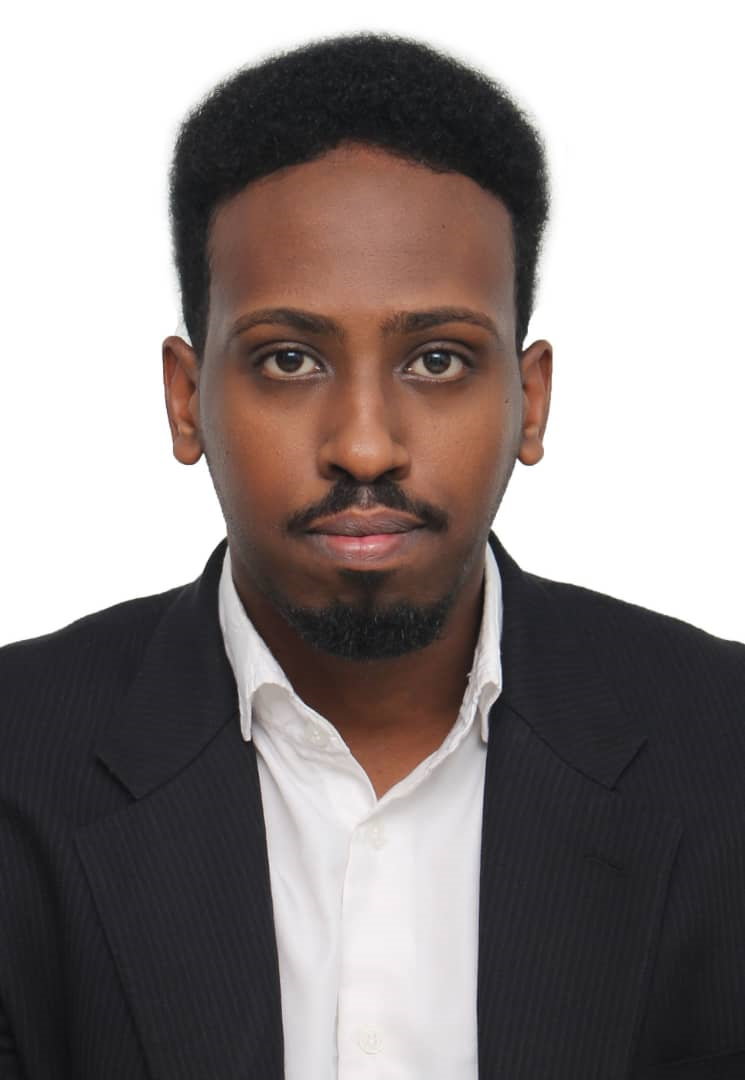}}]{Abdikarim Mohamed Ibrahim}
received the B.Eng. degree in Communication Engineering from the Academy of Engineering Sciences, Sudan, the M.Eng. degree in Electronics and Telecommunications Engineering from Universiti Teknologi Malaysia, and the Ph.D. degree in Computing from Sunway University, Malaysia, where he focused on multi-agent deep reinforcement learning for wireless networks. He is currently a Postdoctoral Research Fellow working on artificial intelligence-driven optimization for non-terrestrial networks (NTNs) targeting 6G systems. His research interests include applied machine learning, deep reinforcement learning, satellite communication, and intelligent wireless systems. He was awarded the Sunway University PhD Studentship by the Jeffrey Cheah Foundation.
\end{IEEEbiography}
\vspace{-33pt}

\begin{IEEEbiography}[{\includegraphics[width=1in,height=1.25in,clip,keepaspectratio]{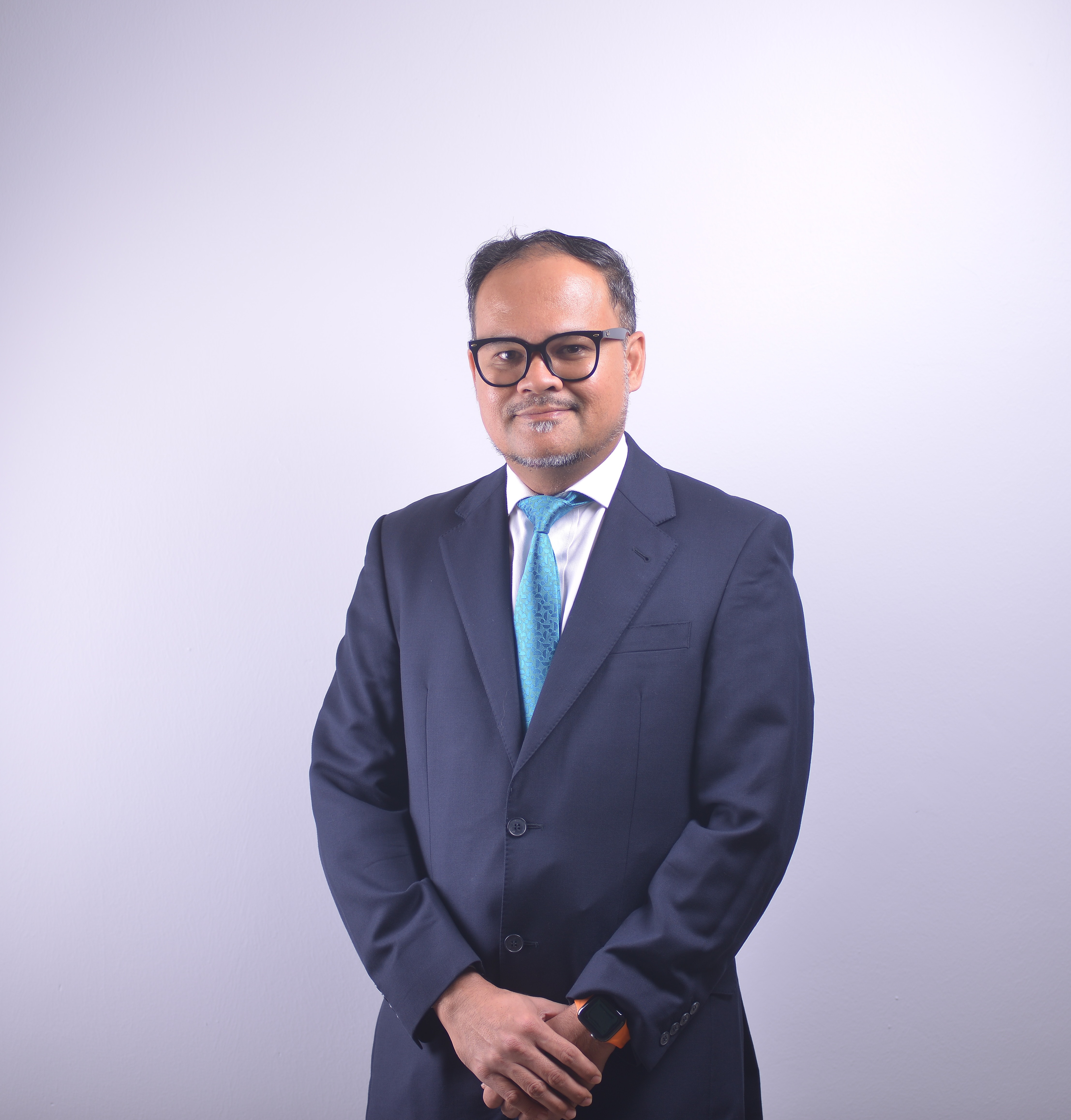}}]{Rosdiadee Nordin (M'10--SM'12)}
received the B.Eng. degree from Universiti Kebangsaan Malaysia in 2001 and the Ph.D. degree from the University of Bristol, U.K. in 2011. He is currently a Professor with the School of Engineering and Technology and Co-Director of Future Cities Research Institute, Sunway University, Malaysia. His research interests include beyond 5G/6G wireless communications, advanced transmission techniques, channel modeling, aerial communications and wireless communications for Internet of Things applications. He received the Leadership in Innovation Fellowship from the Royal Academy of Engineering, U.K., in 2021, and was named a Top Research Scientist Malaysia by the Academy of Sciences Malaysia in 2020.
\end{IEEEbiography}

\vspace{-33pt}

\begin{IEEEbiography}[{\includegraphics[width=1in,height=1.25in,clip,keepaspectratio]{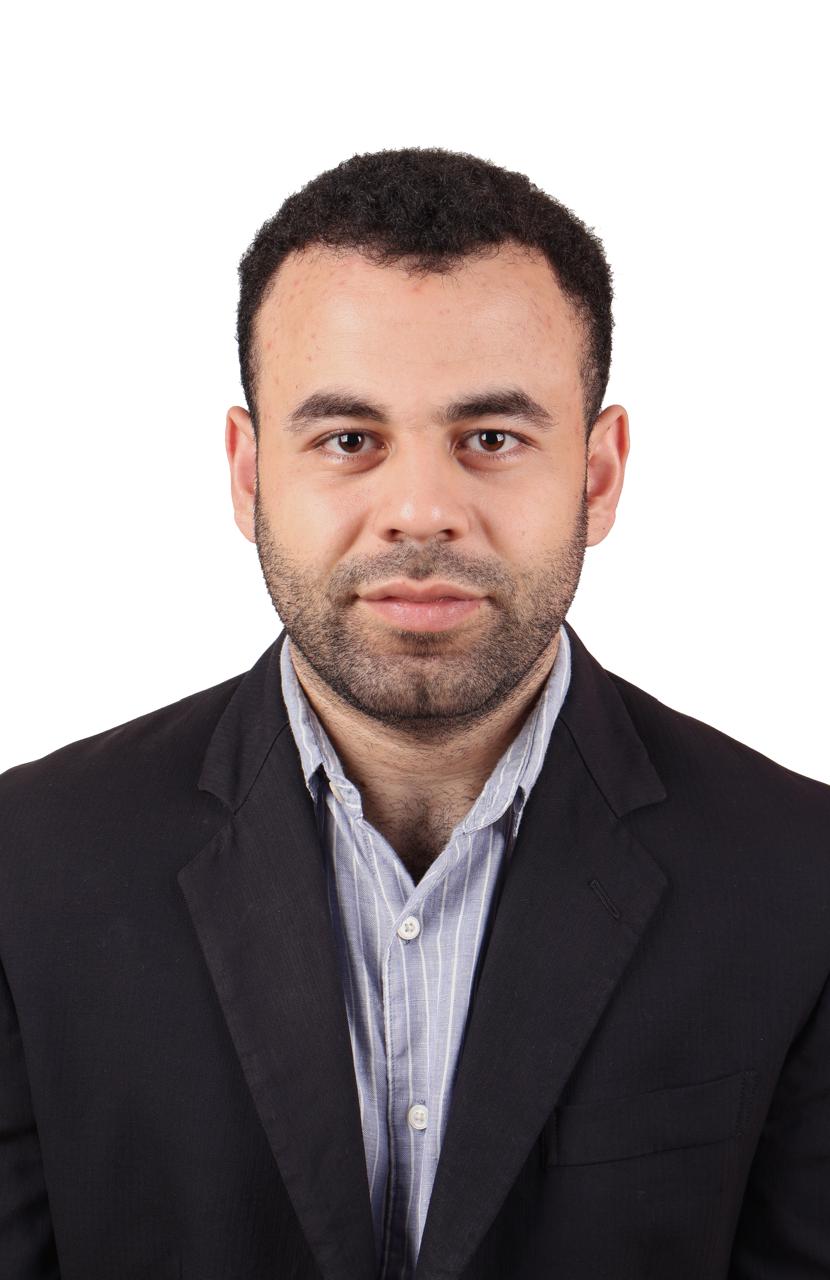}}]{Yahya Khamayseh} is currently a PhD student at Sunway University, Malaysia, where he focuses on Wireless 6G and Artificial Intelligence (AI). He earned his Master’s degree in Computer Science and Engineering from the University Malaysia of Computer Science and Engineering, graduating with honours. Before this, he completed his Bachelor’s degree at An-Najah National University in Nablus, Palestine, where he received the First Place Award in his department in 2018 and graduated with honours. His research interests are centred on integrating AI to enhance the performance, and efficiency of next-generation wireless networks, particularly in the context of 6G. In addition to his academic work, Khamayseh has over five years of experience as a backend engineer, specializing in designing scalable, high-performance systems.
\end{IEEEbiography}

\vspace{-33pt}

\begin{IEEEbiography}[{\includegraphics[width=1in,height=1.25in,clip,keepaspectratio]{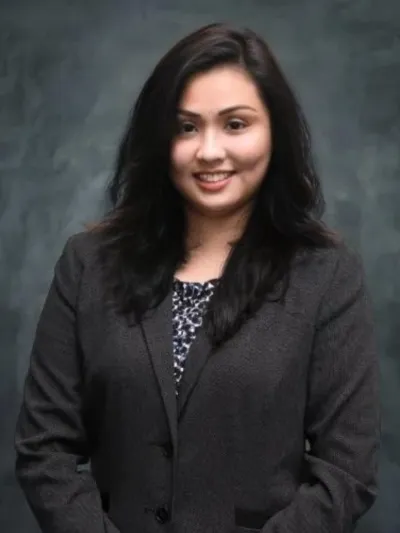}}]{Angela Amphawan}(Senior Member, IEEE) received the Ph.D. degree in optical communications and the Ph.D. degree in optical engineering from the University of Oxford, U.K. She then won the prestigious Fulbright Award to work on optical devices and networks at the Research Laboratory of Electronics and MIT Media Lab, Massachusetts Institute of Technology, USA. She is currently leading the Photonics Research Laboratory, School of Engineering and Technology, Sunway University. Prior to this, she was the Deputy Vice Chancellor of the University Malaysia of Computer Science and Engineering. Her research projects have been funded by U.S. Department of States, German Government, Malaysian Ministry of Education, and Telekom Malaysia. She has won several best paper awards and exhibition medals. She was a recipient of the Fulbright Award at the Research Laboratory of Electronics, Massachusetts Institute of Technology. She was the Publicity Co-Chair of the IEEE Wireless Communications and Networking Conference. She serves on the editorial board of the APL Photonics with the American Institute of Physics and is on the National 5G Task Force.
\end{IEEEbiography}

\vspace{-33pt}

\begin{IEEEbiography}[{\includegraphics[width=1in,height=1.25in,clip,keepaspectratio]{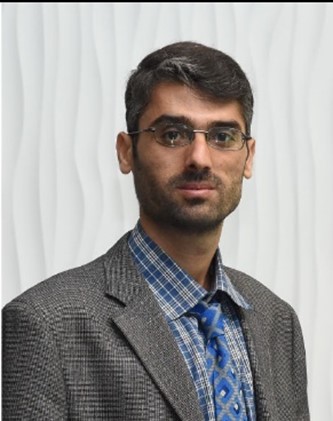}}]{Muhammed Basheer Jasser}
is an associate professor and the program leader of the Master in Advanced Computing in the Department of Data Science and Artificial Intelligence, School of Computing and Artificial Intelligence, Faculty of Engineering and Technology at Sunway University, Malaysia. He has obtained his Ph.D and Master's degree in computer science and software engineering from the University Putra Malaysia (UPM) (A research-intensive university ranked in the top 150 universities worldwide according to the QS ranking). He was granted the Malaysian Technical Cooperation Program scholarship (MTCP) from the Ministry of Higher Education (Malaysia) for his postgraduate studies. Dr. Basheer's major research interests include optimization algorithms, evolutionary computation, model-driven software engineering, formal specification, verification and theorem proving, artificial intelligence, and machine learning. Dr Basheer has extensive experience in research, teaching, and management for more than 15 years. Dr. Basheer is working on several fundamental and industrial research projects in the area of artificial intelligence and software engineering funded by several companies and universities. Several postgraduate (Ph.D \& Masters) students have graduated and have been working under his supervision on these projects. He is also a senior member of several professional academic bodies, including IEEE (Institute of Electrical and Electronics Engineers), the Institute of Electronics, Information and Communication Engineers (IEICE), and the Formal Methods Europe organization. Dr. Basheer has published and presented over 100 research articles (journals and conferences) indexed in Scopus and Web of Science (WoS) and published by prestigious publishers (e.g. Elsevier, Springer, IEEE). He received several awards among which best papers awards at IEEE conferences and the Student Appreciation of Teaching Award. He was also invited to be a keynote speaker at some IEEE conferences. He also serves as an editor and reviewer in several top Q1/Q2 WoS-indexed journals. 
\end{IEEEbiography}

\vspace{-33pt}

\vfill
\end{document}